\documentclass[12pt, a4paper]{article}
\usepackage[utf8]{inputenc}
\usepackage[left=2cm, right=2cm, bottom=2cm, top=2cm]{geometry}
\usepackage{bm}
\usepackage{amsmath, amsfonts}
\usepackage{url}
\usepackage{booktabs}
\usepackage{graphicx}
\usepackage{caption}
\usepackage{subcaption}
\usepackage{hyperref}
\usepackage[capitalise]{cleveref}
\usepackage{csvsimple}
\usepackage{authblk}
\usepackage{mathtools}
\usepackage{float}
\usepackage{xcolor}
\usepackage{nameref}

\usepackage[square, numbers, comma, sort&compress]{natbib}

\title{Impact of complex spatial population structure on early and long-term adaptation in rugged fitness landscapes}
\author{Richard Servajean\textsuperscript{1,2}, Arthur Alexandre\textsuperscript{1,2}, Anne-Florence Bitbol\textsuperscript{1,2,*}}
\affil{\textbf{1} Institute of Bioengineering, School of Life Sciences, École Polytechnique Fédérale de Lausanne (EPFL), CH-1015 Lausanne, Switzerland\

\textbf{2} SIB Swiss Institute of Bioinformatics, CH-1015 Lausanne, Switzerland\
*Email: anne-florence.bitbol@epfl.ch}
\date{}

\begin{document}

\maketitle

\begin{abstract}
We study how rugged fitness landscapes are explored by spatially structured populations with demes on the nodes of a graph, connected by migrations. In the weak mutation and rare migration regime, we find that, in most landscapes, migration asymmetries associated with some suppression of natural selection allow the population to reach higher fitness peaks first. In this sense, suppression of selection can make early adaptation more efficient. However, the time it takes to reach the first fitness peak is then increased. We also find that suppression of selection tends to enhance finite-size effects. Finite structures can adapt more efficiently than very large ones, especially in high-dimensional fitness landscapes. We extend our study to frequent migrations, suggesting that our conclusions hold in this regime. We then investigate the impact of spatial structure with rare migrations on long-term evolution by studying the steady state of the population with weak mutation, and introducing an associated steady-state effective population size. We find that suppression of selection is associated to small steady-state effective population sizes, and thus to small average steady-state fitnesses.
\end{abstract}

\section*{Introduction}

Natural microbial populations often possess complex spatial structures. For instance, pathogens are transmitted between hosts during epidemics~\cite{Bertels19}, and the gut microbiota~\cite{Donaldson16} and the soil microbiota~\cite{Mod21} inhabit environments with complex spatial structures. A simple and generic model of spatially structured populations considers well-mixed demes (i.e.\ subpopulations) connected by migrations~\cite{Wright31,Kimura64,Nagylaki80,Whitlock97,Nordborg02,Sjodin05,Houchmandzadeh11,Houchmandzadeh13,Constable14,Marrec21,yagoobi2021fixation,Yagoobi23,abbara2023frequent}. Spatial structure can impact the way populations evolve, in particular the probability that a mutant fixes, i.e.\ takes over, in the population. If migrations are symmetric enough to maintain the overall mutant fraction, the fixation probability of a mutant is not impacted by spatial structure~\cite{maruyama70, maruyama74}, unless deme extinctions occur~\cite{Barton93,whitlock2003}, and while fixation time can be impacted~\cite{slatkin_fixation_1981,Sudbrack24}. However, more complex spatial structures on graphs can impact mutant fixation probability~\cite{Lieberman05,Antal06}. This effect was evidenced in the framework of evolutionary graph theory, with update rules that specify how individuals divide, die and replace their neighbors on the graph, so that there is always exactly one individual per node~\cite{Lieberman05}. In this framework, a graph under a given update rule is said to amplify natural selection if it enhances the fixation probability of beneficial mutants and reduces that of deleterious ones, compared to a well-mixed population with the same size. A graph is said to suppress natural selection in the opposite case. In particular, the star graph amplifies natural selection under the birth-death update rule with selection on birth, but suppresses selection under the death-birth update rule with selection on birth~\cite{Kaveh15,Hindersin15,Pattni15}. Important impacts of update rules were also found in models with demes where migration is coupled to birth and death to exactly preserve deme size~\cite{Houchmandzadeh11,yagoobi2021fixation,Yagoobi23}. Models of structured populations allowing to decouple growth from migrations were recently developed~\cite{Marrec21,abbara2023frequent}. In these models, when migrations between demes on the nodes of a graph are rare enough not to affect mutant fixation or extinction within a deme, the asymmetry of migrations determines whether the star amplifies or suppresses natural selection~\cite{Marrec21}. Meanwhile, when migrations become more frequent, suppression of selection becomes pervasive in graphs with asymmetric migrations~\cite{abbara2023frequent}. These results do not depend on the choice of an update rule, but on experimentally relevant quantities such as migration intensity and asymmetry. Here, we ask how spatial structure, described by these recent models~\cite{Marrec21,abbara2023frequent}, impacts the evolution of a population, beyond the fate of one mutation. 

In a constant environment, the evolution of a population can be described as an exploration of its fitness landscape, which maps each genotype to a fitness, representing reproductive success~\cite{wright1932,smith1970}. Fitness landscapes are often rugged and possess multiple peaks (i.e., local maxima, corresponding to genotypes that are locally best adapted)~\cite{Dawid10, Bloom10,Kryazhimskiy11b,szendro2013b,Draghi13,Covert13,bank2016,visser2018,Fragata19}, due to interactions between genes known as epistasis~\cite{visser2011,poelwijk2011,visser2014,visser2018}. 
Such topographical properties of fitness landscapes affect the predictability of evolution~\cite{visser2014,bank2016,visser2018}. 
When mutations are rare, a mutation fixes or goes extinct before a new one occurs. Hence, most of the time, only one genotype, called the wild type, exists in the population \cite{visser2009, nowak2015}. 
When the lineage of a mutant fixes, it becomes the new wild type, and the population moves to another point of the fitness landscape. Evolution on a fitness landscape under rare mutations can thus be viewed as a biased random walk, where the fixation of a mutation constitutes an elementary step~\cite{orr2005}. In the strong selection regime~\cite{orr2005,gillespie1983}, which is appropriate for large enough populations~\cite{hartl1998}, only beneficial mutations fix and the population always goes uphill in the fitness landscape. In well-mixed populations, these trajectories, known as adaptive walks~\cite{orr2005}, have been extensively studied~\cite{orr2003minimum, Jain05, schoustra2009properties,Franke11,nowak2015, park2016greedy}. However, in smaller populations, neutral or deleterious mutations may fix~\cite{mccandlish2011,mccandlish2014,Weissman09}, due to strong genetic drift~\cite{ewens1979}. 
It was recently shown that finite population size can foster the early adaptation of a well-mixed population in a rugged fitness landscape~\cite{servajean2023}. Because demes can have small sizes, we hypothesize that these finite-size effects~\cite{servajean2023} may play important roles in the exploration of fitness landscapes by spatially structured populations. 

Historically, the impact of spatial structure on adaptation on rugged fitness landscapes was at the core of Wright's shifting balance theory~\cite{Wright31,wright1932}, which proposed that spatial structure could foster fitness valley crossing, thanks to enhanced genetic drift in individual demes. The crossing of fitness valleys was studied in detail in well-mixed populations~\cite{Weissman09,Weissman10}. The acceleration of fitness valley crossing by population subdivision was then quantified, considering demes all connected with the same migration rates~\cite{Bitbol14}. In more complex spatial structures, while much work has investigated the fixation of a mutant, more long-term evolution, involving several mutations, has only recently begun to be addressed. Fitness valley crossing was recently investigated in the framework of evolutionary graph theory with one individual per node of the graph~\cite{kuo_evolutionary_2024}. 
Also in this framework, Refs.~\cite{sharma_suppressors_2022,sharma_graph-structured_2024} showed that under specific update rules and mutant initializations, some graphs can reach higher or lower mean steady-state fitnesses than well-mixed populations. 
As for mutant fixation, these results are strongly impacted by the choice of the update rule.

Here, we investigate the exploration of rugged fitness landscapes by spatially structured populations with demes on the nodes of a graph, connected by rare migrations. To describe the fixation of each mutation, we employ the model from Ref.~\cite{Marrec21}, which bypasses update rules by making migrations, birth and death independent. We mainly focus on the star graph and on the line graph. An analytical expression of the fixation probability was obtained in Ref.~\cite{Marrec21} for the star, and we derive one for the line.  We show that, in most rugged landscapes, suppression of selection can make early adaptation more efficient. Indeed, starting from a random genotype, the first encountered peak tends to be higher on average with suppression. However, the time it takes to reach this first fitness peak is then increased. Besides, suppression enhances finite-size effects. We find that in some landscapes, early adaptation is more efficient for finite structures than that for very large ones. We extend our study to frequent migrations, using the model from Ref.~\cite{abbara2023frequent}. Our results suggest that these conclusions also hold in this regime where suppression of selection is pervasive. We also investigate the impact of spatial structure on long-term evolution with rare migrations by studying the steady state of the population. We define an effective population size for the steady-state distribution. We find that suppression of selection is associated to reduced steady-state effective population sizes, and thus to smaller mean fitnesses at steady state.

\section*{Model and methods}
\label{sec:model}

\subsection*{Modeling evolution in spatially structured populations}
\label{subsec:framework}

\paragraph{Genotypes and mutations.}
We consider populations of asexual haploid individuals, e.g.\ bacteria. We assume that each genotype possesses a fixed fitness, which defines a fitness landscape~\cite{wright1932,smith1970}. Note that we do not address environment variations or frequency-dependent selection~\cite{visser2018}. 
Genotypes are described by sequences of $L$ binary variables, taking values $0$ or $1$, which represent genetic units, such as nucleotides, amino acids or genes. Considering genotypes as strings of binary variables is a simplification~\cite{zagorski2016} that we make for simplicity, in line with many studies~\cite{szendro2013b, visser2014}. Mutations are modelled as substitutions from $0$ to $1$ or vice versa at one site. In this framework, genotype space is a hypercube with $2^L$ vertices, each of them having $L$ neighboring genotypes that are accessible by a single mutation. We assume that mutations appear in an individual 
chosen uniformly at random. This assumption is realistic for DNA replication errors, and for mutations induced by a stress that is uniform through the structure. 

\paragraph{Evolution in the weak mutation regime.}
We focus on the weak mutation regime, defined by $\mathcal{N}\mu \ll 1$, where $\mathcal{N}$ denotes total population size and $\mu$ denotes mutation rate (i.e., mutation probability per generation). In this regime, mutations are sufficiently rare for the fate of a mutation (either fixation or extinction) in the whole structured population to be set before any new mutation occurs. Thus, the population almost always comprises a single genotype. Note that phenomena like clonal interference~\cite{elana2003,Jain11,Laessig17,good2017} are thus not described. When a mutant genotype takes over (i.e., fixes), it becomes the new wild type. The evolution of the population can be modelled as a biased random walk in genotype space~\cite{orr2005}, and is driven by random mutations, natural selection and genetic drift. A mutation that fixes yields one step of this random walk, whereby the population hops from one genotype to another in the hypercube representing genotype space. This description is known as the origin-fixation approach~\cite{mccandlish2014}.

\begin{figure}[h!]
 \centering
 \includegraphics[width=0.75\textwidth]{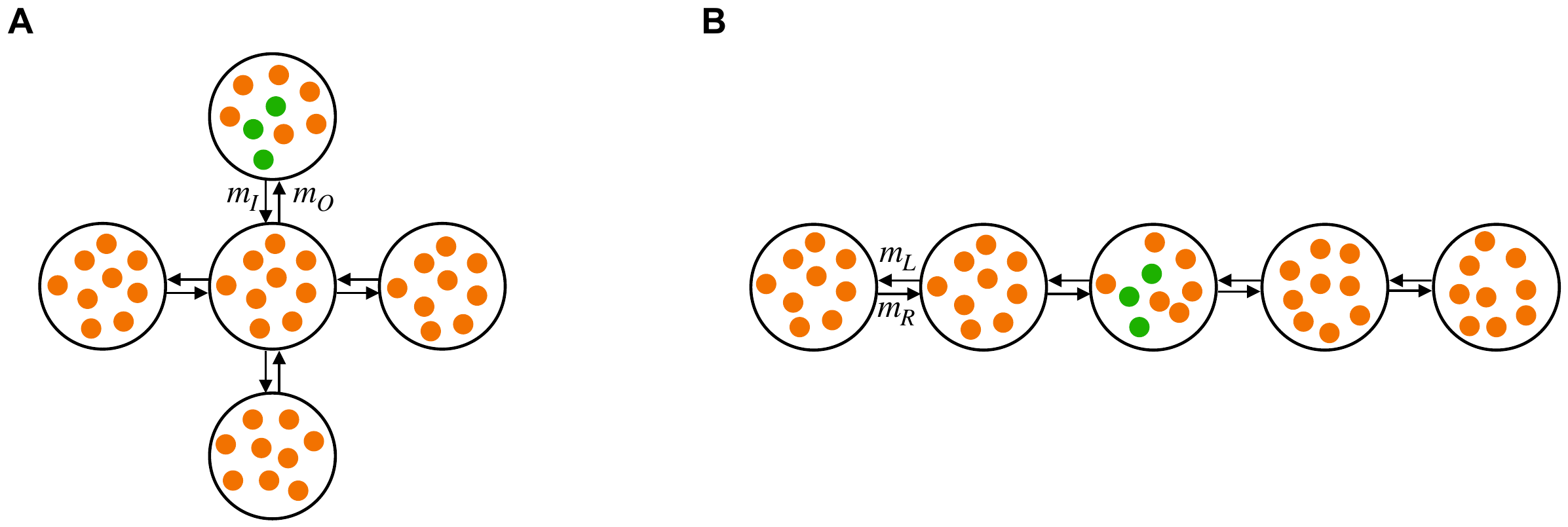}
 \caption{\textbf{Spatial structures investigated.} (A) Star; (B) Line. One well-mixed deme is placed on each node of a graph. Markers represent individuals (wild-type ones in orange, mutants in green). Demes are connected by migrations along the edges of the graphs, represented by arrows. In the star, the incoming and outgoing migration rates per individual to and from the center are denoted by $m_I$ and $m_O$, respectively. In the line, the left and right migration rates per individual are denoted by $m_L$ and $m_R$, respectively. }
\label{fig:structs}
\end{figure}

\paragraph{Modeling spatially structured populations.}
We consider spatially structured populations on graphs, using the model introduced in Refs.~\cite{Marrec21,abbara2023frequent}. This model considers $D$ well-mixed demes (i.e.\ subpopulations), each placed on one node of a graph, and migrations along the edges of this graph. All demes have the same carrying capacity $C$. Each genotype $i$ is characterized by its fitness $f_i$, which represents its division rate in the exponential phase. We only consider selection on birth. Migration events, where an individual moves from one deme to another, occur with rate $m_{ab}$ from deme $a$ to deme $b$. As in Ref.~\cite{Marrec21}, we focus on the rare migration regime, where each mutation either gets extinct or fixes within a deme before any migration event (see below). We then briefly extend our work to more frequent migrations, using the approach introduced in~\cite{abbara2023frequent}. We consider highly symmetric graphs, illustrated in \cref{fig:structs}. The star is a simple prototype of graphs with hubs, while the line is of interest for modeling one-dimensional or quasi-one-dimensional systems. We also briefly extend our work to a structure comprising two demes of different sizes. In these different structures, the fixation probability of a mutant under rare migrations can be expressed analytically, which strongly facilitates our investigations. 

\paragraph{Mutant fixation within a deme.} With demes of equal sizes, which are our main focus, and under uniform mutation, a single mutant appears in a deme chosen uniformly at random. In the rare migration regime, the first step of mutant fixation is fixation in the deme where the single mutant initially appeared. Then, mutants may spread to other demes by migrations.
The separation of timescales between fixation or extinction of a given mutant within a deme and migration phenomena allows to analytically express mutant fixation probabilities in highly symmetric structures. We assume that mutations are more rare than migrations (see below), so that all this process happens before another mutation arises.  
For completeness, we consider two models within demes: (i) an individual-based model that can be approximated by the Moran model~\cite{Marrec21}, and (ii)  the Wright-Fisher model. 

First, in the individual-based model of Ref.~\cite{Marrec21}, individuals with genotype $i$ have a division rate $f_i(1-N/C)$ where $N$ denotes deme size. All individuals have the same death rate $g$. Disregarding migrations, the steady-state size of a deme is $C(1-g/f_i)$, obtained when death rate and division rate are equal. We focus on the regime where deme size fluctuates weakly around this steady-state value, i.e.\ we assume $g\ll f_i$ and $C\gg 1$. As a deme is well-mixed, with a size that weakly fluctuates around a steady-state value (see above), we approximate fixation within a deme by that in a population of constant size equal to its steady-state size, within the Moran model, see \cite{Marrec21}. 
In the Moran model, the fixation probability $\rho_{ij}$ of the lineage of a single mutant with genotype $j$ in a population where all other individuals are wild-types with genotype $i$ reads \cite{ewens1979, moran1958}:
\begin{equation}\label{eq:fix_deme_moran}
\rho_{ij} = \frac{1-(f_i/f_j)}{1-(f_i/f_j)^{N_i}} = \frac{1-(1+s_{ij})^{-1}}{1-(1+s_{ij})^{-N_i}}\,,
\end{equation}
where $s_{ij} = f_j / f_i - 1$ is the relative fitness advantage of the mutant and $N_i = C(1-g/f_i)$ is the steady-state size of the deme composed of $N_i$ individuals. Note that in the regime we consider, $N_i\approx C$.

Second, in the Wright-Fisher model under the diffusion approximation, we have \cite{Kimura1962, crow2009}:
\begin{equation}\label{eq:fix_deme_wf}
\rho_{ij} = \frac{1-e^{-2s_{ij}}}{1-e^{-2N_is_{ij}}}\,.
\end{equation}

\paragraph{Circulations.} Circulations are graphs where, for each deme, the total incoming migration rate is equal to the total outgoing migration rate~\cite{Lieberman05,Marrec21,abbara2023frequent}. One example is the clique, or fully connected graph, or island model~\cite{Wright31}, where each deme is connected to all others and the migration rate is the same between any two demes. In a circulation in the rare migration regime, the fixation probability of the lineage of a single mutant with genotype $j$ in a population where all other individuals are wild-types with genotype $i$ reads~\cite{Marrec21}: 
\begin{equation}\label{eq:clique}
\phi_{ij} = \rho_{ij}\frac{1-\gamma}{1-\gamma^D},
\end{equation}
where we introduced $\gamma = N_i\rho_{ji} / (N_j\rho_{ij})$ with $N_j = C(1-g/f_j)$, and where $\rho_{ij}$ denotes the fixation probability of type $j$ in a deme full of individuals of type $i$ (see paragraph above). 

Note that the fact that the fixation probability is the same for all circulation graphs extends beyond the rare migration regime~\cite{abbara2023frequent}. This is consistent with Maruyama's theorem~\cite{maruyama70,maruyama74} and with the circulation theorem in models with one individual per node of the graph~\cite{Lieberman05}. Furthermore, the fixation probability in \cref{eq:clique} is almost the same as that of a well-mixed population with the same total size~\cite{Marrec21}. In other words, for circulations, spatial structure has almost no effect on mutant fixation probability.

\paragraph{Star.} In a star in the rare migration regime, the fixation probability of one mutant appearing in a deme chosen uniformly at random is~\cite{Marrec21}:
\begin{equation}\label{eq:star}
\phi_{ij} = \rho_{ij} \frac{(1-\gamma^2)[\gamma + \alpha D + \gamma\alpha^2(D-1)]}{D(\alpha+\gamma)[1+\alpha\gamma -\gamma^D(\alpha+\gamma)^{2-D}(1+\alpha\gamma)^{D-1}]},
\end{equation}
where $\alpha$ represents migration asymmetry, and is defined as the ratio $\alpha=m_I/m_O$ of the incoming migration rate $m_I$ from a leaf to the center to the outgoing migration rate $m_O$ from the center to a leaf (see \cref{fig:structs}). The star is a suppressor of selection when $\alpha < 1$, and an amplifier when $\alpha > 1$ \cite{Marrec21}. As a reminder, in a suppressor of selection, the fixation probability of beneficial mutations is smaller than in a well-mixed population with same size, and the opposite holds for deleterious mutations. Meanwhile, in an amplifier of selection, the fixation probability of beneficial mutations is larger than in a well-mixed population with same size, and the opposite holds for deleterious mutations. When $\alpha = 1$, the star is a circulation, with fixation probability given by \cref{eq:clique} \cite{Marrec21, abbara2023frequent}. 

\paragraph{Line.} Let us now turn to a line of demes, which corresponds to the linear stepping stone model~\cite{Kimura64}. Migration rates to the left and to the right can differ, see \cref{fig:structs}. In \cref{sec:fixation_line} of the Supplementary Material, we derive the fixation probability of one mutant appearing in a deme chosen uniformly at random within a line of demes, using Refs.~\cite{miller1994matrix,broom2008analysis}. We obtain:
\begin{align}\label{eq:line}
	&\phi_{ij}= \frac{\rho_{ij}}{D} \left\{\dfrac{1 -{\gamma}/{\alpha} }{1 - \left({\gamma}/{\alpha} \right)^{D}} + \dfrac{1 -\alpha\gamma }{1 - (\alpha\gamma)^{D}} \right. \\ &\left. + \frac{4 }{D^2(1+\alpha)(1+\gamma)}\sum_{k = 1}^{D-2} \sum_{n = 1}^{D-1} \gamma^{1-\frac{n}{2}}\left[ \dfrac{1 -(\alpha\gamma)^{n} }{1 - (\alpha\gamma)^{D}} \alpha^{k+1-\frac{n}{2}} + \dfrac{1 - \left(\frac{\gamma}{ \alpha} \right)^{n}}{1 -  \left(\frac{\gamma}{ \alpha}\right)^{D}} \alpha^{\frac{n}{2}-k}\right] \nonumber\left( t^{(1,n)}_{k,k+1} - t^{(n,1)}_{k,k+1}  \right) \right\}\,,
\end{align}
with \begin{align}
t^{(a,b)}_{k,k+1} 
& = \sum_{r = 1}^{D-1}\sum_{s = 1}^{D-1} \frac{\sin\left(\frac{k r \pi}{D} \right) \sin\left(\frac{a r \pi}{D} \right) \sin\left(\frac{b s \pi}{D} \right) \sin\left(\frac{(k+1) s \pi}{D} \right)}{1- \frac{2 \sqrt{\alpha \gamma}}{(1+ \alpha)(1+\gamma)} \left[ \cos\left(\frac{r \pi}{D} \right) + \cos\left(\frac{s \pi}{D} \right) \right]}\,.
\end{align}
Notations are as above, except that the migration asymmetry is defined as $\alpha = m_R / m_L$, where $m_R$ (resp.\ $m_L$) is the migration rate from a deme to its right (resp.\ left) neighbor. Like the star, when $\alpha = 1$, the line becomes a circulation, with fixation probability given by \cref{eq:clique}. The more $\alpha$ deviates from 1, the more the line suppresses natural selection, see \cref{fig:fixation_probability_line}. Note that by symmetry, the fixation probability is the same for asymmetries $\alpha$ and $1/\alpha$. 

\paragraph{Rare migration regime.} The rare migration regime is such that migrations happen on a slower timescale than the process of mutant fixation within a deme, starting from one mutant destined to fix~\cite{slatkin_fixation_1981,hauert_fixation_2014,Marrec21}. Indeed, if this holds, fixation of mutants within a deme is not perturbed by wild-type bacteria migrating from other demes, and can be described as if the deme was isolated. Then, the subsequent possible spread of mutants to other demes can be described by assuming that each deme is either fully mutant or fully wild-type, without having to consider mixed demes~\cite{slatkin_fixation_1981,hauert_fixation_2014,Marrec21}. This substantially simplifies the theoretical description of the fixation process, and allows for exact analytical calculations of fixation probabilities in highly symmetric graphs~\cite{Marrec21}. Using the Moran process result~\cite{ewens1979}, the average fixation time of a neutral mutant destined to fix in a well-mixed deme of fixed size $N^*$ is $(N^*-1)/g$. In our model, deme sizes are not fixed, but we focus on regimes where they weakly fluctuates around a steady-state value (see above), and thus taking $N^*=N_i=C(1-g/f_i)$ (which is close to $C$ in the regime of interest) gives a good approximation \cite{Marrec21}. In addition, fixation is shorter for non-neutral mutants~\cite{teimouri2019,ewens1979}, meaning that the neutral value is an upper bound. To assess the migration timescale, let us consider the simplest spatial structure of the clique, a circulation composed of $D$ demes connected to one another with identical migration rates $m$ per individual. In the clique, the expected time until a wild-type individual migrates into the deme where mutants are in the process of taking over is $1/[N_i(D-1)m]$. Thus, for incoming migrations not to strongly perturb local fixations, we need $(N_i-1)/g\ll1/[N_i(D-1)m]$, i.e.
\begin{equation}
    \frac{m}{g}\ll\frac{1}{(N_i-1)N_i(D-1)}\,.
    \label{eq:raremigr}
\end{equation}
This equation defines the rare migration regime for the clique. Similar equations can be written for other structures, by replacing the total inflow $N_i(D-1)m$ of migrations to a deme by the maximum one in the structure, namely $N_i\textrm{max}\left[(D-1)m_I,m_O\right]$ in the star, or $N_i(m_R+m_L)$ in the line. 

Note that the condition in \cref{eq:raremigr} is stringent, since most incoming wild-type migrants would not perturb the fixation process, as their lineage would rapidly go extinct in their destination deme. A less stringent condition would be to require that incoming migrations do not strongly perturb local fixation processes, by focusing on lineages of wild-types destined to locally fix in the deme they migrate to~\cite{fruet2024}. However, here, we retain the stringent condition above, to ensure that the description with separated timescales of local fixation and migration is fully valid. 

Importantly, while we need migration timescales to be slower than the fixation process of a mutant lineage destined to fix within a deme, we do not need them to be slower than the time of appearance of a mutant destined to fix. The latter condition would be much more demanding, especially for substantially deleterious mutants. Indeed, the average time until a mutant of type $j$ that is destined to fix in a deme of type $i$ individuals appears is given by $1/(N_iD\mu \rho_{ij})$~\cite{desai_beneficial_2007,Weissman09}. For a mutation with fitness cost $\delta_{ij}=-s_{ij}>0$ such that $N_i\delta_{ij}\gg 1$ but $\delta_{ij}\ll 1$, \cref{eq:fix_deme_moran,eq:fix_deme_wf} show that this timescale is proportional to $e^{AN_i\delta_{ij}}$ with $A=1$ (resp.\ 2) for the Moran (resp.\ Wright-Fisher) model. For substantially deleterious mutants, this time is much longer than the time taken by the fixation process itself, and is the dominant contribution~\cite{desai_beneficial_2007,Weissman09} to the total time taken by the change of fixed genotype in a deme~\cite{kimura_average_1980}. However, migrations during this time have little impact, as they overwhelmingly move wild-type individuals around.

\paragraph{Weak mutation and rare migration regime.} As mentioned above, we assume weak mutations, i.e.\ mutations sufficiently rare for the fate of a mutation in the whole structured population to be set before any new mutation occurs. In addition to the condition $N_i D\mu \ll 1$ (see ``Evolution in the weak mutation regime'' above), this entails another one that compares mutation and migration timescales. Let us consider the clique again for simplicity. The average fixation time of a neutral mutant destined to fix in the whole population is $(D-1)/(Dm)$ under rare migrations~\cite{Bitbol14,MarrecP}, and it is again longer than those in cases with selection. In fact, the average fixation time of a mutant with selective advantage $s$ (relative fitness $1+s$) is identical to that of a mutant with selective disadvantage $s$ (relative fitness $1-s$). This can be shown explicitly by using the general expression of the average fixation time in a clique~\cite{Bitbol14,MarrecP} and the fixation probability within a deme in the diffusion approximation (either for the Moran process or for the Wright-Fisher process), and is reminiscent of the well-mixed population fixation time~\cite{maruyama1974b}. Hence, the neutral case is again an upper bound. To be in the weak mutation regime, we need the additional condition $(D-1)/(Dm)\ll1/(N_iD\mu)$ where $N_iD\mu$ is the total mutation rate in the population. This yields
\begin{equation}
    \frac{\mu}{m}\ll\frac{1}{N_i(D-1)}\,.
\end{equation}

\paragraph{Beyond rare migrations.} In the rare migration regime, the separation of timescales allows one to treat the fixation or extinction of a given mutant within a deme separately from mutation spread across demes. This simplified treatment is not possible for more frequent migrations. In addition, in our individual-based model, under frequent asymmetric migrations, some deme steady-state sizes may exceed $C$, which prevents division there. This may or may not be realistic, and strongly depends on how deme size regulation is implemented (recall that we implemented it on division rate). Wright-Fisher sampling does not lead to such issues, but requires choosing how sampling is coupled to spatial structure. 

Ref.~\cite{abbara2023frequent} introduced a serial dilution model, where demes undergo local exponential growth followed by events of dilution and migration, modeling serial passage with spatial structure. This model is equivalent to a structured Wright-Fisher model when choosing a growth time $t=1$ \cite{abbara2023frequent}. Futhermore, Ref.~\cite{abbara2023frequent} considered different ways of coupling sampling to spatial structure, and showed that they gave similar results for large demes and moderate selection. Thus motivated, we employ the model of~\cite{abbara2023frequent} with growth time $t=1$. We use  multinomial sampling to perform dilution and migration events, and we choose the convention where all demes contribute by $C$ to the next bottleneck (see the Supplementary Material of \cite{abbara2023frequent} for details). Under rare migrations, this model gives results that are fully consistent with that of~\cite{Marrec21}, with the within-deme fixation probability given by \cref{eq:fix_deme_wf} instead of \cref{eq:fix_deme_moran}, see Ref.~\cite{abbara2023frequent}. Analytical expressions of the fixation probability beyond the rare migration regime are only available within the branching process approximation, which holds for slightly beneficial mutants \cite{abbara2023frequent}. Here, we aim to include neutral and deleterious ones. Hence, we estimate the fixation probability for the specific $f_i$ and $f_j$ in our fitness landscapes by performing simulations of the structured Wright-Fisher model.

\subsection*{Quantifying short-term and long-term adaptation of spatially structured populations}
\label{subsec:obs}

\paragraph{Exploration of fitness landscapes by spatially structured populations.} Here, we investigate the biased random walk performed in a fitness landscape by structured populations in the rare migration regime. For a star, the fixation probability of a mutant in the whole population is given by \cref{eq:star}, combined with \cref{eq:fix_deme_moran} if the Moran model is used to describe within-deme fixation (resp.\ with \cref{eq:fix_deme_wf} if the Wright-Fisher model is used). Let us call these walks a Moran star walk and a Wright-Fisher star walk, respectively. Let us use analogous names for the line: Moran line walk and Wright-Fisher line walk. Recall that in Ref.~\cite{servajean2023}, the Moran and the Wright-Fisher walks were studied in well-mixed populations.

As for well-mixed populations \cite{servajean2023}, the walks performed by structured populations on fitness landscapes are discrete-time Markov chains, where each time step corresponds to a mutation event. They are irreducible, aperiodic and positive recurrent (and thus ergodic). Hence, they possess a unique steady-state distribution, describing the probability that the population features a given genotype at steady state, towards which they converge for any initial condition~\cite{norris,aldous2002,sella2005,mccandlish2018}. 

Ref.~\cite{servajean2023} addressed the effect of finite population size $N$ on the exploration of fitness landscapes by well-mixed populations. Here, we will mainly study the impact of the number $D$ of demes, and of the migration asymmetry $\alpha$, which characterize spatial structure and have a significant impact on mutant fixation probability in the star and in the line. Indeed, in the rare migration regime, the star can amplify or suppress natural selection depending on the value of $\alpha$ (see above and Ref.~\cite{Marrec21}), and the intensity of amplification depends on $D$~\cite{Lieberman05,Marrec21}.
We will also investigate the effect of varying the carrying capacity $C$ of each deme, which impacts the amplitude of fluctuations within each deme.

\paragraph{Characterizing early adaptation.} In order to investigate the short-term evolution of a population, we will focus on the height $h$ of a walk, which is the fitness of the first encountered peak~\cite{nowak2015}, as in Ref.~\cite{servajean2023}. We consider its average $\bar{h}$ over initial genotypes chosen uniformly at random. It provides a global characterization of early adaptation by a structured population in the fitness landscape. Note that starting from a random genotype may be relevant to describe evolution after an environmental change, which makes the population ill-adapted to the current environment. 
Furthermore, for ensembles of landscapes in a given class, we study the corresponding ensemble mean heights $\left<\bar{h}\right>$. 

In addition to the height $h$ of a walk, we examine its length $\bar{\ell}$ and time $\bar{t}$, which are defined respectively as the mean number of mutation fixation events, and of mutation events (leading to fixation or not), before the first fitness peak is reached. 

We obtain $\bar{h}$, $\bar{\ell}$ and $\bar{t}$ by two different methods: by numerically solving the equations from a first step analysis (FSA), and by numerical simulations. For details, see \cref{sec:fsa-si} of the Supplementary Material, and Ref.~\cite{servajean2023}. 

\paragraph{Characterizing steady state.} We are also interested in the impact of spatial structure on the long-term evolution of structured populations. As a first step to characterize it, we investigate the steady-state distribution associated to our walks on fitness landscapes. It provides the probability that the population features a given genotype after a sufficiently long time of evolution. Specifically, we define the steady-state effective population size of the population, as the size of a well-mixed population with the same steady-state distribution.

\subsection*{Fitness landscapes}

We investigate how spatially structured populations explore various fitness landscapes, both model ones and experimental ones.

\paragraph{$LK$ fitness landscapes.} 
In the $LK$ model (originally called $NK$ model \cite{kauffman1989}), the fitness of a genotype $\vec{\sigma} = (\sigma_1, \sigma_2, ..., \sigma_L) \in \{0, 1\}^L$ is expressed as:
\begin{equation} \label{eq:LK}
    f(\vec{\sigma}) = \sum_{i=1}^L f_i\left(\{\sigma_j\}_{j\in \nu_i}\right)\,,
\end{equation}
where $f_i$ is the contribution of site $i$, and $\nu_i$ comprises $i$ and its epistatic partners, which are the sites that interact with $i$. In this model, the first parameter $L$ is the number of (binary) units of the genome, i.e.\ the dimension of the hypercube of genotypes. The second parameter $K$ is the number of epistatic partners of each site~\cite{kauffman1989}. Thus, $\nu_i$ comprises $K+1$ elements, for all $i$. For each site, we consider that its epistatic partners are chosen uniformly at random among other sites: each site is said to have a `random neighbourhood' \cite{nowak2015}. Increasing $K$ increases epistatis, and yields more rugged landscapes. If $K = L-1$, the $LK$ model reduces to the House of Cards model~\cite{Kauffman1987, kingman1978} where the fitness values of each genotype are independent and identically distributed. Indeed, then, any mutation affects the contribution of all sites in \cref{eq:LK}. Conversely, when $K = 0$, landscapes are additive and display a single peak. The $LK$ fitness landscape model allows for easy tuning of ruggedness. Note that we draw the $f_i$ in a uniform distribution between 0 and 1.

\paragraph{Other fitness landscapes.} We consider two generalizations of the $LK$ model: the $LK$ model with non-binary genotype states~\cite{zagorski2016}, and a modified $LKp$ model \cite{barnett1998} where the contributions $f_i$ in \cref{eq:LK} have a probability $p$ to be set to a positive constant value to include neutral mutations (the constant is zero in the classic version; our modification avoids zero overall fitnesses). We also investigate the Rough Mount Fuji model \cite{aita2000, szendro2013b}, the Eggbox model \cite{ferretti2016}, the Ising model \cite{diu1989, ferretti2016}, and tradeoff-induced landscapes \cite{das2020, das2022driven}. Details about these models are given in the Supplementary Material of Ref.~\cite{servajean2023}. Finally, we also consider experimentally measured landscapes from \cite{hall2020} and \cite{lozovsky2009}. 

\section*{Results}
\label{sec:results}

\subsection*{Early adaptation on small $LK$ rugged landscapes}\label{subsec:small_landscapes}

\begin{figure}[h!]
 \centering
 \includegraphics[width=0.95\textwidth]{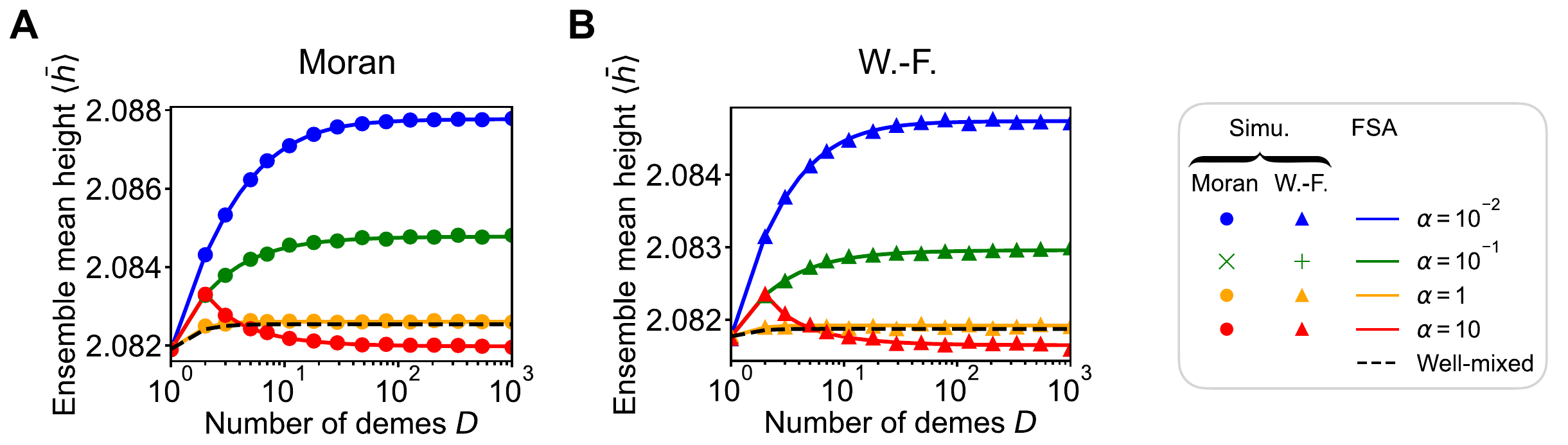}
 \caption{\textbf{Impact of the number of demes $D$ and migration asymmetry $\alpha$ of a star on early adaptation in $LK$ landscapes with $L=3$ and $K=1$.} (A) Ensemble mean height $\left\langle \bar{h} \right\rangle$ of the first fitness peak reached when starting from a uniformly chosen initial genotype versus the number of demes $D$ for the Moran star walk with rare migrations, with various values of migration asymmetry $\alpha$. Lines: numerical resolutions of the FSA \cref{eq:systemfsa,eq:barh} for each landscape; markers: simulation results averaged over $10$ walks per starting genotype in each landscape. In both cases, the ensemble average is performed over $2\times 10^5$ landscapes. Dashed black line: well-mixed case with population size $DC(1-g/f_W)$ where $f_W$ is the fitness of the wild type, shown as reference. (B) Same as (A) but for the Wright-Fisher star walk. Parameter values (both panels): $C = 20$ and $ g = 0.01$.}
\label{fig:mean_h_vs_D}
\end{figure}

\paragraph{Average trends of early adaptation over many small fitness landscapes.} 
In \cref{fig:mean_h_vs_D}, we study the mean height of biased random walks performed by star-structured populations, averaged over an ensemble of $2\times 10^5$ small $LK$ landscapes with $L=3$ and $K=1$. To assess the impact of spatial structure on early adaptation, we vary the number of demes $D$ in the star, for a fixed deme carrying capacity $C$. In \cref{fig:mean_h_vs_D}, we observe that the way the ensemble mean height $\left\langle \bar{h} \right\rangle$ varies with $D$ strongly depends on migration asymmetry $\alpha$. This shows that the star structure has a notable impact on early adaptation in rugged landscapes. For $\alpha=1$, i.e.\ when the star is a circulation, results are very close to those for the well-mixed population with the same total size. Indeed, the mutant fixation probabilities in circulations are extremely similar to those in well-mixed populations~\cite{Marrec21}. In \cref{fig:mean_h_vs_D}, we further observe that, for moderate to large $D$, the ensemble mean height $\left<\bar{h}\right>$ in the star is larger when migration asymmetry $\alpha$ is smaller. Thus, smaller values of $\alpha$ make early adaptation more efficient for a star-structured population, on average. This is \textit{a priori} a surprising observation: the star suppresses natural selection for $\alpha<1$, thus making beneficial mutations less likely to fix, and this is precisely where we find that the highest peak is most likely to be reached first. To further investigate this, we will next analyze early adaptation in specific small landscapes. 

\begin{figure}[h!]
 \centering
 \includegraphics[width=\textwidth]{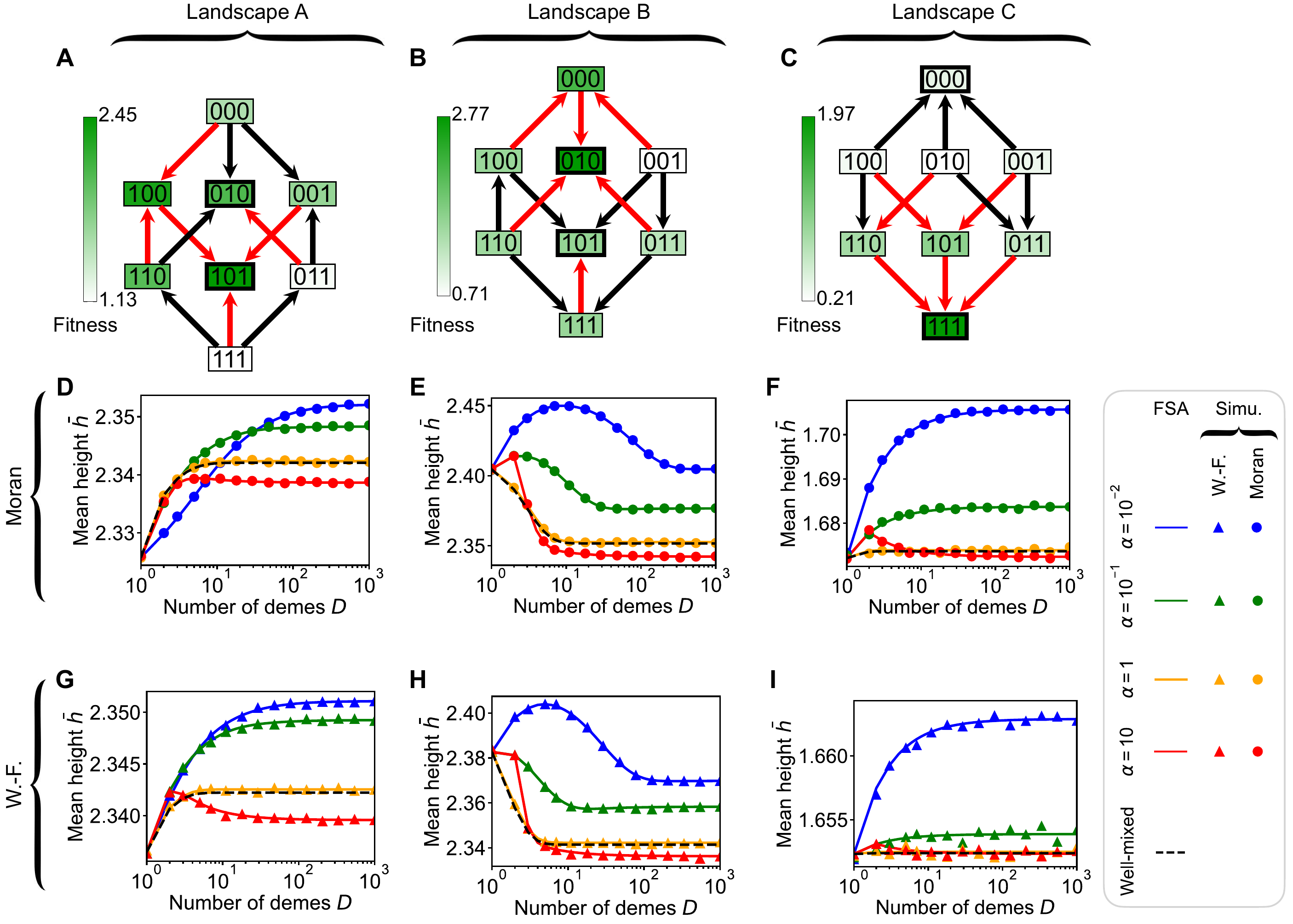}
 \caption{\textbf{Impact of the number of demes $D$ and migration asymmetry $\alpha$ of a star on early adaptation in  specific  landscapes.}  (A-C) Graphical representation of 3 $LK$ landscapes with $L=3$ and $K=1$, called landscapes A, B and C.  Genotypes with thick edges are peaks; arrows point towards fitter neighbors; red arrows point toward fittest neighbors. (D-F) Mean height $\bar{h}$ of the Moran star walk with rare migrations versus $D$ in landscapes A, B and C. (G-I) Same as (D-F) but for the Wright-Fisher star walk with rare migrations.  Lines: numerical resolutions of the FSA \cref{eq:systemfsa,eq:barh}; markers: simulation results averaged over at least $10^5$ walks per starting genotype.  Dashed black line: well-mixed case with population size $DC(1-g/f_W)$ where $f_W$ is the fitness of the wild type, shown as reference. Parameter values (all panels): $C = 20$ and $ g = 0.01$.}
\label{fig:h_vs_D}
\end{figure}

\paragraph{Early adaptation in specific small landscapes.} 
We give particular attention to three specific landscapes from this ensemble, denoted by A, B and C. These fitness landscapes are depicted in \cref{fig:h_vs_D}(A-C), see \cref{table:landscapes} and \cref{fig:landscape_fig2_magellan} for details. Each of these landscapes comprises two fitness peaks. Landscapes A and B were randomly chosen among the landscapes with more than one peak obtained when generating $LK$ landscapes. The adaptation of well-mixed populations on these two landscapes was studied in Ref.~\cite{servajean2023}, which facilitates comparison. Landscape C was chosen because it comprises a very high fitness peak (genotype 111) and a much lower one (genotype 000), which is reminiscent of a (rough) Mount Fuji landscape \cite{aita2000, szendro2013b}, see \cref{fig:h_vs_D}(C). 

In \cref{fig:h_vs_D}, we observe that, in each of our three example landscapes, for moderate to large number of demes $D$, the mean height $\bar{h}$ in the star is larger when migration asymmetry $\alpha$ is smaller. This is consistent with our observation above that the ensemble mean height $\left<\bar{h}\right>$ in the star is larger when migration asymmetry $\alpha$ is smaller. 
\cref{fig:h_vs_alpha} shows these heights versus $\alpha$ for $D=5$, and confirms that the highest values are obtained for intermediate $\alpha$ values satisfying $\alpha<1$. We highlight again that this is a striking observation, since the star suppresses natural selection for $\alpha<1$, and thus makes beneficial mutations less likely to fix. Our present focus on specific small landscapes allows us to further dissect the causes of this trend. 

\paragraph{Suppression of selection can allow to reach higher peaks.} Small migration asymmetry $\alpha$ in the star gives suppression of selection, and the range of small-effect mutations that are effectively neutral increases when $\alpha$ is decreased (see \cref{fig:p_fix_vs_s}). Importantly, among beneficial mutations, the fixation of weakly beneficial ones is most strongly suppressed, giving a relative advantage to strongly beneficial ones, see \cref{sec:suppr-si} of the Supplementary Material.

A genotype in landscape A, with two accessible beneficial mutations, is studied in \cref{fig:p_fix_vs_alpha}. In this case, the most beneficial mutation is most favored for $\alpha\approx 10^{-2}$, as it is then weakly suppressed, while the weakly beneficial one is strongly suppressed. 
This most beneficial mutation is on the path to the highest peak. Hence, the mean height $\bar{h}$ increases when $\alpha$ is decreased down to $10^{-2}$ in landscape A for large $D$, see \cref{fig:h_vs_D}(D,G) and \cref{fig:hi}. More generally, favoring the most beneficial mutations relative to the other ones often promotes reaching higher peaks. This rationalizes our striking observation that small $\alpha$ values, giving suppression of selection in the star, generally make early adaptation more efficient for large $D$, see \cref{fig:h_vs_D}. 

Note however that further decreasing $\alpha$ in the example of \cref{fig:p_fix_vs_alpha} 
reduces the bias toward the most beneficial mutation, as its fixation also becomes suppressed. Accordingly, \cref{fig:h_vs_alpha} shows that the most efficient early adaptation is generally obtained for small but intermediate $\alpha$ values. Besides, there are exceptions to the general trend that biasing toward the most beneficial mutations promotes reaching higher peaks, see e.g.\ \cref{fig:beyond_LK_L=3}(B,H).

\paragraph{Suppression of selection can enhance finite-size effects.} 
In \cref{fig:h_vs_D}, we observe that $\bar{h}$ often increases with $D$, which corresponds to the average behavior observed in \cref{fig:mean_h_vs_D}. Nevertheless, in landscape B, $\bar{h}$ versus $D$ displays a broad maximum for $\alpha = 10^{-2}$, see \cref{fig:h_vs_D}(E,H). In this case, there is a finite number of demes $D$ that yields a more efficient overall search for high fitness peaks than the large-$D$ limit. This is reminiscent of the maximum of $\bar{h}$ observed for finite well-mixed populations in this landscape in Ref.~\cite{servajean2023}. However, this maximum is obtained for a larger total size here than in the well-mixed case. This suggests that population structure can increase the range of population sizes where finite size effects matter. Note that the value of $D$ giving this maximum is slightly smaller for the Wright-Fisher star walk than for the Moran star walk. As in well-mixed populations, this comes from a factor of 2 that differs in the Moran and Wright-Fisher fixation probability formulas in the diffusion regime \cite{servajean2023}. 

To better understand the impact of suppression of selection on finite-size effects, we analyze a concrete example from landscape B in \cref{sec:suppr-si} of the Supplementary Material. For $\alpha = 10^{-2}$, while the mean height $\overline{h_i}$ of the walk is constant or increases with $D$ for most starting genotypes $i$, for one of them, $\overline{h_i}$ decreases with $D$, see \cref{fig:hi}(I,M). This gives rise to the maximum observed when averaging over $i$. From the genotype with decreasing $\overline{h_i}$, all mutations are deleterious, except the one leading to the small peak, which is beneficial. Hence, only the small peak can be reached when $D$ is large enough for the fixation of deleterious mutations to be negligible. However, suppression of selection entails that the convergence to this limit  occurs for larger $D$ when $\alpha$ is small. Analyzing the standard deviation of $h$ starting from uniformly chosen genotypes (see  \cref{fig:sigma_h_vs_D}) further shows that the maximum of $\bar{h}$ observed for $\alpha = 10^{-2}$ in landscape B is associated with a higher predictability.

Beyond these rich specific features, in \cref{fig:h_vs_D}, we notice in all landscapes that large-$D$ plateaus are reached for larger values of $D$ when $\alpha$ is smaller. Again, in the star, small $\alpha$ leads to suppression of selection. This enhances finite-size effects, since larger values of $D$ are then needed to obtain significant bias toward more beneficial mutations.

Here, we focused on the impact of spatial structure on the height $h$ of the first fitness peak reached. For completeness, \cref{sec:time} of the Supplementary Material shows the impact of spatial structure on the time it takes to reach this first peak. We find that this time tends to grow with suppression of selection, as fixation probabilities of beneficial mutations are reduced (see \cref{fig:t_vs_D}). Thus, while suppression of selection
often allows to reach higher peaks first, this tends to require more time.

\paragraph{Impact of the carrying capacity $C$.} \cref{fig:h_vs_C} shows the ensemble mean height $\left<\bar{h}\right>$, and the mean height $\bar{h}$ in landscapes A, B and C, versus deme carrying capacity $C$ for various values of $\alpha$ in the star, with $D=5$. We observe that all curves converge to the limiting result for a large well-mixed population (dashed horizontal black lines). 
Indeed, when demes are large, deleterious mutations cannot fix within their deme of origin. In the rare migration regime, this entails that they cannot fix in the structured population, just as in the case of large well-mixed populations. Similarly, if a beneficial mutation fixes in the deme where it appeared, the wild type, which is less fit than the mutant, cannot re-invade this deme, and mutant fixation in the structured population is thus assured. Thus, for rare migrations and in the limit of large demes, the fate of a mutation in the whole population is entirely determined by its fate in its deme of origin. Therefore, in this limit, spatial structure does not impact mutant fixation probabilities or the mean height of a walk. \cref{fig:h_vs_C} also shows that smaller values of $\alpha$ lead to a later convergence of $\left<\bar{h}\right>$ and $\bar{h}$ versus $C$. Thus, once again, suppression of selection enhances finite-size effects.

\begin{figure}[h!]
 \centering
 \includegraphics[width=\textwidth]{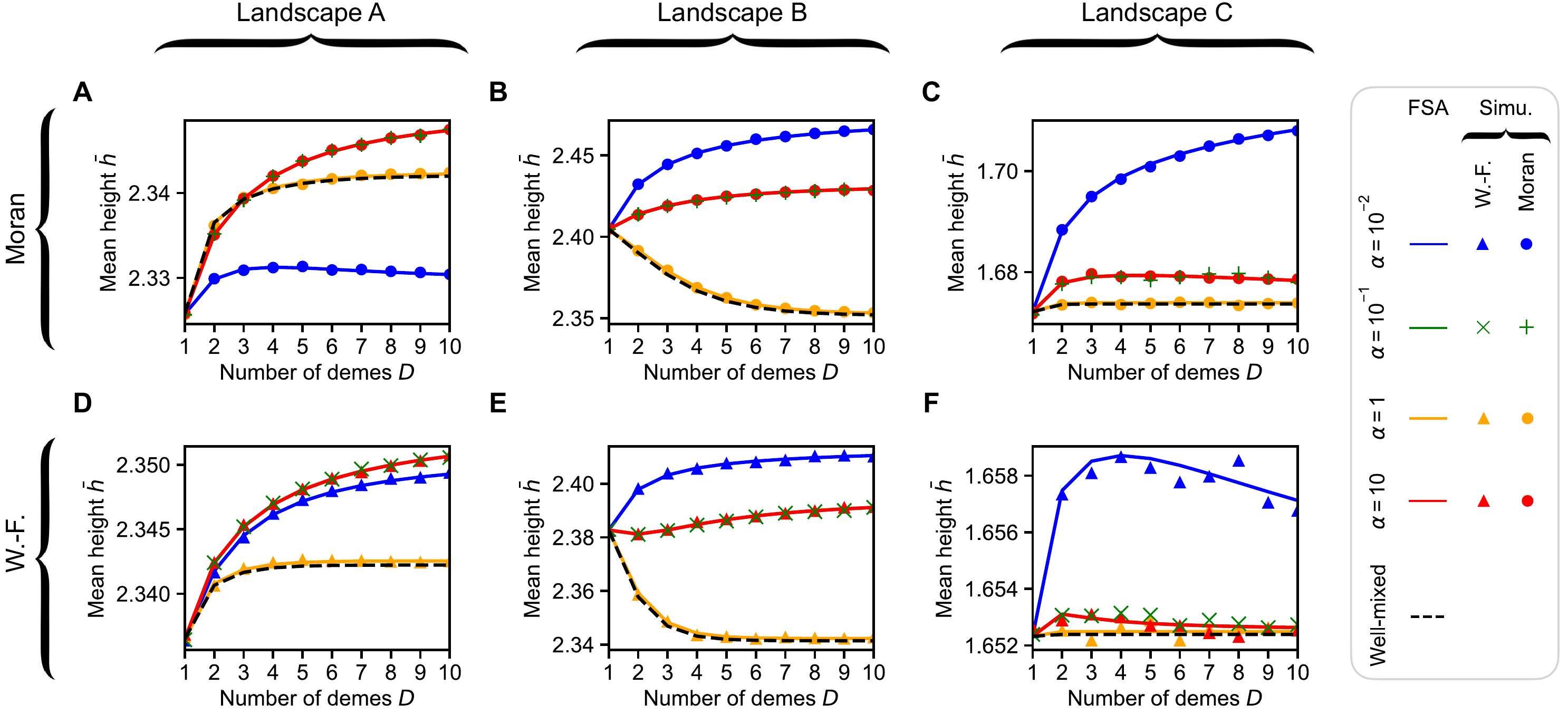}
 \caption{\textbf{Impact of the number of demes $D$ and migration asymmetry $\alpha$ of a line on early adaptation in 3 $LK$ landscapes with $L=3$ and $K=1$.} (A-F) Mean height $\bar{h}$ of the first fitness peak reached when starting from a uniformly chosen initial genotype versus the number of demes $D$ for the Moran line walk (A-C) and for the Wright-Fisher line walk (D-F) with rare migrations for various values of migration asymmetry $\alpha$, in landscapes A (A,D), B (B,E), and C (C,F) (see \cref{fig:h_vs_D}(A-C)). Dashed black line: well-mixed case with population size $DC(1-g/f_W)$, where $f_W$ is the fitness of the wild-type genotype, shown as reference. Lines: numerical resolutions of the FSA~\cref{eq:systemfsa,eq:barh} for each landscape; markers: simulation results averaged over at least $10^5$ walks per starting genotype in each landscape. Parameter values (all panels): $C = 20$ and $ g = 0.01$.}
\label{fig:line}
\end{figure}

\paragraph{Another spatial structure: the line.} While we focused on the star so far, it is interesting to assess the impact of the number of demes $D$ and of migration asymmetry $\alpha$ in other structures. \cref{fig:line} shows that the main conclusions obtained for the star also hold for the line, which is represented in \cref{fig:structs}(B). Namely, the height $\bar{h}$ of the first peak reached usually increases with $D$, with exceptions, and for large $D$, small $\alpha$ usually leads to larger $\bar{h}$, with exceptions too. Recall that, for the line, there is suppression of selection for all $\alpha\neq 1$, and a symmetry such that  asymmetries  $\alpha$ and $1/\alpha$ are equivalent (see \cref{sec:fixation_line} of the Supplementary Material) -- which is why the same results are obtained for $\alpha=10$ and $\alpha=10^{-1}$. 

In landscape A, we observe that $\bar{h}$ is smaller for $\alpha = 10^{-2}$ than for $\alpha=10^{-1}$ or $\alpha=1$ when considering the Moran line walk (\cref{fig:line}(A)), while the opposite holds when considering the Wright-Fisher line walk (\cref{fig:line}(D)). This unusual difference between the two types of walks arises from the factor of 2 which differs in the Moran and Wright-Fisher fixation probability formulas in the diffusion regime. Indeed, when $\alpha$ is decreased, fixation probabilities of beneficial mutations decrease due to suppression of selection. For a given mutation in a given type of walk, this decrease occurs below a given threshold value of $\alpha$. In landscape A, for specific relevant mutations, $\alpha=10^{-2}$ happens to be either below or above that threshold depending on the type of walk, explaining the difference between \cref{fig:line}(A) and (D)  (see \cref{sec:suppr-si} of the Supplementary Material, in particular \cref{fig:p_fix_vs_alpha}).

\paragraph{Extension to demes of different sizes.} While our focus is on graphs with identically-sized demes on each node, our model of spatially structured populations based on~\cite{Marrec21} allows to treat the more general case of demes with different sizes. As a first step, we consider the doublet, composed of a small deme and a larger one, connected by migrations that can be asymmetric, see \cref{fig:doublet_structure}. When deme sizes are sufficiently different, this model resembles the continent-island model~\cite{haldane_mathematical_1930}. In the rare migration regime, the fixation probability in the doublet was analytically determined in Ref.~\cite{Marrec21}, and can thus be used for our simulations and our numerical calculations. We investigate the impact of the ratio of the sizes of the demes and of migration asymmetry on the mean height $\bar{h}$ of the first fitness peak reached in landscapes A, B and C. Here too, $\bar{h}$ depends on the ratio of deme sizes for intermediate values in a way that is impacted by migration asymmetry. However, $\bar{h}$ converges to a value that does not depend on migration asymmetry when the ratio of deme sizes becomes large, see \cref{fig:h_vs_D_doublet}. This appears to differ from the results we obtained when all demes have the same size. However, here, as the large deme grows larger, we reach the point where the fate of a mutant in it fully determines that in the whole doublet. This is akin to the effect of increasing $C$ when all demes have the same size.

\paragraph{Beyond the rare migration regime.} So far, we considered the rare migration regime~\cite{Marrec21}. How does spatial structure impact early adaptation for more frequent migrations? Even in our symmetric structures, no general analytical expressions of fixation probabilities are known with frequent migrations. Thus, we estimate fixation probabilities in a structured Wright-Fisher model which is a particular case of the recent serial dilution model developed in Ref.~\cite{abbara2023frequent}, see ``Model and methods''. 
We checked that, for small migration rates in the star, our results from this model are consistent with those we obtained above in the rare-migration regime, assuming the Wright-Fisher model in the diffusion approximation within demes, see \cref{fig:landscape_with_small_s}. 

\begin{figure}[h!]
 \centering
 \includegraphics[width=0.9\textwidth]{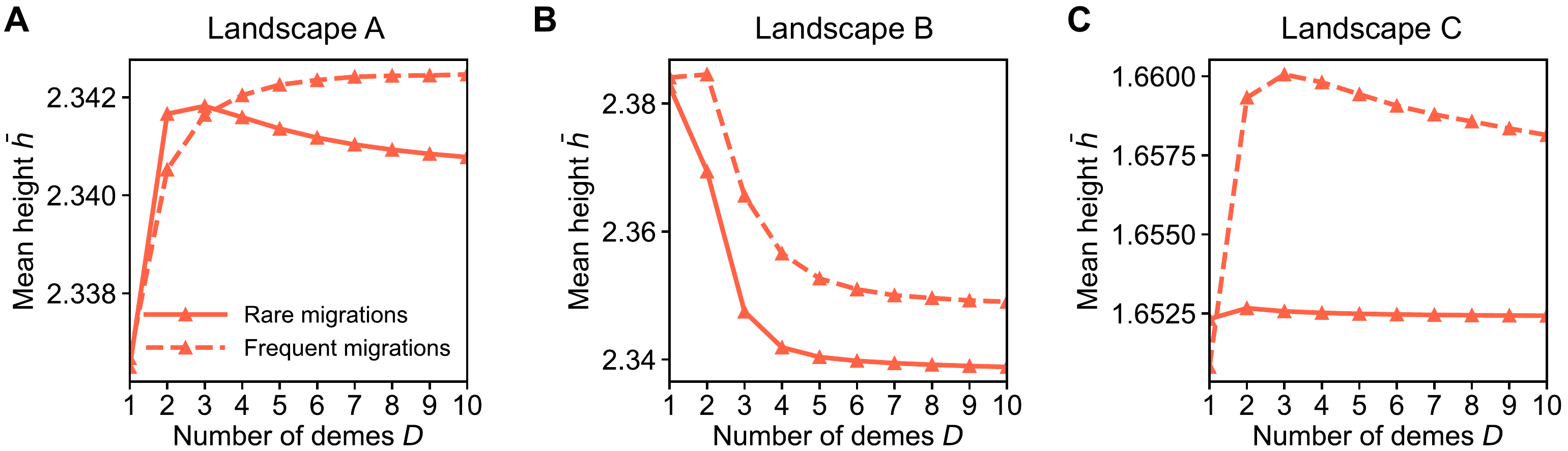}
 \caption{\textbf{Impact of the number of demes $D$ of a star on early adaptation in three $LK$ landscapes with $L=3$ and $K=1$ beyond the rare migration regime.} (A) Mean height $\bar{h}$ of the first fitness peak reached when starting from a uniformly chosen initial genotype of landscape A versus the number of demes $D$ in a star with $\alpha = 4$, for frequent and rare migrations. (B-C) Same as (A) but in landscapes B and C. All values of $\bar{h}$ are obtained by numerically solving the FSA~ \cref{eq:systemfsa,eq:barh}. With rare migrations, we use our analytical expressions of fixation probabilities (see ``Model and methods"). With frequent migrations, we use fixation probabilities estimated from simulations of our structured Wright-Fisher model (at least $10^7$ replicates). In all cases,  $C = 20$. For frequent migrations, $m_I = 0.2$ and $m_O = 0.05$. For rare migrations, $g = 0.01$. Markers are results and lines are guides for the eye.}
\label{fig:serial_dilution}
\end{figure}

In \cref{fig:serial_dilution}, we show the mean height $\bar{h}$ versus the number of demes $D$ in landscapes A, B and C, for the star with migration asymmetry $\alpha = 4$, both with frequent migrations and with our usual approach that assumes rare migrations. For frequent migrations, Ref.~\cite{abbara2023frequent} showed that all non-circulation graphs are strict suppressors of selection in the branching process approximation. This includes stars with migration asymmetry $\alpha > 1$, while they amplify selection in the rare-migration regime. In \cref{fig:serial_dilution}, we observe that $\bar{h}$ is higher for frequent migrations than in the rare migration regime, and that the value of $D$ from which the curve plateaus is larger with frequent migrations. Both of these results match what we observed in the rare migration regimes for structures that suppress natural selection (namely the star with $\alpha<1$ and the line with $\alpha \neq 1$). 

In \cref{fig:serial_dilution_C=200}, we provide similar results as in \cref{fig:serial_dilution}, but for a larger carrying capacity $C$. Strikingly, the height $\bar{h}$ depends on $D$ for frequent migrations in this case too, while this carrying capacity is too large for structure to affect fixation probabilities or the mean height $\bar{h}$ in the rare-migration regime in our small landscapes A, B and C (see paragraph above on the impact of $C$). This suggests that increasing the amount of migrations within spatially structured populations enhances finite size effects.            

\newpage
\subsection*{Early adaptation in larger landscapes}

So far, we focused on small fitness landscapes with $L=3$ genetic units. It allowed us to link features of early adaptation to the probabilities of specific transitions in the fitness landscapes, and shed light on the impact of suppression of selection on early adaptation in concrete cases. However, real genotypes have many units (genes or nucleotides). It is thus important to extend our study to larger landscapes. For all this analysis, we focus on early adaptation in star-structured populations in the rare migration regime.

\begin{figure}[h!]
 \centering
 \includegraphics[width=0.9\textwidth]{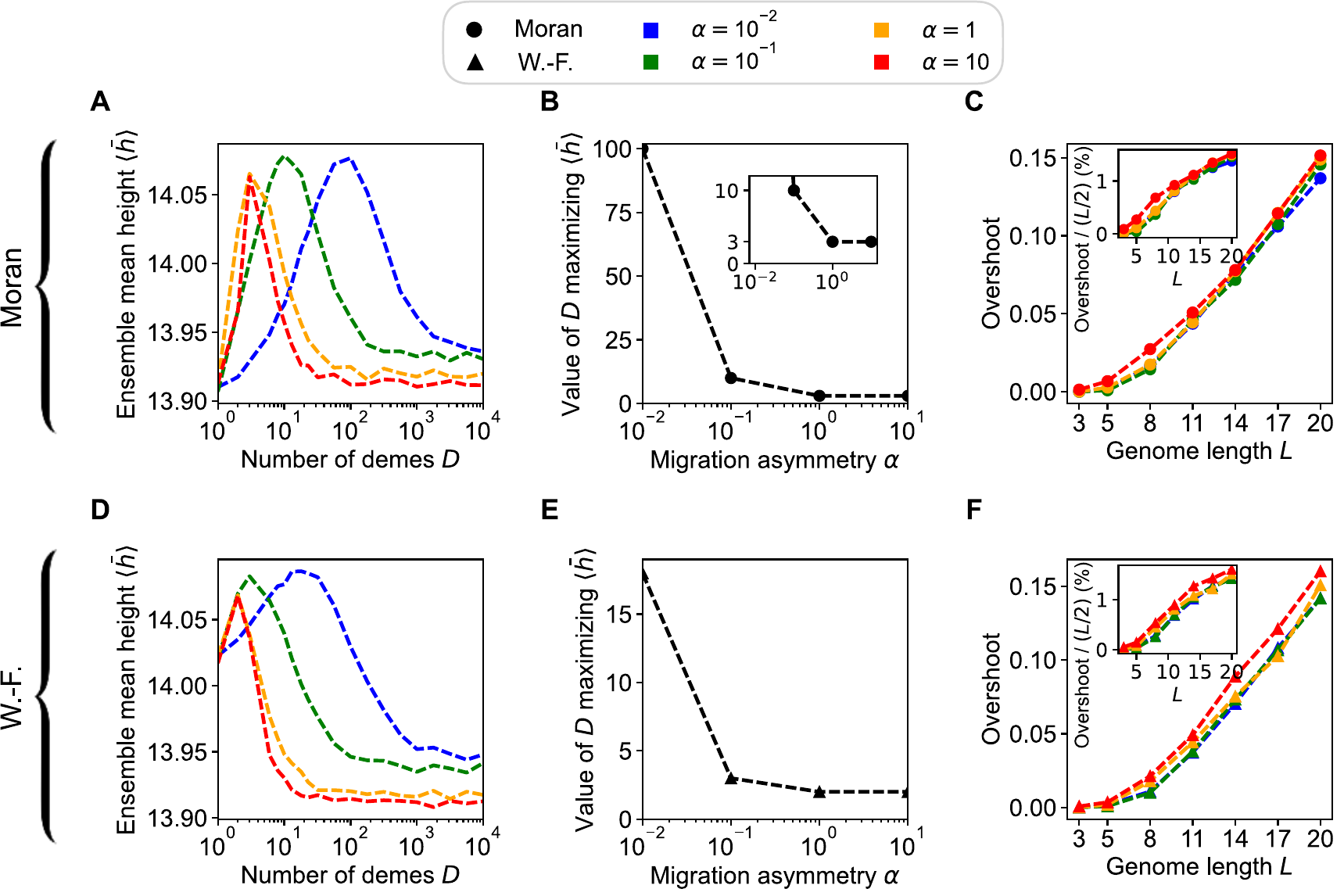}
 \caption{\textbf{Impact of genome dimension $L$ on early adaptation in $LK$ landscapes with $K = 1$.} (A) Ensemble mean height $\left<\bar{h}\right>$ of the first fitness peak reached when starting from a uniformly chosen initial genotype in $LK$ fitness landscapes with $L = 20$ and $K=1$, versus the number $D$ of demes for the Moran star walk with rare migrations with various values of migration asymmetry $\alpha$. (B) Value of $D$ maximizing the ensemble mean height $\left<\bar{h}\right>$ of a Moran star walk versus migration asymmetry $\alpha$, for $LK$ fitness landscapes with $L = 20$ and $K=1$. (C) Overshoot, defined as the maximum value of $\left<\bar{h}\right>(D)$ minus the large-$D$ limit of $\left<\bar{h}\right>$ (evaluated at $D = 10^5$ in practice), versus $L$. Inset: same, but the overshoot is divided by $L/2$ and expressed in percentage. (D-F) Same as (A-C), but for the Wright-Fisher star walk with rare migrations. In all cases, dashed lines and markers represent simulation results (obtained using at least $\sim\!10^5$ simulations per data point). Because of memory constraints associated to large genotypes space, landscapes were generated along the way and reinitialized for each walk, meaning that each walk takes place in a different landscape. Exception: the results for $L = 3$, were obtained in the $2\times 10^5$ landscapes used in \cref{fig:mean_h_vs_D}. Note that memory constraints are also the reason why FSA results are not shown here. Parameter values (all panels): $C = 20$ and $ g = 0.01$. }
\label{fig:vs L}
\end{figure}

\paragraph{Early adaptation in star-structured populations in larger model landscapes.} First, we consider $LK$ landscapes, and we analyze the impact of genome length $L$ and number of epistatic partners $K$ on early adaptation. \cref{fig:vs L}(A,D) shows that the ensemble mean height $\left<\bar{h}\right>$ reached in $LK$ landscapes with $L=20$ and $K=1$ displays a maximum for a finite value of $D$ for all values of $\alpha$ considered. This means that the most efficient early adaptation is obtained for stars with a finite number of demes, and not in the limit of very large stars. Recall however that demes need to be small enough for structure to matter in the rare migration regime (see above). In \cref{fig:vs L}(B,E), we further observe that the position of this maximum shifts towards larger values of $D$ when $\alpha$ decreases, i.e.\ when suppression of selection is stronger. This confirms that suppression of selection enhances finite-size effects. Finally, \cref{fig:vs L}(C,F) shows that the size of the overshoot of $\left<\bar{h}\right>$, with respect to its large-$D$ limit, increases with $L$, and \cref{fig:impact_of_K} shows that it decreases with $K$. These results are in line with those obtained for well-mixed populations \cite{servajean2023}. Thus, both in structured populations and in well-mixed ones, the adaptive advantage of some finite populations over large ones becomes stronger for larger genomes with few epistatic partners per genetic unit (gene or nucleotide). Importantly, this case is highly relevant in nature. For instance, pairwise epistasis has been shown to describe protein sequences well~\cite{Weigt09,Marks11,Morcos11}. 

In \cref{sec:beyondsmallLK} of the Supplementary Material, we further extend our study to large model fitness landscapes beyond the $LK$ model (see \cref{fig:beyond_LK}). 
Again, in most cases, smaller $\alpha$ yield larger values of $\bar{h}$ for large $D$. In addition, the value of $D$ from which $\bar{h}$ plateaus becomes larger as $\alpha$ decreases, i.e.\ suppression of selection enhances finite-size effects. This illustrates the broad validity of our main findings.

\begin{figure}[h!]
 \centering
 \includegraphics[width=0.793\textwidth]{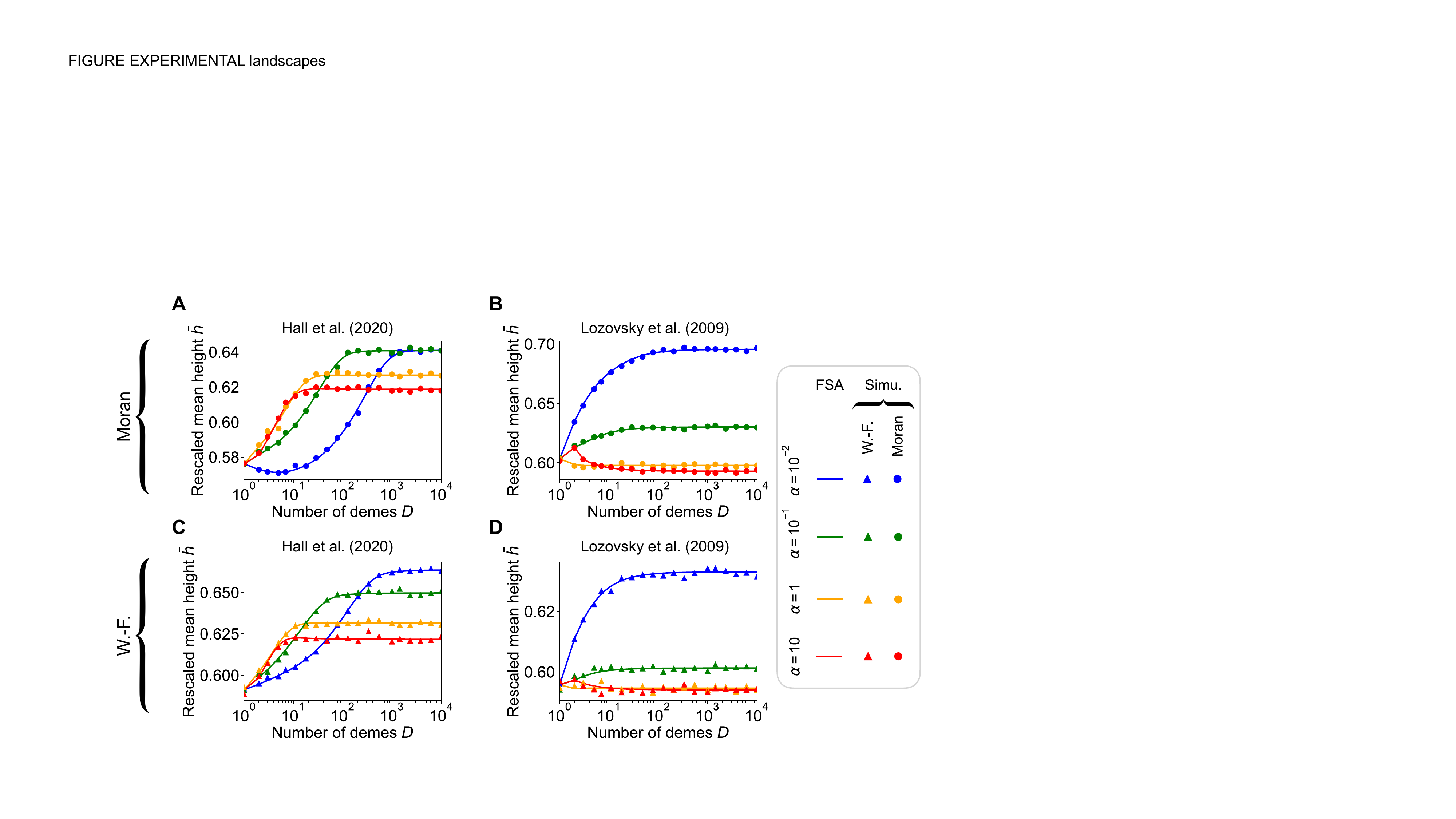}
 \caption{\textbf{Impact of the number of demes $D$ and migration asymmetry $\alpha$ of a star on early adaptation in two experimental landscapes.} (A) Rescaled mean height $\bar{h}$ of the first fitness peak reached when starting from a uniformly chosen initial genotype versus the number $D$ of demes for the Moran star walk under rare migrations, with various values of migration asymmetry $\alpha$, in an experimental landscape from Ref.~\cite{hall2020}. The rescaled mean height is defined as $(\bar{h}-h_\textrm{min})/(h_\textrm{max}-h_\textrm{min})$, where $h_\textrm{min}$ (resp.\ $h_\textrm{max}$) denotes the fitness of the lowest peak (resp.\ highest peak). Note that $h_\textrm{min}<\bar{h}<h_\textrm{max}$. The fitness landscape corresponds to Fig.~2E of Ref.~\cite{hall2020} ($L = 7$; organism: \textit{Escherichia coli}; fitness proxy: half-maximal effective concentration $\text{EC}_{50}$ using chloramphenicol). (B) Same as (A) but in the experimental landscape given in Table~S2 from Ref.~\cite{lozovsky2009} ($L = 4$, organism: \textit{E. coli} carrying dihydrofolate reductase from the malaria parasite \textit{Plasmodium falciparum}; proxy for fitness: relative growth rate in the absence of pyrimethamine). (C-D) Same as (A-C) but for the Wright-Fisher star walk with rare migrations. In all panels, lines are numerical resolution of the FSA  \cref{eq:systemfsa,eq:barh}, while markers are simulation results averaged over at least $10^3$ walks per starting genotype. Parameter values (all panels): $C = 20$ and $ g = 0.01$.}
\label{fig:experimental_landscapes}
\end{figure}

\paragraph{Experimental fitness landscapes.}
Finally, we consider experimentally-measured fitness landscapes. \cref{fig:experimental_landscapes} shows that the general features observed in model landscapes also hold for these experimental fitness landscapes. Namely, small $\alpha$ generally leads to larger values of $\bar{h}$ for large $D$, and enhances finite-size effects. Moreover, for experimental landscapes, we observe that the rescaled $\bar{h}$ features a large amplitude of variation when $D$ is varied. 

\newpage
\subsection*{Steady-state probabilities and steady-state effective population sizes}\label{subsec:steady state}

So far, we investigated early adaptation by focusing on the first fitness peak reached. However, the long-term properties of adaptation are also of interest, for instance if the environment remains constant for a long time. The steady-state probability that a population possesses each given genotype was studied in well-mixed populations in Ref.~\cite{sella2005}. How does spatial structure with demes on nodes of a graph impact these steady-state probabilities?

We derive analytically the steady-state probability of each genotype in the rare migration regime for circulations, and for stars and lines with extreme migration asymmetries. Derivations are presented in the Supplementary Material \cref{sec:effective_pop}. For all these spatial structures, we obtain a steady-state probability distribution that has the same form as that associated to a well-mixed population \cite{sella2005, servajean2023}. This motivates us to define a steady-state effective population size $N_e$ as the size of the well-mixed population that possesses the same steady-state distribution as the structured population considered. Our analytical results for these steady-state effective population sizes are given in \cref{table:effective_pop}. For circulations, the steady-state effective population size is close to the actual population size. For a star with a very small $\alpha$, as well as for a line with either a very small or very large $\alpha$, all featuring strong suppression of selection, the effective population size is the size of a deme. Conversely, when $\alpha$ is very large in the star, yielding amplification of selection, the effective population size is roughly twice the actual total population size. In the Supplementary Material, we compute numerical estimates of steady-state effective population sizes, which agree well with our analytical results for extreme migration asymmetries, and are intermediate for less extreme ones, see \cref{fig:Ne_vs_alpha}.

\begin{table}[h!]
\centering
\begin{tabular}{c c c} 
 \hline
Structure & Population size & Effective population size \\
\hline
 Circulation & $CD$ & $(C-1)D+1$ \\ 
 Star ($\alpha \rightarrow 0$) & $CD$ & $C$ \\
 Star ($\alpha \rightarrow \infty$) & $CD$ & $(C-1)(2D-3)+1$ \\ 
 Line ($\alpha \rightarrow 0$) & $CD$ & $C$ \\
 Line ($\alpha \rightarrow \infty$) & $CD$ & $C$ \\
 \hline
\end{tabular}
\caption{\textbf{Steady-state effective population size of different structured populations.} The rare migration regime is considered. In addition, for the star and the line, the focus is on extreme migration asymmetries $\alpha$. 
}
\label{table:effective_pop}
\end{table}

Our results show that suppression of selection is associated to reduced steady-state effective population sizes compared to actual sizes, while amplification is associated to larger ones. Smaller steady-state effective sizes entail smaller mean fitnesses at steady state, see \cref{subsec:effsize} of the Supplementary Material and Ref.~\cite{servajean2023}. Therefore, our results show  that, on average, suppression of selection does not allow to reach higher fitness in the very long term.

While our steady-state effective population size has interesting implications about the long-term trends of fitness, it is worth emphasizing two limitations. 
First, the time needed to reach a steady state can be extremely long. Going beyond the first fitness peak involves crossing fitness valleys. In the weak mutation regime we focus on, this requires fixing deleterious mutations~\cite{Weissman09,Bitbol14}. The average time until the appearance of a mutant destined to fix in a deme is very long for a substantially deleterious mutation (see ``Model and methods'', end of the ``Rare migration regime'' paragraph). Hence, we expect early adaptation, which is our main object of study here, to be more relevant in practice. 
Second, our steady-state effective population size pertains to the steady-state distribution in the rare migration regime. More generally, the effective size of a structured population is the size of a well-mixed population that gives the same value of a quantity of interest, which may depend on the quantity considered~\cite{ewens1979}, or may not exist~\cite{Sjodin05}. Spatially structured populations have been characterized e.g.\ by the inbreeding effective size~\cite{Nagylaki80,Whitlock97}, the coalescent effective size~\cite{Nordborg02,Sjodin05}, and the variance effective population size~\cite{Cherry03}. 
In particular, the inbreeding effective size can be calculated using identity by descent, following~\cite{Whitlock97}, for our structured populations with demes on a graph. However, this approach assumes that migration is fast with respect to drift~\cite{Nagylaki80,Whitlock97}, in contrast with the rare migration regime we considered. Furthermore, while several definitions of effective population sizes coincide under frequent migrations, they primarily focus on neutral evolution, and may not suffice to describe evolution with selection~\cite{Barton93, Whitlock97,Nordborg02}, which is important on fitness landscapes.

\section*{Discussion}

In this paper, we investigated the impact of spatial structure on early adaptation in rugged fitness landscapes, in the weak mutation and rare migration regime. 
We showed that, in most landscapes, migration asymmetries yielding suppression of selection allow the population to reach higher fitness peaks first, when the number of demes is large. Thus, perhaps surprisingly, suppression of selection can make early adaptation more efficient. We showed that this occurs because suppression of selection increases the bias toward the most beneficial mutations, when weakly beneficial mutations are more suppressed than strongly beneficial ones. However, with suppression, more time is then needed to reach this first fitness peak. We also found that migration asymmetries associated with suppression of selection enhance finite-size effects. 
In particular, for some fitness landscapes, early adaptation is most efficient for a finite number of demes. This tends to be the case for large genomes with few epistatic partners per site, which are relevant in practice. 
We presented extensions to a simple structure with different deme sizes, and to frequent migrations, illustrating the generality of our approach. Our extension to frequent migrations suggests that our key result that suppression of selection can foster early adaptation still holds in this regime where suppression of selection is pervasive, and that more frequent migrations can enhance the impact of spatial structure on early adaptation. 
We further investigated the impact of spatial structure on long-term adaptation in the weak mutation and rare migration regime, by considering the steady-state distribution, and introducing a steady-state effective population size that describes it. We found that suppression of selection is associated to small steady-state effective sizes, and thus to small steady-state mean fitnesses. 
Hence, suppression of selection can foster early adaptation, but leads to smaller steady-state mean fitnesses in the rare migration regime. The first result has more practical relevance than the second, since it can take a long time to get beyond the first fitness peak with rare mutations.

A limitation of our results is that, at least in the rare migration regime, the impact of spatial structure on early adaptation relies on finite-size effects, and thus requires relatively small demes. How small depends on the fitness landscape, the graph, and the migration asymmetry. Importantly, migration asymmetries that yield strong suppression of selection enhance finite-size effects. 
However, in the weak mutation and rare migration regime, for a given graph with a given migration asymmetry and number of demes, and on a given fitness landscape, there exists a deme size beyond which spatial structure has no effect on early adaptation. Indeed, for sufficiently large demes with rare migrations, the fate of a mutant in one deme determines its fate in the whole population. Our results are relevant despite this limitation, as natural spatially structured populations often involve relatively small demes. In particular, host-associated bacterial populations are spatially structured at several scales. First, each host can be viewed as a deme of bacteria. For \textit{Caenorhabditis elegans} nematodes, colonization of a new food source (e.g.\ a fruit) gives rise to a population of up to $10^4$ worms~\cite{frezal_thenatural_2015}. In monoculture colonization of immunocompetent and initially germ-free nematodes, each worm’s gut was found to harbor a population of $20$ to $2\times 10^4$ bacteria, depending on the bacterial species~\cite{ortiz_interspecies_2021}. Second, within each host, bacterial populations are further subdivided. For instance, the mouse intestine comprises about $10^5$ crypts~\cite{casteleyn_surface_2010}, each containing about 100 to 400 bacteria~\cite{pedron_acrypt_2012,dekaney_expansion_2007}.

The impact of complex spatial structures on the evolutionary trajectories of populations, involving several mutations, has only recently begun to be addressed. Recent works, performed in the framework of evolutionary graph theory models with one individual per node of a graph, have studied steady-state mean fitnesses~\cite{sharma_suppressors_2022,sharma_graph-structured_2024}. They showed that amplification of selection and suppression of selection lead respectively to increased or decreased steady-state mean fitnesses. Our findings on long-term adaptation therefore generalize these results to populations with demes on graph nodes. Ref.~\cite{sharma_suppressors_2022,sharma_graph-structured_2024} further evidenced suppression or amplification of fixation effects, which can lead respectively to increased or decreased steady-state mean fitnesses. However, these effects involve a coupling between update rules and mutant initialization which is not present under our assumptions where growth and migrations are decoupled, and mutant individuals appear uniformly at random. Besides, Refs.~\cite{sharma_suppressors_2022,sharma_graph-structured_2024} did not consider rugged fitness landscapes, which makes our study of early adaptation very different from these works. 

It would be very valuable to study more complex graphs than the highly symmetric ones we considered here, and to assess the impact of frequent migrations in more detail. A challenge in both cases is that fixation probabilities are generally not known analytically. In our extension to frequent migrations, we bypassed this challenge by using simulations to estimate fixation probabilities. However, developing more general descriptions of complex structured populations would allow more progress. Other possible directions include taking into account frequency-dependent selection \cite{ayala1974frequency, kojima1971there}, which impacts fixation probabilities (see e.g.\ \cite{moawad2024evolution}), as well as genotype-dependent migration rates~\cite{MarrecP}. Moreover, while we focused on early and long-term adaptation, it would be interesting to consider intermediate timescales. In particular, studying adaptation after the first fitness peak is reached would involve fitness valley crossing, a process which is itself impacted by spatial structure \cite{Weissman09,Bitbol14,kuo_evolutionary_2024}. Finally, it would be of interest to consider more frequent mutations~\cite{szendro2013a}. Note that this could lead to tunneling in fitness valley crossing~\cite{Weissman09}. In the framework of biased walks on fitness landscapes, this can be seen as a population hopping from a genotype to another one that is not among its nearest neighbors. Frequent mutations would also lead to multiple types being present in the population, and to clonal interference~\cite{elana2003,Jain11,Laessig17,good2017}, which could have an interplay with population spatial structure.

\section*{Code availability}
Our code is freely available at \url{https://github.com/Bitbol-Lab/fitness-landscapes/}. 

\section*{Acknowledgments}
The authors thank Alia Abbara for helpful discussions about the serial dilution model, and Claude Loverdo and Cecilia Fruet for useful discussions about spatial structure in host-associated populations. This research was partly funded by the European Research Council (ERC) under the European Union’s Horizon 2020 research and innovation programme (grant agreement No.~851173, to A.-F.~B.), by the Swiss National Science Foundation (SNSF) (grant No.~315230\_208196, to A.-F.~B.), and by the Chan-Zuckerberg Initiative (CZI).

\clearpage
\newpage

\begin{center}
 {\LARGE \bf Supplementary Material}   
\end{center}

\tableofcontents

\setcounter{section}{0}
\renewcommand{\thefigure}{S\arabic{figure}}
\setcounter{figure}{0}
\renewcommand{\thetable}{S\arabic{table}}
\setcounter{table}{0} 
\renewcommand{\theequation}{S\arabic{equation}}
\setcounter{equation}{0} 

\section{Fixation probability of a mutant in a line with rare migrations}\label{sec:fixation_line}

Let us consider a line of $D$ demes, which may comprise individuals of two types, denoted by $i$ and $j$. Each individual has a migration rate $m_{R}$ to the right and $m_{L}$ to the left. We focus on the rare migration regime, such that migrations are slower than extinction or fixation events within demes~\cite{Marrec21}. Thus, each deme only comprises individuals of one type, at the timescale relevant for migrations. In this section, we derive the fixation probability of the lineage of one mutant of type $j$ within a line initially composed of individuals of (wild) type $j$ in the rare migration regime.

\begin{figure}[h!]
 \centering
 \includegraphics[width=0.5\textwidth]{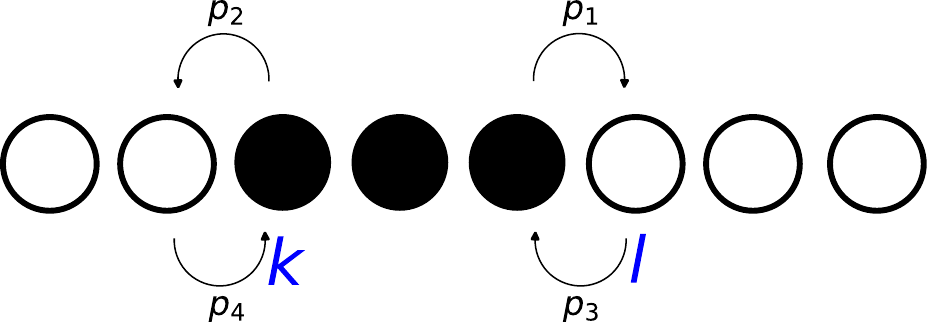}
\caption{\textbf{State of a line graph in the rare migration regime.} We show a schematic of a line graph, where white demes only comprise wild-type individuals of type $i$, while black demes only comprise mutant individuals of type $j$. Arrows denote the transition probabilities given in \cref{eq:p_i_def}. Indices $k$ and $l$ correspond to the position of the first mutant (type $j$) deme and the position of the first wild type (type $i$) deme, respectively. }
 \label{fig:schema_line}
\end{figure}

For mutant fixation to occur in the line in the rare migration regime, local fixation first needs to occur in the deme where the mutant appeared, and then mutants have to spread to other demes by successive migration and fixation steps~\cite{Marrec21}. Therefore, at all times, there is one contiguous cluster of demes of mutant type $j$. Let us denote by $\varphi_{k, l}$ the probability that mutants fix in the line starting from a cluster of $l-k$ adjacent fully mutant demes. Here, $k\in[0,D]$ indicates the position of the first mutant deme along the line, while $l\in[0,D]$ is the position of the first wild-type deme along the line, with the constraint that $k\leq l$, see \cref{fig:schema_line}. Note that demes are numbered from $0$ to $D-1$, that $k =l$ corresponds to the case where there is no mutant deme (including $k =l =D$) and that $l= D$ (resp.\ $k =0$) means that there is no wild type deme on the right-hand side (resp.\ left-hand side) of the mutant cluster.

The only migration steps that change the state of the population are those where either a mutant migrates to a wild-type deme and fixes there, or the opposite. Because migrations are to nearest neighbouring demes, these events happen at the edges of the cluster of mutant demes. Discriminating on the first migration step gives the following equation on $\varphi_{k,l}$: 
\begin{align}
	\varphi_{k, l} & = T_{(k,l) \rightarrow (k,l +1) }\varphi_{k,l +1} + T_{(k,l) \rightarrow (k-1,l) }\varphi_{k-1,l} +
	 T_{(k,l) \rightarrow (k,l-1) }\varphi_{k,l-1} +
	  T_{(k,l) \rightarrow (k+1,l) }\varphi_{k+1,l} \nonumber\\ &+ \left( 1 -  T_{(k,l) \rightarrow (k,l+1) }-  T_{(k,l) \rightarrow (k-1,l) } - T_{(k,l) \rightarrow (k,l-1) } -  T_{(k,l) \rightarrow (k+1,l) } \right)\varphi_{k, l}\,,
   \label{eqvarphi}
\end{align}
with the transition probabilities
\begin{align}
	 T_{(k,l) \rightarrow (k,l+1) }& = \frac{m_R N_j \rho_{ij}}{Z_{k,l} },\quad
	 T_{(k,l) \rightarrow (k-1,l ) } = \frac{m_L N_j \rho_{ij}}{ Z_{k,l} }, \\
 T_{(k,l) \rightarrow (k,l-1) }& = \frac{m_L N_i \rho_{ji}}{	Z_{k,l} },\quad
	 T_{(k,l) \rightarrow (k+1,l ) } = \frac{m_R N_i \rho_{ji}}{Z_{k,l}},
\end{align}
where $Z_{k,l}$ is a normalisation constant. \cref{eqvarphi} can be rewritten as follows:
\begin{equation}
	\varphi_{k, l} = p_1\varphi_{k,l +1} + p_2\varphi_{k-1,l} +
	p_3\varphi_{k,l-1} +
	p_4\varphi_{k+1,l},  \label{eq:2d_markov}
\end{equation}
with 
\begin{align}
	p_1 &=\frac{\alpha }{(1+\alpha)(1+\gamma)},\quad p_2 =   \frac{ 1}{(1+\alpha)(1+\gamma)}, \nonumber\\
	p_3 &= \frac{\gamma}{(1+\alpha)(1+\gamma)},\quad
	p_4 = \frac{\alpha\gamma }{(1+\alpha)(1+\gamma)}, \label{eq:p_i_def}
\end{align}
where we introduced the dimensionless coefficients $\alpha = m_R / m_L$ and $\gamma =  N_i \rho_{ji} / N_j \rho_{ij} $, where $\rho_{ij}$ is the fixation probability of the lineage of one mutant with genotype $j$ within a deme of individuals with genotype $i$. 

In addition to \cref{eq:2d_markov}, the following boundary conditions hold: 
\begin{align}
		\varphi_{k, k} &= 0,\quad
		\varphi_{0, D} = 1,\label{bc_2d_markov1}\\
		\varphi_{k, D}& = U_{k},\quad
		\varphi_{0, k} = V_{k}, \label{bc_2d_markov2}
\end{align}
where $U_k$ and $V_k$ will be computed below. 

\begin{figure}[h!]
 \centering
 \includegraphics[width=\textwidth]{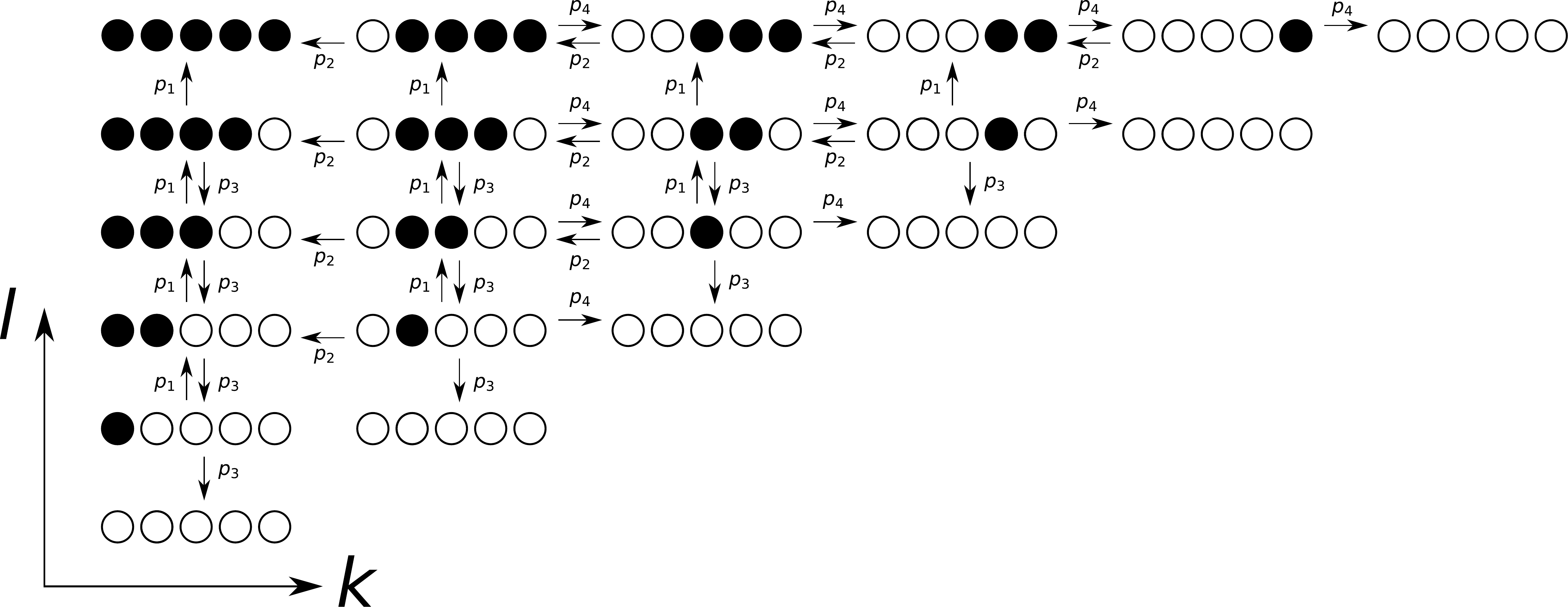}
\caption{\textbf{Markov chain for a line graph in the rare migration regime.} We show a schematic of all different possible states for a line of $D= 5$ demes. The arrows correspond to the possible transitions, and are annotated with their probabilities. This figure is adapted from Fig.~5 of Ref. \cite{broom2008analysis}. }
 \label{fig:schema_2d_markov}
\end{figure}

In \cref{fig:schema_2d_markov}, we illustrate the Markov chain described here, with the different states of the line and the possible transitions between them. 

\newpage
To find the expression of $\varphi_{k, l}$, we need the boundary conditions $U_k$ and $V_k$. Let us  first determine $U_k = \varphi_{k, D}$. It satisfies
\begin{align}
	U_k &= W_{k\rightarrow k-1} U_{k-1 } +  W_{k\rightarrow k+1} U_{k+1 } + (1-  W_{k\rightarrow k-1}-  W_{k\rightarrow k+1}) U_k\,,
 \label{eqUk}
\end{align}
with the following transition probabilities:
\begin{align}
	 W_{k\rightarrow k-1} = \frac{m_L N_j \rho_{ij}}{L_k}, \quad
	 W_{k\rightarrow k+1} = \frac{m_R N_i \rho_{ji}}{L_k}, 
\end{align}
where  $L_k$ is a normalisation constant. 
We can be rewrite \cref{eqUk} as:
\begin{align}
 0 &=  \alpha\gamma U_{k+1 } -(1+\alpha\gamma) U_k +  U_{k-1 }\,.
\end{align}
The solution of this equation has the form
\begin{equation}
	U_k =  \frac{A}{\left(\alpha\gamma \right)^{k}} + B\,,
\end{equation}
where the constants $A$ and $B$ are determined by the boundary conditions:
\begin{align}
	U_0 &= 1\,,  \\
	U_{D} &= 0\,.
\end{align}  
This gives
\begin{equation}
	U_k = \dfrac{\dfrac{1}{(\alpha\gamma)^{k}}  - \dfrac{1}{(\alpha\gamma)^{D}} }{1 - \dfrac{1}{(\alpha\gamma)^{D}}}. 
\end{equation}

Let us now perform a similar calculation for $V_k = \varphi_{0,k}$, which satisfies
\begin{align}
	V_k &= R_{k\rightarrow k-1} V_{k-1 } +  R_{k\rightarrow k+1} V_{k+1 } + (1-  R_{k\rightarrow k-1}-  R_{k\rightarrow k+1}) V_k,
\end{align}
with the following transition probabilities:
\begin{equation}
	R_{k\rightarrow k-1}  = \frac{m_L N_i \rho_{ji}}{M_k},\quad
	R_{k\rightarrow k+1}  = \frac{m_R N_j \rho_{ij}}{M_k} \,,
\end{equation}
where $M_k$ is another normalisation constant. This leads to:
\begin{align}
	0 &= \alpha V_{k+1 } -( \alpha +\gamma ) V_k + \gamma V_{k-1 }\,.
\end{align}
We obtain
\begin{equation}
	V_k = C_1\left(\frac{\gamma}{\alpha} \right)^k  + C_2\,,
\end{equation}
with constants $C_1$ and $C_2$ that are determined by the following boundary conditions:
\begin{align}
	V_0 &= 0\,,  \\
	V_{D} &= 1\,.
\end{align}  
This gives
\begin{equation}
	V_k = \dfrac{1 - \left(\dfrac{\gamma}{\alpha} \right)^{k}}{1 -  \left(\dfrac{\gamma}{\alpha} \right)^{D}}\,.
\end{equation}

Now that \cref{eq:2d_markov} together with \cref{bc_2d_markov1,bc_2d_markov2} constitute a fully posed problem, we can determine $\varphi_{k,l}$. In Ref.~\cite{miller1994matrix}, a general solution of \cref{eq:2d_markov} was obtained for a two-dimensional square lattice. It reads:
\begin{equation}
	\varphi_{k,l} = \sum_{n = 1}^{D-1} \left[ p_1 P_{n,D} T^{(n,D-1)}_{k,l} + p_2 P_{0,n} T_{k,l}^{(1,n)} + p_3 P_{n,0} T^{(n,1)}_{k,l} + p_4 P_{D, n} T^{(D-1,n)}_{k,l}\right],
\end{equation}
where the $P_{i,j}$ terms correspond to the four boundary conditions at each edge of the square and the transition probability $T^{(a,b)}_{ k,l}$ reads
\begin{align}
T^{(a,b)}_{k,l} & = \frac{4}{D^2} \left( \frac{p_2}{p_4}\right)^{(k-a)/2}\left( \frac{p_3}{p_1}\right)^{(l-b)/2} \sum_{r = 1}^{D-1}\sum_{s = 1}^{D-1} \frac{\sin\left(\frac{k r \pi}{D} \right) \sin\left(\frac{a r \pi}{D} \right) \sin\left(\frac{b s \pi}{D} \right) \sin\left(\frac{l s \pi}{D} \right)}{1- 2 \sqrt{p_2 p_4} \cos\left(\frac{r \pi}{D} \right) - 2 \sqrt{p_1 p_3} \cos\left(\frac{s \pi}{D} \right)}\nonumber \\
& = \frac{4}{D^2} \left( \frac{1}{\alpha\gamma}\right)^{(k-a)/2}\left( \frac{\gamma}{\alpha }\right)^{(l-b)/2} \sum_{r = 1}^{D-1}\sum_{s = 1}^{D-1} \frac{\sin\left(\frac{k r \pi}{D} \right) \sin\left(\frac{a r \pi}{D} \right) \sin\left(\frac{b s \pi}{D} \right) \sin\left(\frac{l s \pi}{D} \right)}{1- \frac{2 \sqrt{\alpha \gamma}}{(1+ \alpha)(1+\gamma)} \left[ \cos\left(\frac{r \pi}{D} \right) + \cos\left(\frac{s \pi}{D} \right) \right]}.
\end{align}
However, this solution does not directly apply to our problem since the boundary conditions are different. Indeed, we have $P_{n,D} = U_n $ and $P_{0,n} = V_n $ but we do not have $P_{D,n} $ or $P_{n,0}$. Instead, we have the following condition on the diagonal $k=l$: $\varphi_{k, k} = 0$. To address this difference, we generalize the method developed in Ref.~\cite{broom2008analysis} in the context of models with one individual per node of the line to the present case, and in particular to asymmetric transition rates. 
The method of Ref.~\cite{broom2008analysis} consists in finding the missing boundary conditions $P_{n,0}$ and $P_{D,n} $ by using the constraint $\varphi_{k,k} = 0$. To do so, we first note that 
\begin{equation}
	T^{(a,b)}_{k,k} = \frac{\left(\frac{p_2}{p_4}\right)^{b/2} \left(\frac{p_3}{p_1}\right)^{a/2}}{\left(\frac{p_2}{p_4}\right)^{a/2} \left(\frac{p_3}{p_1}\right)^{b/2}} T^{(b,a)}_{k,k} = \gamma^{a-b}T^{(b,a)}_{k,k}\,.
\end{equation}
Using this property, we obtain
\begin{align}
	\varphi_{k, k} = 0 = &\sum_{n = 1}^{D-1} T^{(n, D-1)}_{k,k} \left[ p_1 U_n  + p_4 P_{D,n} \gamma^{D-1-n} \right] + \sum_{n = 1}^{D-1} T^{(1,n)}_{k,k}\left[ p_2 V_{n}  + p_3 P_{n, 0}\gamma^{n-1}\right].
\end{align}
This equality is satisfied if
\begin{align}
	p_1 U_n  + p_4 P_{D,n} \gamma^{ D-1-n}&= 0\,, \\
	p_2 V_{n}  + p_3 P_{n, 0}\gamma^{n-1}  &= 0\,.
\end{align}
This yields
\begin{align}
	 P_{D,n} & = -  \gamma^{n-D} U_n\,,\\
	P_{n,0} & = - \gamma ^{-n} V_n\,.
\end{align}
Finally, the solution of \cref{eq:2d_markov} with the boundary conditions in \cref{bc_2d_markov1,bc_2d_markov2} reads 
\begin{align}
	\varphi_{k,l} &= \sum_{n = 1}^{D-1} p_1 U_n \left[ T^{(n,D-1)}_{k,l} - \gamma^{n-D+1} \:T^{(D-1,n)}_{k,l} \right]  + p_2V_n \left[ T^{(1,n)}_{k,l} - \gamma^{1-n} \:T^{(n,1)}_{k,l}  \right],
\end{align}
which finally leads to:
\begin{align}
	\varphi_{k,l} =\frac{1 }{(1+\alpha)(1+\gamma)} \sum_{n = 1}^{D-1} &\left\{ \alpha\:\dfrac{(\alpha\gamma)^{-n}  -(\alpha\gamma)^{-D} }{1 - (\alpha\gamma)^{-D}}  \left[ T^{(n,D-1)}_{k,l} - \gamma^{n-D+1}\: T^{(D-1,n)}_{k,l} \right] \right. \nonumber \\&\left. + \dfrac{1 - \left(\gamma / \alpha \right)^{n}}{1 -  \left(\gamma / \alpha\right)^{D}}  \left[ T^{(1,n)}_{k,l} - \gamma^{1-n}\:T^{(n,1)}_{k,l}  \right]\right\}.
\end{align}

The fixation probability in the line starting from one mutant in deme $k$ can be expressed as $\rho_{ij}\varphi_{k, k+1}$. By averaging out over all demes, assuming that the mutant appears in a deme chosen uniformly at random, we can write the overall mutant fixation probability as
\begin{equation}
\phi_{ij}= \frac{\rho_{ij}}{D} \sum_{k = 0}^{D-1}	\varphi_{k, k+1}, \label{eq:p_fixation_line}
\end{equation}
where $\varphi_{0,1} = V_1$, $\varphi_{D-1,D} = U_{D-1}$ and, $\forall k\in \left[1,D-2 \right]$, we have
\begin{align}
	\varphi_{k,k+1} =\frac{4 }{D^2(1+\alpha)(1+\gamma)} \sum_{n = 1}^{D-1} &\left\{\dfrac{(\alpha\gamma)^{-n}  -(\alpha\gamma)^{-D} }{1 - (\alpha\gamma)^{-D}} \frac{\gamma^{(n-D+2)/2}}{\alpha^{(2k-D-n)/2}} \left[ t^{(n,D-1)}_{k,k+1} -  t^{(D-1,n)}_{k,k+1} \right] \right. \nonumber \\&\left. +  \dfrac{1 - \left(\gamma / \alpha \right)^{n}}{1 -  \left(\gamma / \alpha\right)^{D}} \frac{\gamma^{(2-n)/2}}{\alpha^{(2k-n)/2}}  \left[ t^{(1,n)}_{k,k+1} - t^{(n,1)}_{k,k+1}  \right]\right\},
\end{align}
with
\begin{align}
t^{(a,b)}_{k,k+1} 
& = \sum_{r = 1}^{D-1}\sum_{s = 1}^{D-1} \frac{\sin\left(\frac{k r \pi}{D} \right) \sin\left(\frac{a r \pi}{D} \right) \sin\left(\frac{b s \pi}{D} \right) \sin\left(\frac{(k+1) s \pi}{D} \right)}{1- \frac{2 \sqrt{\alpha \gamma}}{(1+ \alpha)(1+\gamma)} \left[ \cos\left(\frac{r \pi}{D} \right) + \cos\left(\frac{s \pi}{D} \right) \right]}\,.
\end{align}

Using the symmetry properties of the line, the overall mutant fixation probability can be rewritten as
\begin{align}\label{eq:p_fixation_line_expanded}
	&\phi_{ij}= \frac{\rho_{ij}}{D} \left\{\dfrac{1 -{\gamma}/{\alpha} }{1 - \left({\gamma}/{\alpha} \right)^{D}} + \dfrac{1 -\alpha\gamma }{1 - (\alpha\gamma)^{D}} \right. \\ &\left. + \frac{4 }{D^2(1+\alpha)(1+\gamma)}\sum_{k = 1}^{D-2} \sum_{n = 1}^{D-1} \gamma^{1-\frac{n}{2}}\left[ \dfrac{1 -(\alpha\gamma)^{n} }{1 - (\alpha\gamma)^{D}} \alpha^{k+1-\frac{n}{2}} + \dfrac{1 - \left(\frac{\gamma}{ \alpha} \right)^{n}}{1 -  \left(\frac{\gamma}{ \alpha}\right)^{D}} \alpha^{\frac{n}{2}-k}\right] \nonumber\left( t^{(1,n)}_{k,k+1} - t^{(n,1)}_{k,k+1}  \right) \right\},
\end{align}

\begin{figure}[h!]
 \centering
 \includegraphics[width=0.7\textwidth]{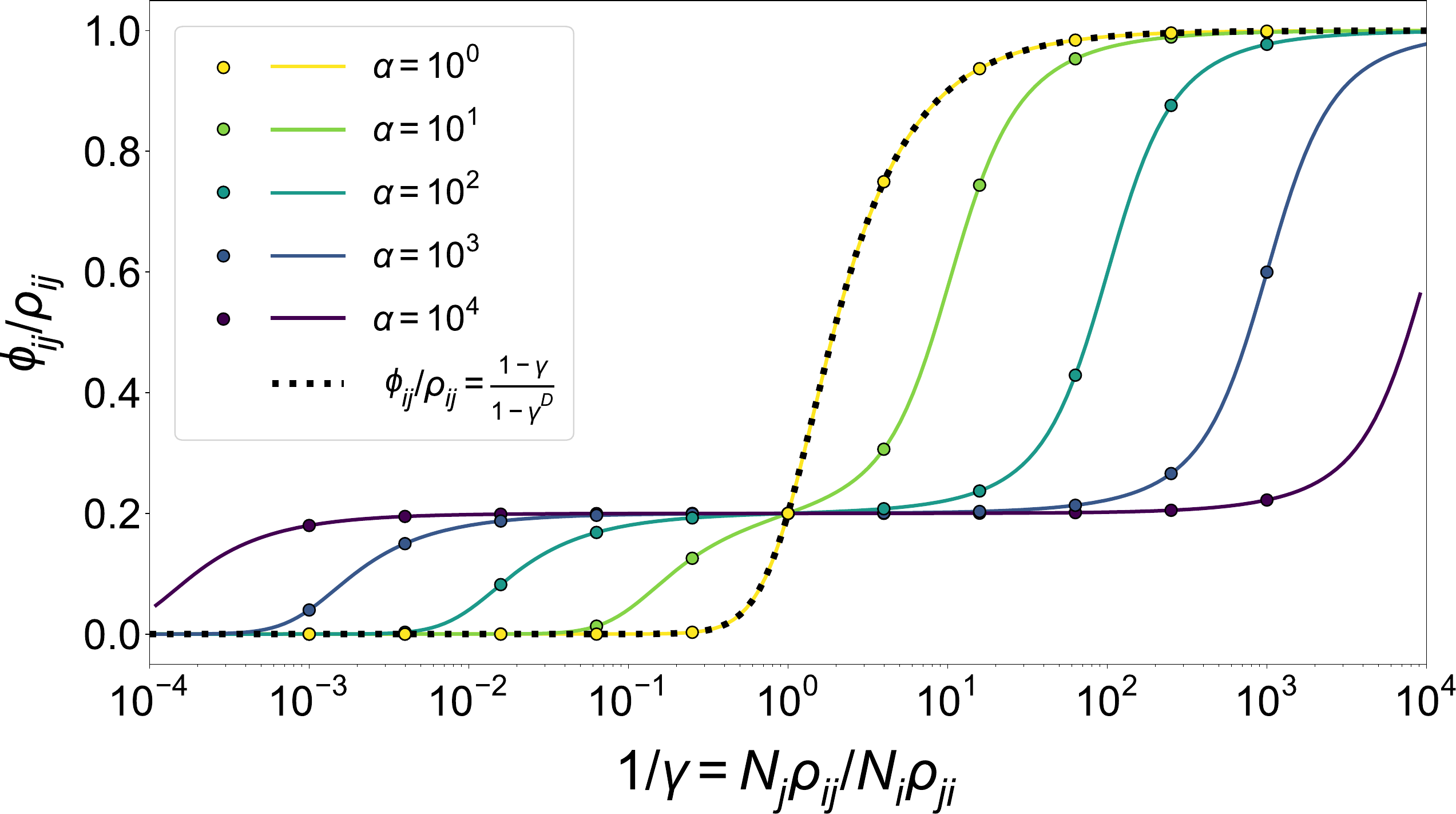}
\caption{\textbf{Mutant fixation probability in the line with rare migrations, starting from a fully mutant deme.} The mutant fixation probability starting from a fully mutant deme, obtained as $\phi_{ij} / \rho_{ij}$, is plotted versus $1/\gamma$ for different values of $\alpha$, with $D=5$ demes. Coloured lines correspond to our analytical prediction in \cref{eq:p_fixation_line_expanded}. The  black dotted line shows the fixation probability for circulation graphs (e.g.\ clique or cycle graphs, see~\cite{Marrec21,abbara2023frequent}). Markers correspond to the numerical simulation of the Markov chain given in \cref{eq:2d_markov}, and are obtained from $10^7$ simulation replicates.}
 \label{fig:fixation_probability_line}
\end{figure}

The ratio of the fixation probability given in \cref{eq:p_fixation_line_expanded} to that in one deme ($\rho_{ij}$) is plotted in \cref{fig:fixation_probability_line} for different values of migration asymmetry $\alpha$. This ratio represents the fixation probability in the line starting from one fully mutant deme chosen uniformly at random, in the rare migration regime. For $\alpha = 1$, we recover the fixation probability found for a circulation graph (see \cref{eq:clique} in the main text). For other values of $\alpha$, we observe in \cref{fig:fixation_probability_line} that the line graph with asymmetric migrations is a strong suppressor of selection compared to circulation graphs. 
In addition, we have the symmetry property $\phi_{ij}(\alpha, \gamma) =\phi_{ij}(1/\alpha, \gamma)$ and in the limit of strong migration rate asymmetries, we find 
\begin{equation}\label{eq:symmetry_line}
    \lim\limits_{\alpha \rightarrow 0} \frac{\phi_{ij}}{\rho_{ij}} = \lim\limits_{\alpha \rightarrow +\infty} \frac{\phi_{ij}}{\rho_{ij}} =\frac{1}{D}
\end{equation}
which corresponds to the strongest suppression of selection, where the fixation probability reduces to that of a neutral mutant whatever the fitness of the mutant.

\section{Studying the first fitness peak reached}
\label{sec:fsa-si}

\subsection{First step analysis (FSA)}

A first step analysis (FSA) can provide systems of equations that give the properties of the first fitness peak reached. We follow the approach presented in Ref.~\cite{servajean2023} for well-mixed populations (section~2(b) of the main text and section~S2 of the Supplementary Material). We adapt it to spatially structured populations, and present it here for completeness. 

\paragraph{Mean height $\bar{h}$.} To express the mean height $\bar{h}$ of the first fitness peak reached, we consider the first hitting times of the different peaks of the landscape~\cite{aldous2002}. Denoting by $M$ the set of genotypes that are peaks and by $T_j$ the first hitting time of $j\in M$ (i.e.\ the time to first reach genotype $j$), consider the probability $P_i\left(T_j = \min \left[T_k, k \in M\right]\right)$ that a walk starting from genotype $i$ hits $j$ before any other peak. Discriminating over all possible first steps of the walk yields
\begin{equation}
  P_i\left(T_j = \min \left[T_k, k \in M\right]\right) =
    \begin{cases}
      1 & \text{if } i = j\,,\\
      0 & \text{if } i \in M \text{ and } i \neq j\,,\\
      \sum_{l \in H_i} \tilde{\phi}_{il} \,P_l\left(T_j = \min \left[T_k, k \in M\right]\right) & \text{otherwise}\,,
    \end{cases}   
\label{eq:systemfsa}
\end{equation}
where $H_i$ is the set of neighbors of $i$ (i.e.\ the $L$ genotypes that differ from $i$ by only one mutation), while $\tilde{\phi}_{il}=\phi_{il}/\sum_{q \in H_i}\phi_{iq}$, where $\phi_{il}$ is the fixation probability of the mutation from $i$ to $l$ in the spatial structure considered. These fixation probabilities are analytically known in the rare migration regime for the star and the line, see \cref{eq:star,eq:line}. For more frequent migrations, we rely on simulations of the fate of single mutants to estimate fixation probabilities. Solving \cref{eq:systemfsa} yields the first hitting probabilities, and allows to calculate
\begin{equation} \label{eq:barh}
    \bar{h} = \frac{1}{2^L}\sum_{i \in H} \bar{h}_i \,,\,\,\,\,\text{with}\,\,\,\,\bar{h}_i =\sum_{j \in M}f_j\, P_i\left(T_j = \min \left[T_k, k \in M\right]\right)\,,
\end{equation}
where $H$ is the set of all genotypes (i.e.\ all vertices of the hypercube). We solve \cref{eq:systemfsa} numerically and use \cref{eq:barh} to obtain the FSA lines shown in our figures.

\paragraph{Mean time $\bar{t}$ and length $\bar{\ell}$.} The time $t$ of a walk is the total number of mutations that occur before the first fitness peak is reached (whether they fix or not). Meanwhile, the length $\ell$ of a walk is the number of successful fixations that occur before the first fitness peak is reached. 

In a given fitness landscape, the mean time $\bar{t}$ (resp.\ length $\bar{\ell}$) of a walk starting from a uniformly chosen genotype can be expressed as the average over all starting genotypes of the mean time $\bar{t_{i}}$ (resp.\ length $\bar{\ell_{i}}$) to reach the set $M$ of all peaks starting from genotype $i$:
\begin{equation}
    \bar{t} = \frac{1}{2^L}\sum_{i \in H} \bar{t_{i}} \,\,\,\textrm{and}\,\,\, \bar{\ell} = \frac{1}{2^L}\sum_{i \in H} \bar{\ell_{i}}\,.
\label{eq:l_and_t}
\end{equation}

To calculate $\bar{t}_i$, we use the transition probabilities $\phi_{il}/L$ to go from genotype $i$ to genotype $l$ upon a given mutation event, where $1/L$ is the probability that the mutation yields the neighbor $l$ of $i$, while $\phi_{il}$ is the fixation probability of this mutation. Discriminating over all possibilities for the first mutation yields
\begin{equation}
  \bar{t}_i =
    \begin{cases}
      0 & \text{if } i \in M\,,\\
      1 + \sum_{l \in H_i} \frac{1}{L} \phi_{il} \,\bar{t}_l + \sum_{l \in H_i} \frac{1}{L} (1-\phi_{il}) \,\bar{t}_i& \text{otherwise}\,.
    \end{cases}   
\label{eq:systemfsa_t}
\end{equation}

The same approach is used to calculate $\bar{\ell}_i$, but considering $\tilde{\phi}_{il}$ satisfying $\sum_{l\neq i} \tilde{\phi}_{il}=1$ for all $i$ (see above), instead of $\phi_{il}/L$. It gives
\begin{equation}
  \bar{\ell}_i =
    \begin{cases}
      0 & \text{if } i \in M\,,\\
      1 + \sum_{l \in H_i} \tilde{\phi}_{il} \,\bar{\ell}_l & \text{otherwise}\,.
    \end{cases}   
\label{eq:systemfsa_l}
\end{equation}

\subsection{Stochastic simulations} 

In the rare migration regime, we also perform direct stochastic simulations of walks, using a Monte Carlo procedure based on the fixation probability of each mutation. For this, we use the analytical expressions of fixation probabilities, see \cref{eq:star,eq:line} for the star and the line. Note that to avoid rejected moves, we simulate the embedded version of these Markov chains, where the transition probabilities to all neighbors of the current genotype are normalized to sum to one except to evaluate the mean time $\bar{t}$ of the walks.

\section{Impact of spatial structure on early adaptation in specific small landscapes}

\subsection{Specific small fitness landscapes considered}

\begin{figure}[h!]
 \centering
 \includegraphics[width=0.96\textwidth]{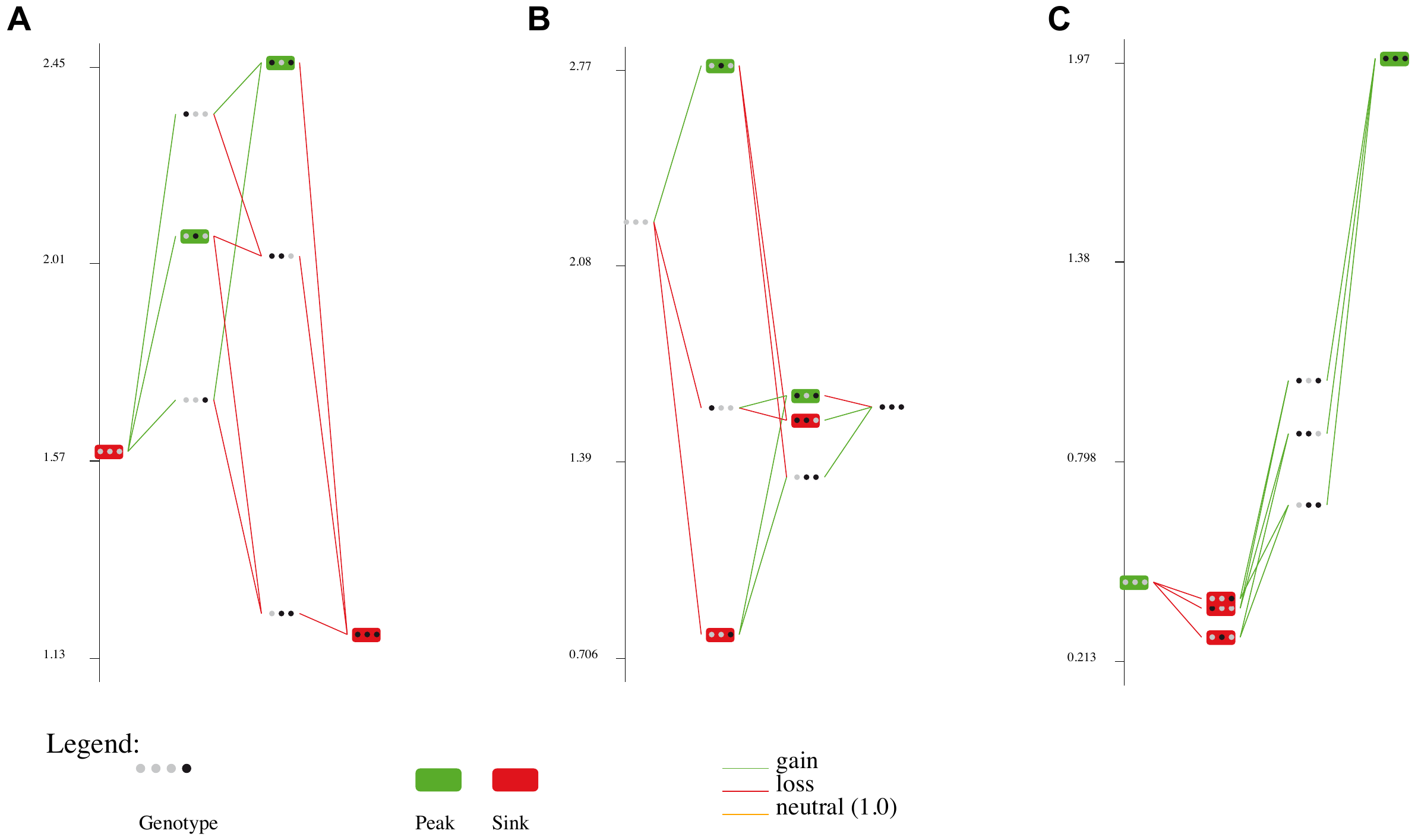}
\caption{\textbf{MAGELLAN representation~\cite{Brouillet2015} of the fitness landscapes A, B and C from \cref{fig:h_vs_D}(A-C).} The vertical axis denotes fitness, while the horizontal axis corresponds to the Hamming distance to the wild type. At each site, a gray marker denotes a wild-type genetic unit (nucleotide, amino acid or gene), while a black one denotes a mutated one.}
 \label{fig:landscape_fig2_magellan}
\end{figure}

We consider the three fitness landscapes A, B and C shown in \cref{fig:h_vs_D}(A-C). These three landscapes were generated within the $LK$ model with $L = 3$ and $K = 1$. Epistatic partners were chosen randomly (in the random neighborhood scheme \cite{nowak2015}) and fitness contributions were drawn from a uniform distribution between 0 and 1 (see \cref{eq:LK} in the main text). Fitness values are given in \cref{table:landscapes}, and the MAGELLAN representation is provided in \cref{fig:landscape_fig2_magellan}.

\begin{table}[h!]
\small
\centering
\begin{tabular}{c c c c} 
 \hline
Genotype & Fitness (landscape A) & Fitness (landscape B) & Fitness (Landscape C) \\ [0.5ex] 
 \hline
 (0, 0, 0) & 1.555 & 2.199 & 0.377 \\  [0.5ex] 
 (0, 0, 1) & 1.674 & 0.706 & 0.328 \\ [0.5ex] 
 (0, 1, 0) & 2.051 & 2.767 & 0.213 \\ [0.5ex] 
 (0, 1, 1) & 1.181 & 1.273 & 0.613 \\ [0.5ex] 
 (1, 0, 0) & 2.332 & 1.524 & 0.299 \\ [0.5ex] 
 (1, 0, 1) & 2.452 & 1.569 & 0.988 \\ [0.5ex] 
 (1, 1, 0) & 2.004 & 1.481 & 0.828 \\ [0.5ex] 
 (1, 1, 1) & 1.134 & 1.527 & 1.967 \\ [0.5ex] 
 \hline
\end{tabular}
\caption{\textbf{Specific $LK$ fitness landscapes with $L=3$ and $K=1$ used throughout this work.} The first column shows genotypes, while the next ones display the corresponding fitness values (rounded to 3 decimal places) in the three landscapes A, B and C shown in \cref{fig:h_vs_D}(A-C). }
\label{table:landscapes}
\end{table}

\newpage

\subsection{Impact of migration asymmetry on early adaptation}

\begin{figure}[h!]
 \centering
 \includegraphics[width=0.7\textwidth]{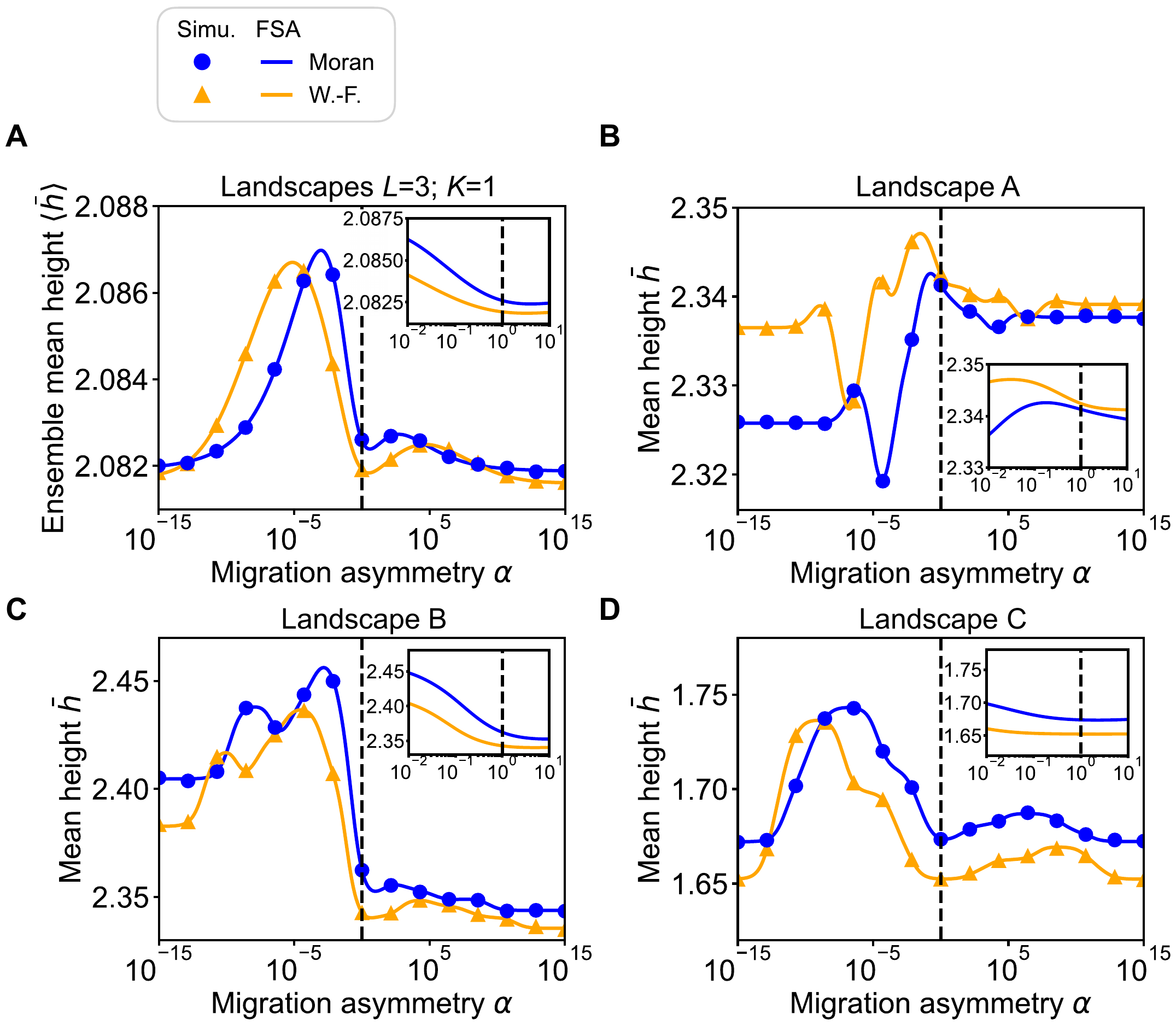}
 \caption{\textbf{Impact of the migration asymmetry $\alpha$ of a star on early adaptation in $LK$ landscapes with $L=3$ and $K=1$.} (A) Ensemble mean height $\left<\bar{h}\right>$ of the first fitness peak reached when starting from a uniformly chosen initial genotype versus $\alpha$. We consider the Moran and Wright-Fisher star walks under rare migrations. (B-D) Mean height $\bar{h}$ versus $\alpha$ in landscapes A, B and C, shown in \cref{fig:h_vs_D}(A-C). Solid lines correspond to numerical resolutions of the FSA \cref{eq:systemfsa,eq:barh}, while markers are simulation results averaged over $2\times10^5$ landscapes and 100 walks per starting genotype in each landscape for (A), and $10^5$ walks per starting genotype for (B-D). In all cases, parameter values are $C = 20$, $D = 5$ and $g = 0.01$.}
 \label{fig:h_vs_alpha}
\end{figure}

In \cref{fig:h_vs_alpha}, we report how migration asymmetry $\alpha$ in the star impacts the mean height $\bar{h}$ of the first peak that is reached. In particular, we observe that the most efficient early adaptation is generally obtained for small but intermediate $\alpha$ values.

\newpage

\subsection{Suppression of selection can lead to higher peaks and enhance finite-size effects} 
\label{sec:suppr-si}

\begin{figure}[h!]
 \centering
 \includegraphics[width=0.8\textwidth]{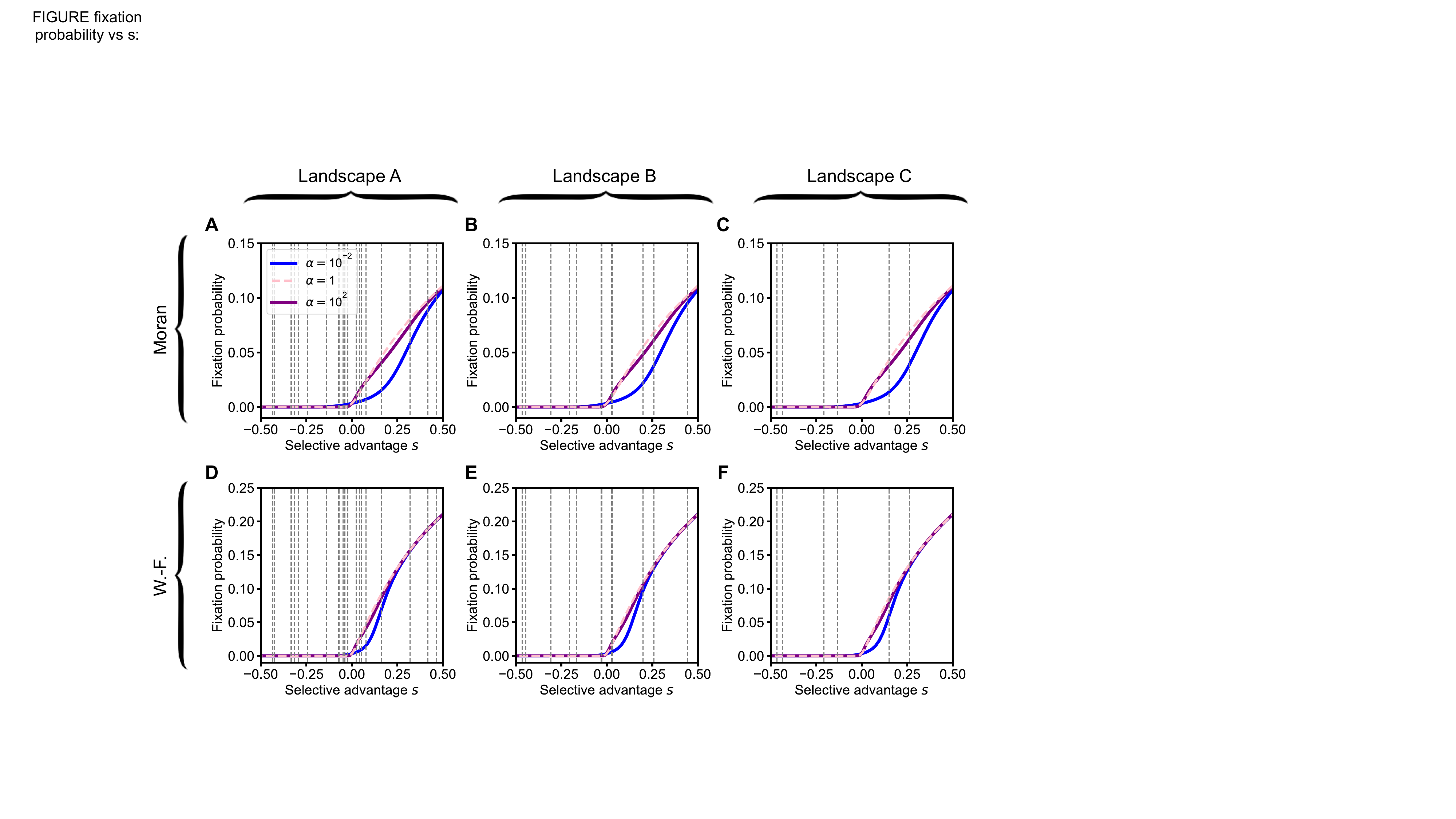}
 \caption{\textbf{Fixation probability of mutations within a star-structured population with rare migrations.} (A) Fixation probability of a mutation in a (Moran) star versus its relative selective advantage $s$ for different values of migration asymmetry $\alpha$ in a star with rare migrations~\cite{Marrec21}. The values of $s$  ranging from $-0.5$ to $0.5$  in landscape A (see \cref{table:landscapes}) are indicated by vertical gray dashed lines.  Outside this range of $s$, all relative differences between curves are smaller than 4 $\%$.  (B-C) Same as (A) but for landscapes B and C, respectively (same curves, but different range and different vertical gray dashed lines). (D-F) Same as (A-C) but for the Wright-Fisher star. In all cases, parameter values are $D = 5$, $C = 20$, and $g = 0.01$.}
 \label{fig:p_fix_vs_s}
\end{figure}

To understand the impact of $\alpha$ in more detail, let us examine how it affects the probability of fixation of available mutations. \cref{fig:p_fix_vs_s} shows the probability of fixation of a mutant versus its relative selective advantage $s$ for different values of migration asymmetry $\alpha$ in the star, highlighting the values of $s$ that exist in the three landscapes considered. Small $\alpha$ gives suppression of selection, and the range of small-effect mutations that are effectively neutral increases when $\alpha$ is decreased. Several mutations in the fitness landscapes considered fall in this range for $\alpha=10^{-2}$ (see \cref{fig:p_fix_vs_s}), and their fixation is thus expected to be particularly impacted. 

\begin{figure}[h!]
 \centering
 \includegraphics[width=0.85\textwidth]{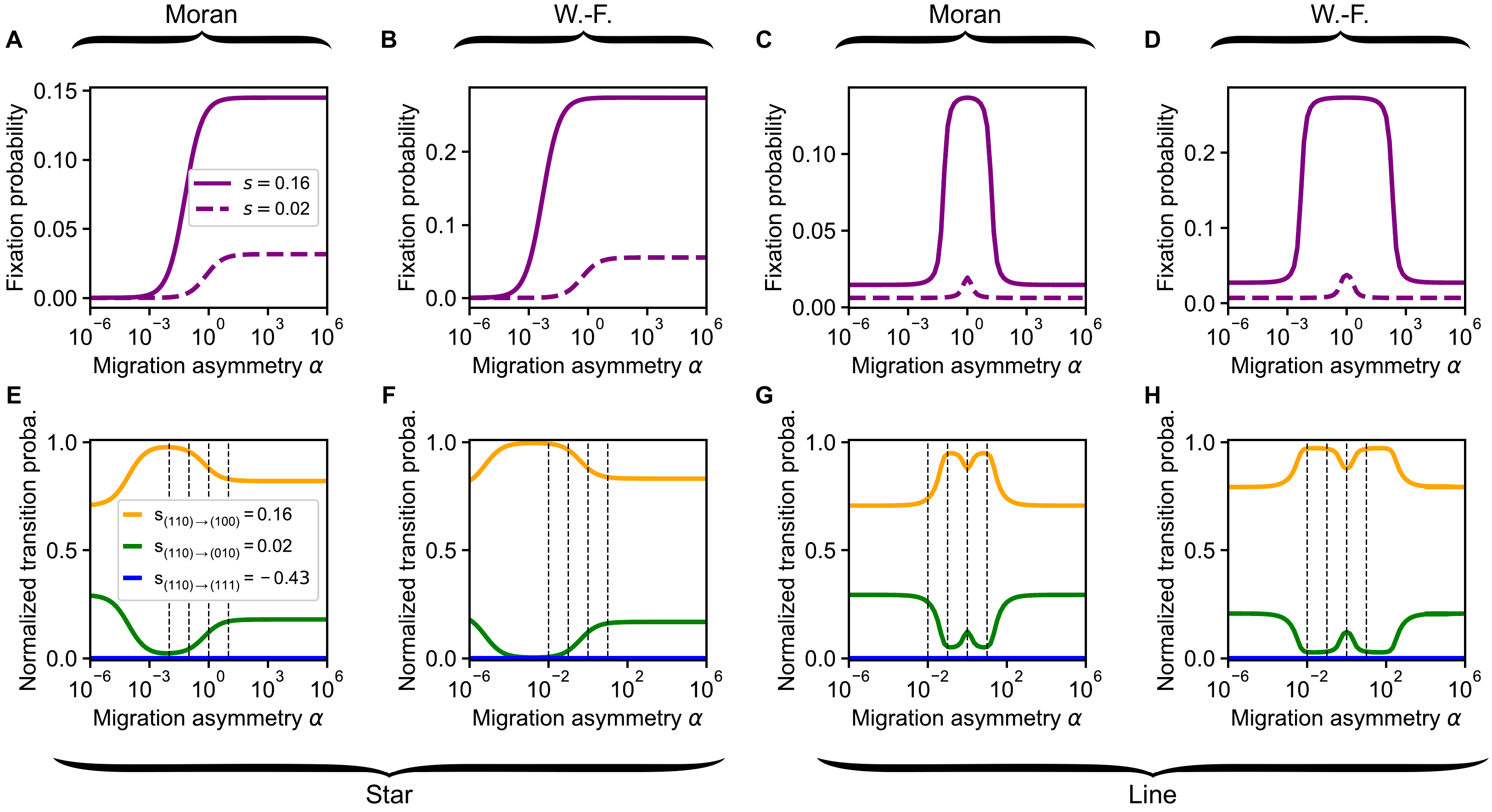}
 \caption{\textbf{Impact of the migration asymmetry $\alpha$ of a star and of a line on the fixation probability of different mutations.} (A) Fixation probability of mutations with respective relative fitness advantage values $s = 0.16$ and $s = 0.02$ in a Moran star for rare migrations versus migration asymmetry $\alpha$, for $D = 10^3$. These relative fitness advantage values correspond respectively to the transitions $(110) \rightarrow (100)$ and $(110) \rightarrow (010)$ within landscape A. (B) Same as (A) but for the Wright-Fisher star. (C-D) Same as (A-B) but for the line with $D = 10$. (E) Probabilities of transitions $(110) \rightarrow (100)$, $(110) \rightarrow (010)$ and $(110) \rightarrow (111)$ within landscape A, normalized so that they sum to 1, versus $\alpha$, for a Moran star with rare migrations, for $D = 10^3$. (B) Same as (A) but for the Wright-Fisher star. (C-D) Same as  (A-B) but for the line with $D = 10$. In all cases, parameter values are $C = 20$ and $g = 0.01$.}
 \label{fig:p_fix_vs_alpha}
\end{figure}

\begin{figure}[h!]
 \centering
 \includegraphics[width=0.87\textwidth]{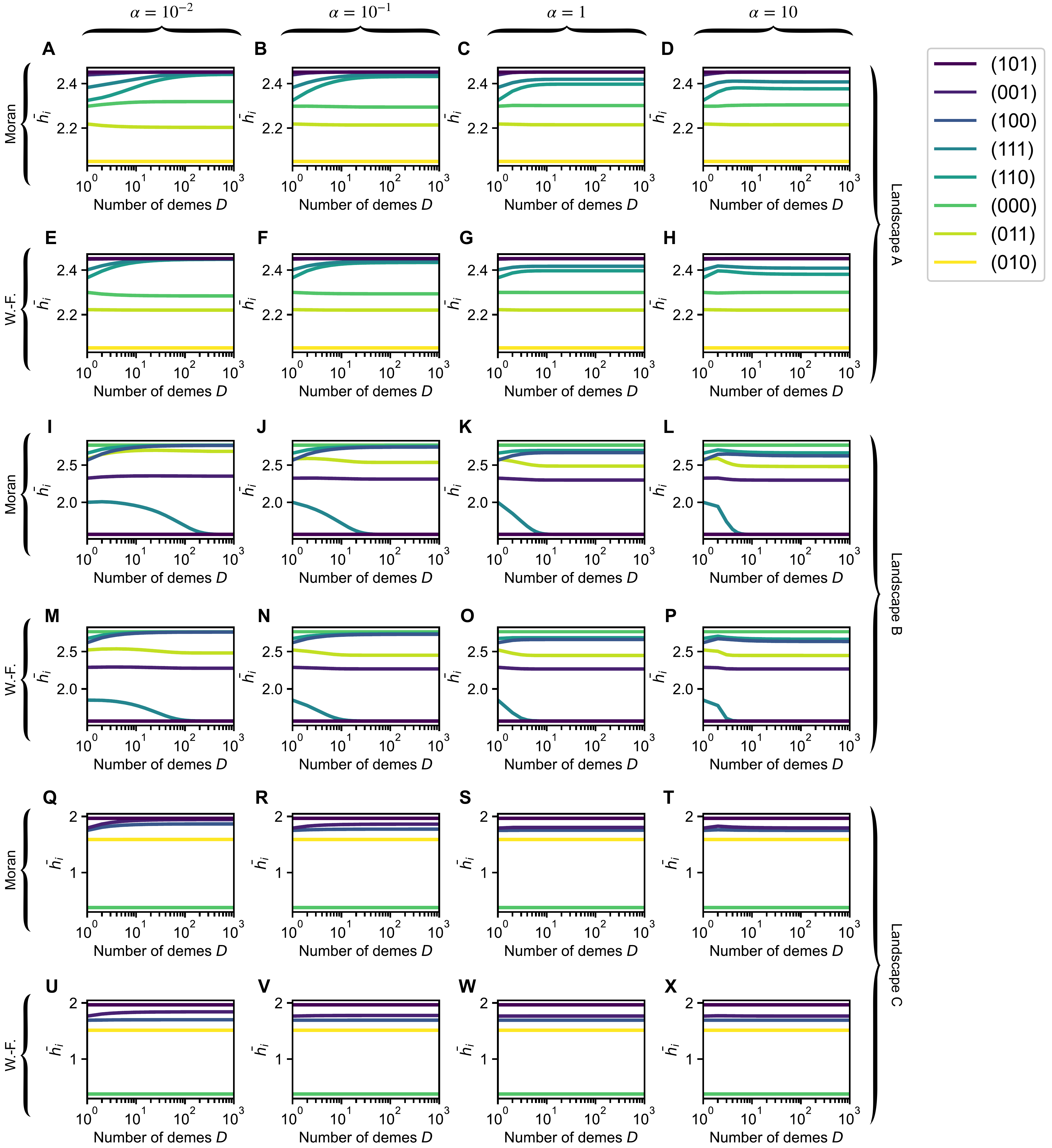}
 \caption{\textbf{Impact of the number of demes $D$ and migration asymmetry $\alpha$ of a star on early adaptation starting from a specific genotype $i$, in $LK$ landscapes with $L=3$ and $K=1$.} Mean height $\bar{h_i}$, starting from a specific genotype $i$, versus $D$ in landscape A (A-H), landscape B (I-P), and landscape C (Q-X), for the Moran star walk with rare migrations (first row for each landscape) and the Wright-Fisher one (second row), for different migration asymmetry values (different columns). Landscapes A, B and C are shown in \cref{fig:h_vs_D}(A-C) and \cref{fig:landscape_fig2_magellan}, and the corresponding fitness values are provided in \cref{table:landscapes}. Lines are numerical resolutions of the FSA \cref{eq:systemfsa,eq:barh} for each landscape. Colors denote the starting genotype. In landscape A, $h_{(101)}$, $h_{(001)}$ and $h_{(100)}$ are indistinguishable. In landscape B, $h_{(010)}$ and $h_{(000)}$ are indistinguishable. In landscape C, $h_{(111)}$, $h_{(011)}$, $h_{(101)}$ and $h_{(110)}$ are indistinguishable. In all cases, parameter values are $C = 20$ and $g = 0.01$, as in \cref{fig:h_vs_D}.}
 \label{fig:hi}
\end{figure}

As a concrete example, \cref{fig:p_fix_vs_alpha} focuses on the three transitions accessible from genotype 110 within landscape A, which are two beneficial mutations, one with a larger $s$ than the other, and one deleterious mutation. Fixation of the latter can be neglected when considering large values of $D$ (\cref{fig:p_fix_vs_alpha}(A,B) and (E,F) use $D=10^3$). Then, the key question is which of the two beneficial mutations fixes. 
\cref{fig:p_fix_vs_alpha}(A,B) shows that the fixation probability of each of them increases when $\alpha$ increases, as the star shifts from suppression to amplification of selection. However, this increase requires larger $\alpha$ values for the least beneficial mutation of the two, which is more subject to suppression (see \cref{fig:p_fix_vs_alpha}(A,B) and \cref{fig:p_fix_vs_s}). To see how the choice between the two mutations is impacted, we normalize these two fixation probabilities by their sum for each $\alpha$. \cref{fig:p_fix_vs_alpha}(E,F) shows that the resulting transition probabilities vary non-monotonically with $\alpha$, with the most beneficial mutation being most favored for $\alpha\approx 10^{-2}$, as the weakly beneficial one is then still strongly suppressed. Further increasing $\alpha$ reduces the bias toward this most beneficial mutation. 
In this example, the most beneficial mutation yields genotype 100, from which the only beneficial mutation leads to the highest peak (genotype 101). Thus, the more favored this transition, the larger $\bar{h}$. This is one of the reasons why the mean height $\bar{h}$ is larger as $\alpha$ is decreased down to $10^{-2}$ in landscape A for large $D$, see \cref{fig:h_vs_D}(D,G) and \cref{fig:hi}. Favoring the most beneficial mutations relative to the other ones often promotes reaching higher peaks. This rationalizes our striking observation that small $\alpha$ values, giving suppression of selection in the star, generally make early adaptation more efficient for large $D$, see \cref{fig:h_vs_D}. 
Among beneficial mutations, suppression of selection most strongly suppresses the fixation of weakly beneficial ones, giving a relative advantage to strongly beneficial ones.

\begin{figure}[h!]
 \centering
 \includegraphics[width=0.77\textwidth]{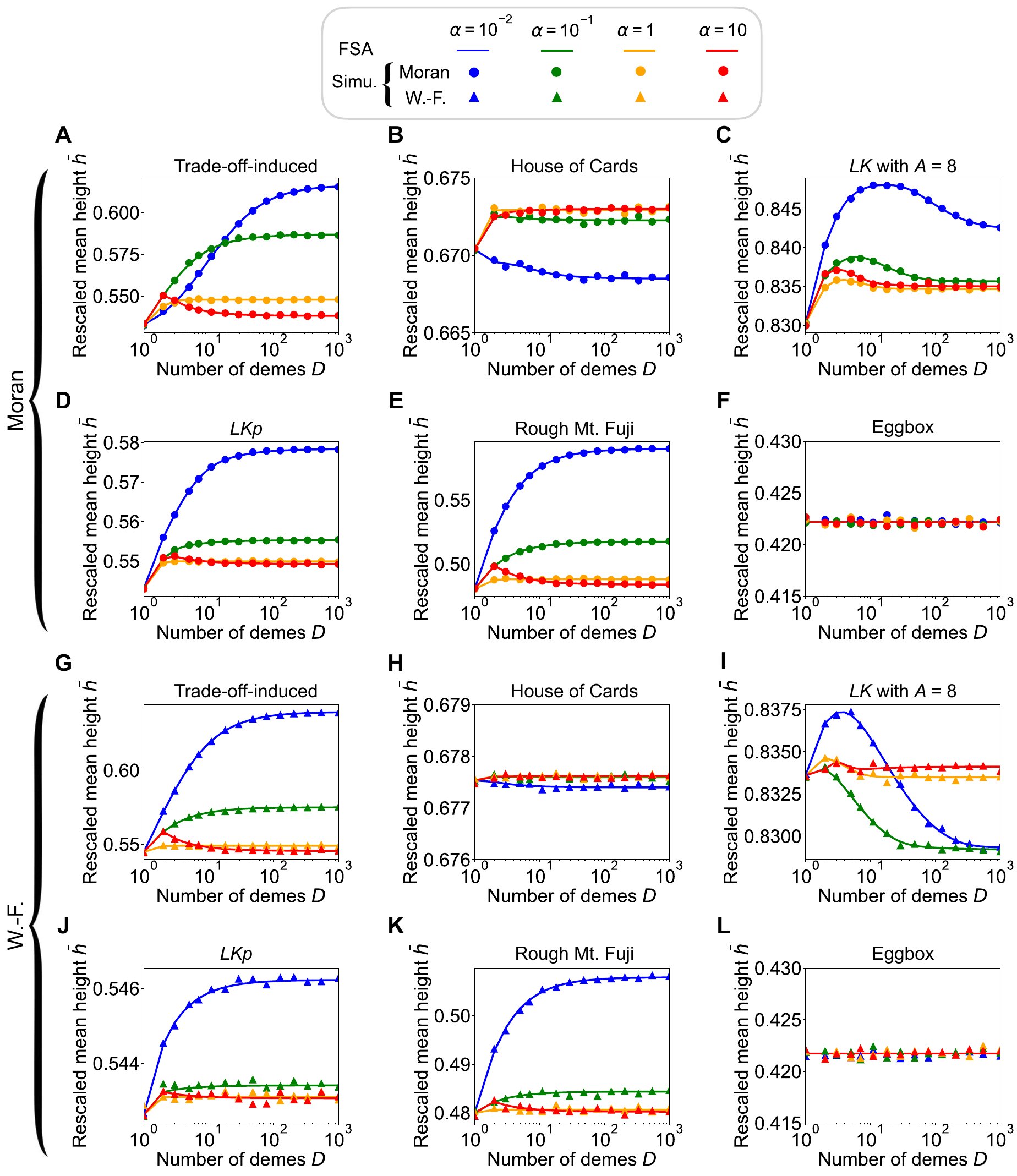}
 \caption{\textbf{Impact of the number of demes $D$ of a star on early adaptation in various model landscapes with $L = 3$.} (A) The rescaled mean height $\bar{h}$ of the first fitness peak reached when starting from a uniformly chosen initial genotype is shown versus $D$ for the Moran star walk under rare migrations for various values of $\alpha$, in a landscape generated by the trade-off-induced landscape model \cite{das2020, das2022driven}. The rescaled mean height is defined as $(\bar{h}-h_\textrm{min})/(h_\textrm{max}-h_\textrm{min})$, where $h_\textrm{min}$ (resp.\ $h_\textrm{max}$) denotes the fitness of the lowest peak (resp.\ highest peak). Recall that $h_\textrm{min}<\bar{h}<h_\textrm{max}$. Landscape parameters \cite{das2020, das2022driven}: dimensionless antibiotic concentration $c = 50$; exponent of the Hill function $a = 1$; $r_i$ and $m_i$ drawn from the distribution given in Eq.~(S16) of the supplementary material of Ref.~\cite{servajean2023}. (B-F) Same as (A) but in a landscape generated by (B) the House of Cards model \cite{Kauffman1987, kingman1978}, (C) the $LK$ model with non-binary genotypes (with cardinal of the alphabet $A = 8$) \cite{zagorski2016}, (D) the $LKp$-like model with $p = 0.5$ and $q = 0.1$ \cite{barnett1998}, (E) the Rough Mount Fuji model \cite{aita2000, szendro2013b} with $f_0' = 5$, $C = 0.8$ and epistatic contributions drawn from a standard normal distribution, (F) the Eggbox model \cite{ferretti2016} with fitnesses drawn from a Gaussian of mean $ f_0 \pm \mu_E/2$, with $f_0 = 4$, $\mu_E = 4$, and a standard deviation of 1. See section~S4 in the supplementary material of Ref.~\cite{servajean2023} for more details on these models. (G-L) Same as (A-F) but for the Wright-Fisher star walk with rare migrations. In all panels, lines are numerical resolution of the FSA \cref{eq:systemfsa,eq:barh}, while markers are simulation results averaged over $10^5$ to $10^7$ walks per starting point ($5\times10^2$ to $5\times10^3$ in panels (C) and (I) for the non-binary $LK$ landscape that includes more genotypes than others). In all cases, $L = 3$, $C = 20$ and $g = 0.01$.}
\label{fig:beyond_LK_L=3}
\end{figure}

Note that varying $\alpha$ can have various impacts. In our previous example, decreasing $\alpha$ below $10^{-2}$ reduces the bias toward the most beneficial mutation, as its fixation also becomes suppressed, see \cref{fig:p_fix_vs_alpha}(A,B) and (E,F). Accordingly, \cref{fig:h_vs_alpha} shows that the most efficient early adaptation is generally obtained for small but intermediate $\alpha$ values. Besides, for each genotype, the values of $\alpha$ that impact the bias between possible transitions depend on the fitness values of neighboring genotypes (and on the values of $C$ and $D$). In particular, we argued that favoring the most beneficial mutations often promotes reaching higher peaks, but there are fitness landscapes in which this does not hold. For instance, \cref{fig:beyond_LK_L=3}(B,H) features a House of Cards landscape where decreasing $\alpha$ reduces $\bar{h}$ for large $D$. Detailed analysis reveals that, in this landscape, a mutation leading to the highest peak has a relatively small fitness advantage, and smaller $\alpha$ values yield biases against it. 

\cref{fig:p_fix_vs_alpha}(C,D) shows the fixation probabilities of the two beneficial mutations available from genotype 110 in landscape A, this time in the line. They are largest for $\alpha = 1$, as the line suppresses selection for all $\alpha\neq 1$. Furthermore, the decay of the fixation probability when $\alpha$ becomes larger or smaller than 1 occurs at more extreme $\alpha$ values for the most beneficial mutation. Normalizing these two fixation probabilities by their sum thus yields a non-monotonic dependence on $\alpha$, see \cref{fig:p_fix_vs_alpha}(G,H). 
For the Moran line walk, there is far less bias against the least beneficial transition among the two when $\alpha = 10^{-2}$ than when $\alpha=10^{-1}$. This least beneficial mutation leads to the small peak (010), explaining the results in \cref{fig:line}(A).

\clearpage
\newpage
\cref{fig:hi}(I,M) shows the mean height $\overline{h_i}$ of the star walk for each starting genotype $i$ for $\alpha = 10^{-2}$ in landscape B. It reveals that the presence of the maximum of $\bar{h}$ versus $D$ in landscape B for $\alpha = 10^{-2}$ is due to the fact that, starting from genotype 111, the transition towards the small peak becomes more favored when increasing $D$. Indeed, $\overline{h_{111}}$ decreases with $D$. Together with the constant or increasing trends of the other $\overline{h_i}$, which plateau for smaller $D$, this gives rise to the maximum observed when averaging over $i$. For larger $\alpha$, $\overline{h_{111}}$ also decreases, but plateaus for smaller values of $D$, see \cref{fig:hi}(I-P). These observations can be explained as follows. Genotype 111 has neighbors with similar fitnesses, but only the mutation to the small peak is beneficial, see \cref{fig:h_vs_D}(B). As $D$ increases, the walk becomes more biased toward beneficial mutations, and this transition is more favored. However, in the star, small $\alpha$ values lead to suppression of selection (see \cref{fig:p_fix_vs_s}), and this transition then occurs for larger $D$. 

In \cref{fig:beyond_LK_L=3}, we study finite size effects beyond $LK$ landscapes. We generally observe similar trends as in $LK$ landscapes, except in \cref{fig:beyond_LK_L=3}(B,H), as discussed above.

\begin{figure}[h!]
 \centering
 \includegraphics[width=0.85\textwidth]{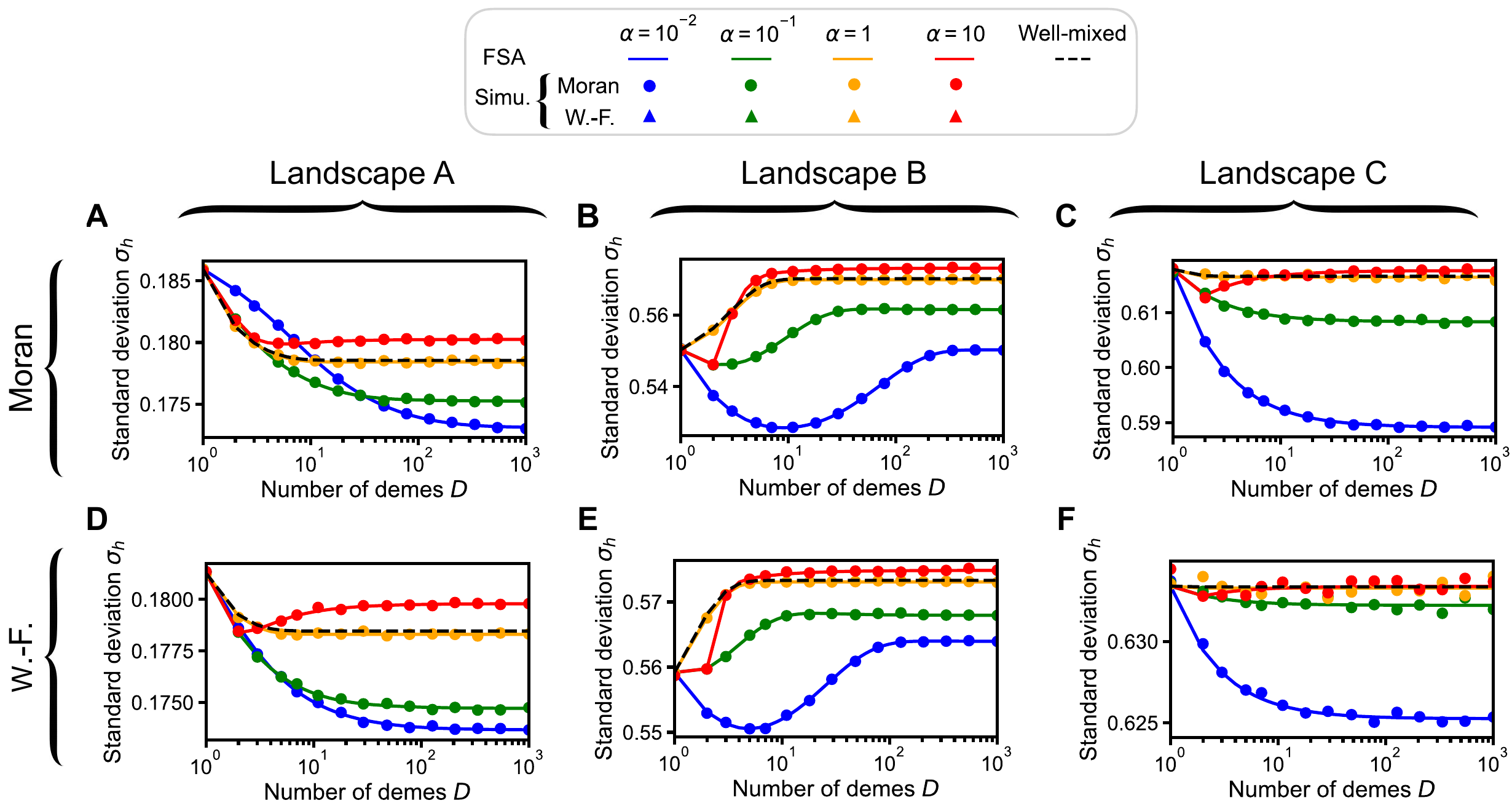}
\caption{\textbf{Impact of the number of demes $D$ of a star on the standard deviation of the height $h$ of a walk, in $LK$ landscapes with $L=3$ and $K=1$.} (A,B,C) The standard deviation of the height $h$ of a walk, starting from a uniformly chosen initial genotype, is shown versus $D$ for the Moran star walk with rare migrations in landscapes A, B and C, respectively. (D,E,F) Same as (A,B,C), but for the Wright-Fisher star walk with rare migrations. In all cases, simulation results are averaged over at least $10^5$ walks per starting genotype. Parameter values: $C = 20$ and $g = 0.01$.}
\label{fig:sigma_h_vs_D}
\end{figure}

\begin{figure}[h!]
 \centering
 \includegraphics[width=0.85\textwidth]{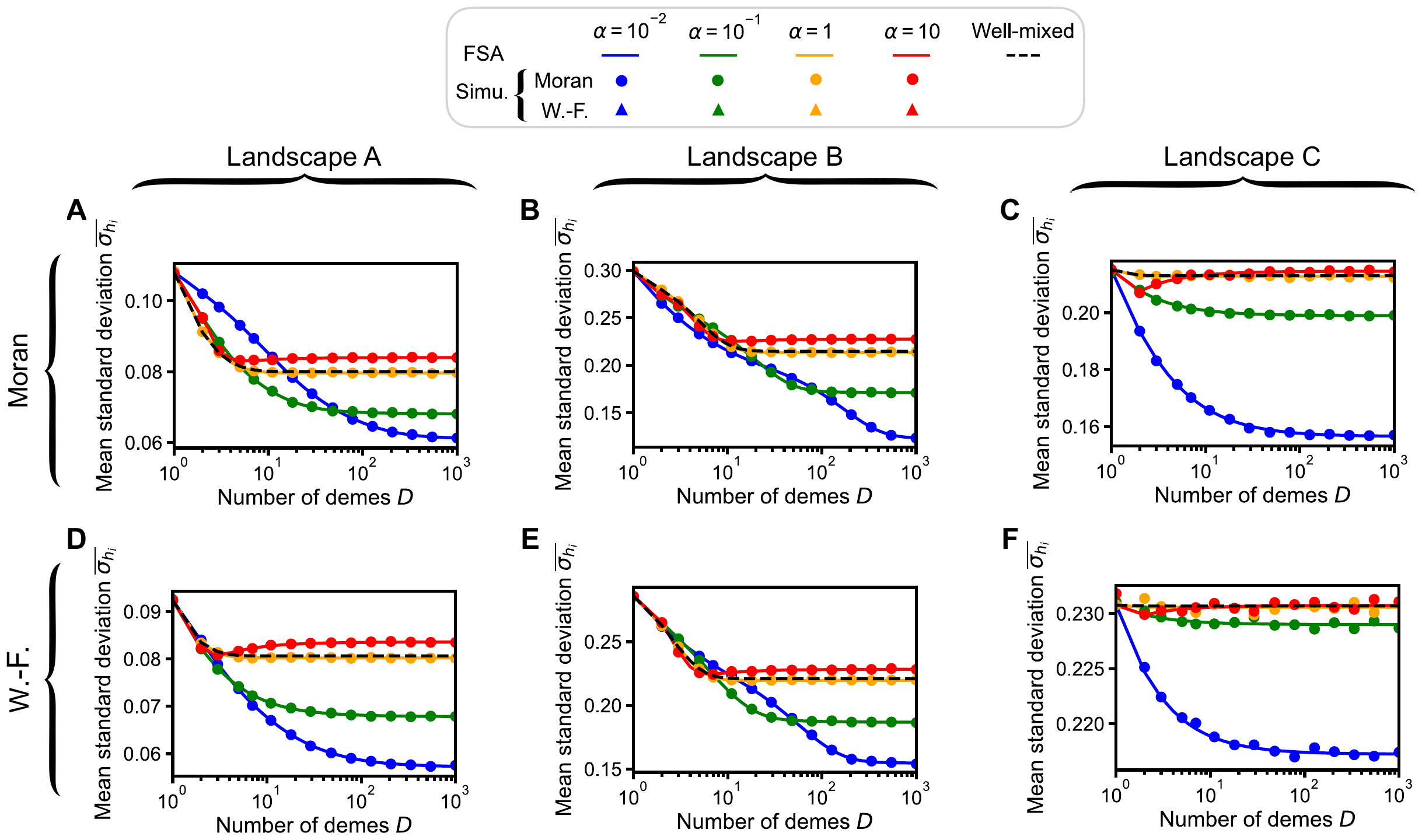}
\caption{\textbf{Impact of the number of demes $D$ of a star on the mean standard deviation of the height $h_i$ of a walk, in $LK$ landscapes with $L=3$ and $K=1$.} Same as \cref{fig:sigma_h_vs_D} but for the mean standard deviation $\overline{\sigma_{h_i}}$ of the height $h_i$ of the first fitness peak reached when starting from genotype $i$, averaged over all possible starting genotypes $i$.}
\label{fig:sigma_hi_vs_D}
\end{figure}

\cref{fig:sigma_h_vs_D} further shows that the standard deviation of $h$, starting from uniformly chosen genotypes, essentially mirrors $\bar{h}$ when plotted versus $D$. This holds for all values of $\alpha$, and is in line with observations for well-mixed populations \cite{servajean2023}. Hence, the maximum of $\bar{h}$ observed for $\alpha = 10^{-2}$ in landscape B is associated with a higher predictability. Finally, \cref{fig:sigma_hi_vs_D} shows that the mean standard deviation $\overline{\sigma_{h_i}}$ (averaged over the starting points $i$), mainly decreases with $D$, which is consistent with the idea that larger values of $D$ yield more biased walks and more reproducibility. 

\newpage

\subsection{Time to reach the first peak}
\label{sec:time}

So far, we mainly investigated the impact of spatial structure on the height $h$ of the first fitness peak reached. It is also interesting to ask how long it takes to reach this first peak.

\begin{figure}[h!]
 \centering
 \includegraphics[width=\textwidth]{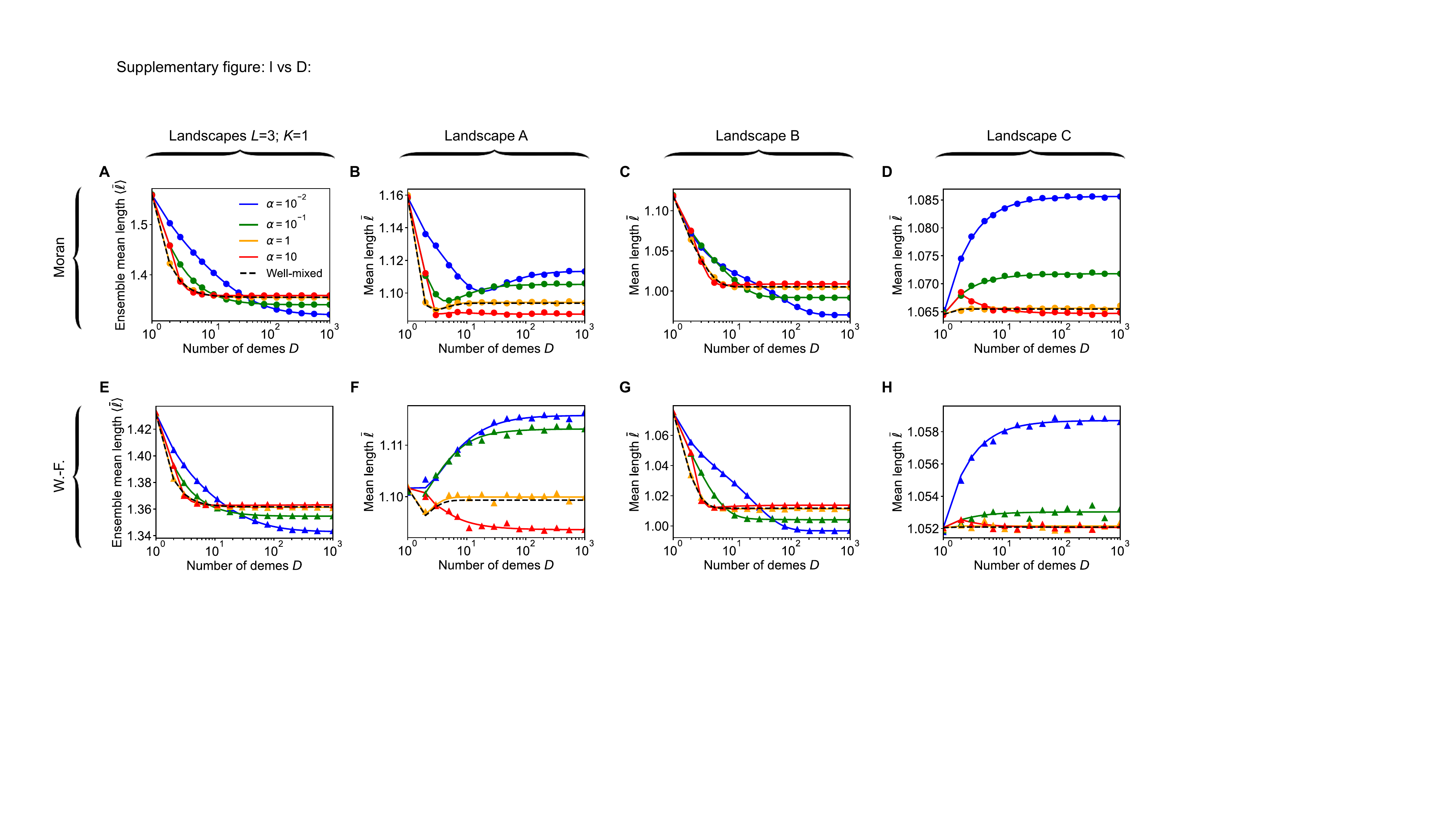}
 \caption{\textbf{Impact of the number of demes $D$ of a star on the length $\ell$ of a walk, in $LK$ landscapes with $L=3$ and $K=1$.} (A)  The ensemble mean length $\left\langle \bar{\ell} \right\rangle$ of a walk (number of successful mutations before reaching the first fitness peak), when starting from a uniformly chosen initial genotype, is shown versus $D$ for the Moran star walk with rare migrations. Different values of migration asymmetry $\alpha$ are considered. The well-mixed case is shown for reference (dashed black line). Lines: numerical resolutions of the FSA equations (see \cref{eq:systemfsa_l,eq:systemfsa_t,eq:l_and_t}) for each landscape; markers: simulation results averaged over $10$ walks per starting genotype in each landscape. The ensemble average is then performed over $2\times 10^5$ landscapes. (B,C,D) The mean length $\bar{\ell}$ is shown versus $D$ for the Moran star walk with rare migrations in landscapes A, B and C from \cref{fig:h_vs_D}(A-C). Simulation results are averaged over at least $10^5$ walks per starting genotype. Same colors, markers and lines as in (A). (E,F,G,H) Same as (A,B,C,D) but for the Wright-Fisher star walk with rare migrations. In all cases, parameter values are $C = 20$ and $g = 0.01$.}
\label{fig:l_vs_D}
\end{figure}

Let us consider the length $\ell$ of a walk, which represents the number of mutation fixation events that occur before a fitness peak is reached. For this, we focus on the star structure. \cref{fig:l_vs_D}(A,E) shows that the ensemble mean length $\left<\bar{\ell}\right>$ decreases when $D$ increases.  
Indeed, larger values of $D$ give a stronger bias towards the most beneficial mutations to the walk, so fewer steps are generally needed to reach a peak. Nevertheless, specific landscapes can lead to a non-monotonic (landscape A, see \cref{fig:l_vs_D}(B,F)) or increasing (landscape C, see \cref{fig:l_vs_D}(D-H)) behavior of $\bar{\ell}$ versus $D$. 
Recall that landscape C comprises a high peak (genotype 111) and a much lower peak (genotype 000), see \cref{fig:h_vs_D}(C). Consider a population that starts from one of the neighboring genotypes of 000. Increasing $D$ increases the bias towards the fittest neighbors. This makes the population more likely to mutate to a neighbor of the high peak 111, instead of hitting the low peak 000, which leads to an increase of both $\ell$ and $h$.

\begin{figure}[h!]
 \centering
 \includegraphics[width=\textwidth]{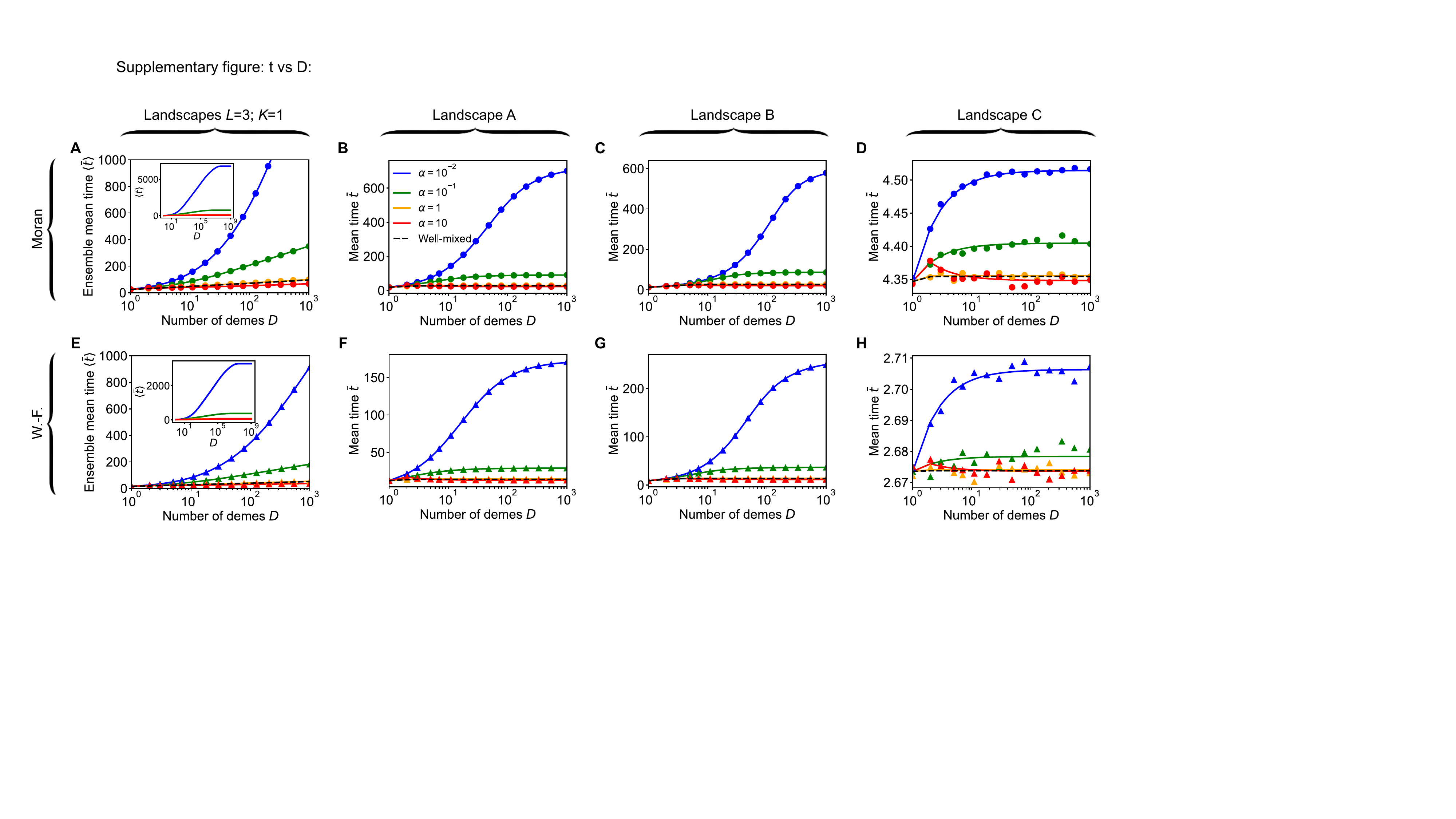}
\caption{\textbf{Impact of the number of demes $D$ of a star on the time $t$ of a walk, in $LK$ landscapes with $L=3$ and $K=1$.} Same as \cref{fig:l_vs_D} but for the time $t$, which represents the number of all mutations (successful or not) before reaching a fitness peak (see main text).}
\label{fig:t_vs_D}
\end{figure}

While $\ell$ describes the number of successful mutations, it is also interesting to consider the number $t$ of all mutations that occurred before a peak is reached, whether they fixed or not. \cref{fig:t_vs_D} shows that it increases with $D$. Indeed, increasing $D$ reduces all fixation probabilities, which usually increases the number of unsuccessful mutations. Note however that, in landscape C, $\bar{t}$ decreases when $D$ increases above 2 when $\alpha = 10$, see \cref{fig:t_vs_D}(D,H). As explained above, the specific features of this landscape entail that $\bar{\ell}$ and $\bar{t}$ have a similar dependence on $D$ as $\bar{h}$. Another striking result is that smaller values of $\alpha$ lead to substantially larger values of $\bar{t}$, see \cref{fig:t_vs_D}. This is due to suppression of selection, which results in more mutations being effectively neutral and having small fixation probabilities. Thus, while suppression of selection often allows to reach higher peaks first, this tends to require more time.

\clearpage

\subsection{Impact of the carrying capacity of a deme}

\begin{figure}[h!]
 \centering
 \includegraphics[width=\textwidth]{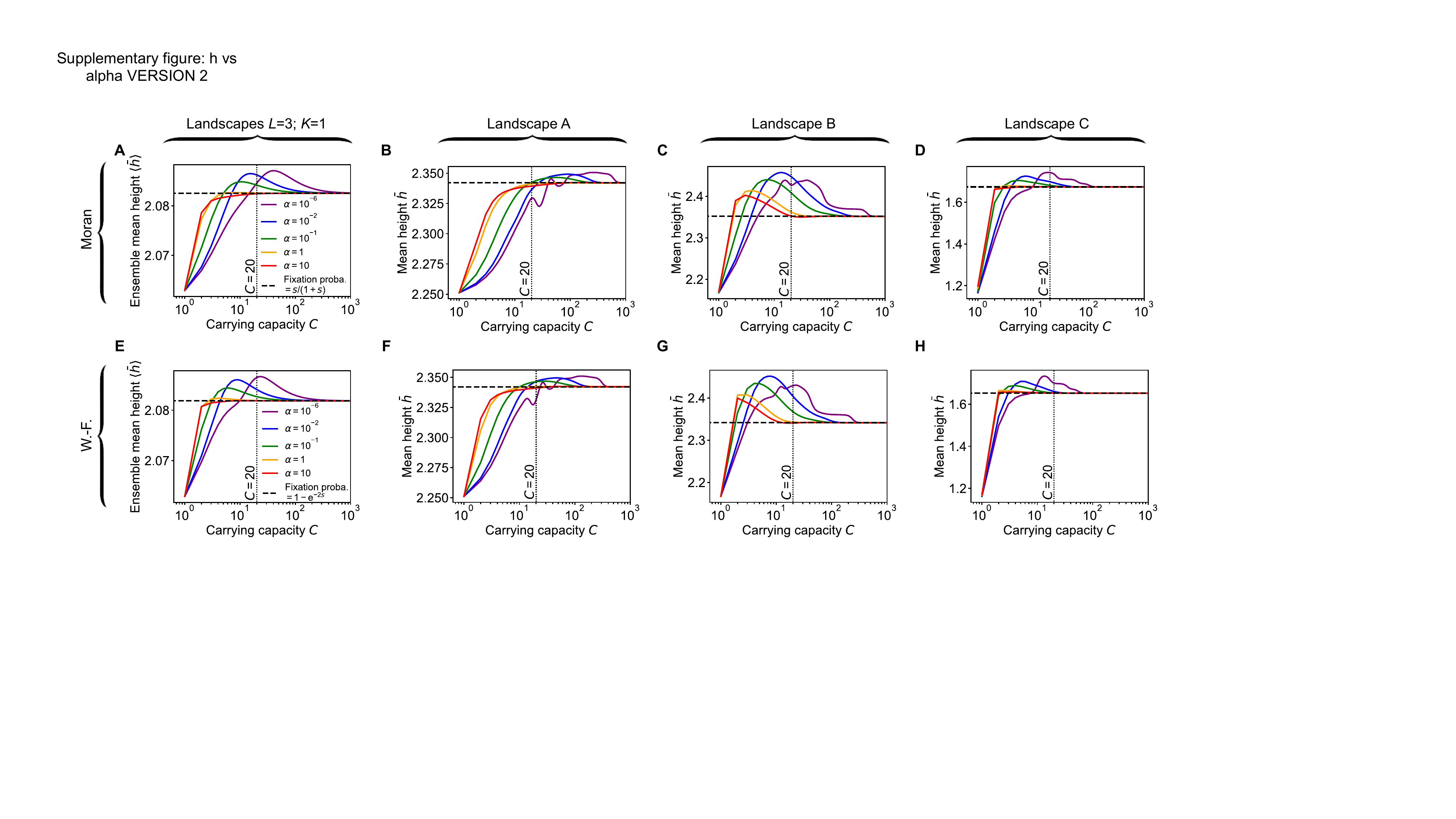}
 \caption{\textbf{Impact of the carrying capacity $C$ of the demes of a star on early adaptation in $LK$ landscapes with $L=3$ and $K=1$.} (A) Ensemble mean height $\left\langle \bar{h} \right\rangle$ of the first peak reached, starting from a uniformly chosen initial genotype, versus $C$ for the Moran star walk with rare migrations with various migration asymmetry values $\alpha$. Lines: numerical resolutions of the FSA equations for each landscape. The ensemble average is then performed over $2\times 10^5$ landscapes. (B,C,D) $\bar{h}$ versus $C$ for the Moran star walk in landscapes A, B and C, respectively. Lines: numerical resolutions of the FSA~\cref{eq:systemfsa,eq:barh}. (E,F,G,H) Same as (A,B,C,D) but for the Wright-Fisher star walk with rare migrations. In all panels, the vertical dotted line indicates $C = 20$, and parameter values are $D = 5$ and $g = 0.01$. The horizontal dashed lines correspond to $\bar{h}$ in the large-size limit of a well-mixed population either under the Moran model (left column) or the Wright-Fisher model (right column) (see Ref.~\cite{servajean2023}).}
 \label{fig:h_vs_C}
\end{figure}

In \cref{fig:h_vs_C}, we report the mean height $\bar{h}$ versus deme carrying capacity $C$ for various values of migration asymmetry $\alpha$ in the star. In particular, we observe that all curves converge to the limiting result for a large well-mixed population (dashed horizontal black lines). As explained in the main text, when demes are large,
deleterious mutations cannot fix within their deme of origin. Meanwhile, if a beneficial mutant fixes in the (large) deme where it appeared, its fixation in the whole structure is ensured.

\section{Extension to demes of different sizes: the doublet}
\label{sec:doublet}

\subsection{The doublet structure}

\begin{figure}[h!]
 \centering
 \includegraphics[width=0.25\textwidth]{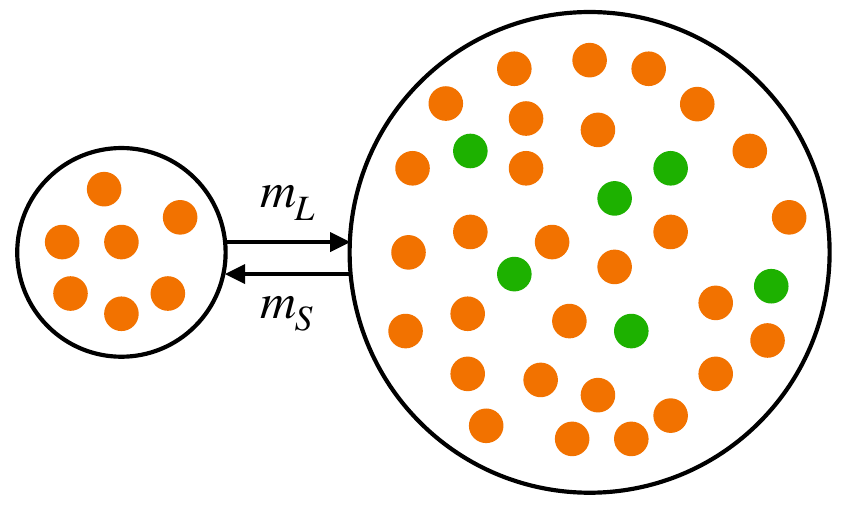}
\caption{\textbf{Doublet structure.} Two demes of different sizes are connected by migrations. $m_L$ denotes the migration rate per individual from the small deme to the large deme (right) and $m_S$ the migration rate in the opposite direction. 
}
 \label{fig:doublet_structure}
\end{figure}

As mentioned in the main text, our model, based on Refs.~\cite{Marrec21,abbara2023frequent}, allows to treat the case of demes with different sizes. Here, we consider the doublet, composed of a small deme and a larger one, connected by migrations that can be asymmetric, illustrated in \cref{fig:doublet_structure}. For ease of comparison with other structures discussed here, we assume that the small (resp.\ large) deme has a carrying capacity $C$ (resp.\ $C(D-1)$). Hence, the total carrying capacity of the doublet is $CD$, as in other structures considered here. The ratio of the size of the large deme to that of the small deme is given by $R = D-1$.

\subsection{Early adaptation with the doublet}

As done throughout this work, we assume that mutations appear in an individual chosen uniformly at random. With demes of different sizes, this entails that a single mutant appears in a deme with probability proportional to its size. In the rare migration regime, the fixation probability of such a single mutant in the doublet was analytically determined in Ref.~\cite{Marrec21}, and can thus be used in our simulations and our numerical calculations. 

\begin{figure}[h!]
 \centering
 \includegraphics[width=\textwidth]{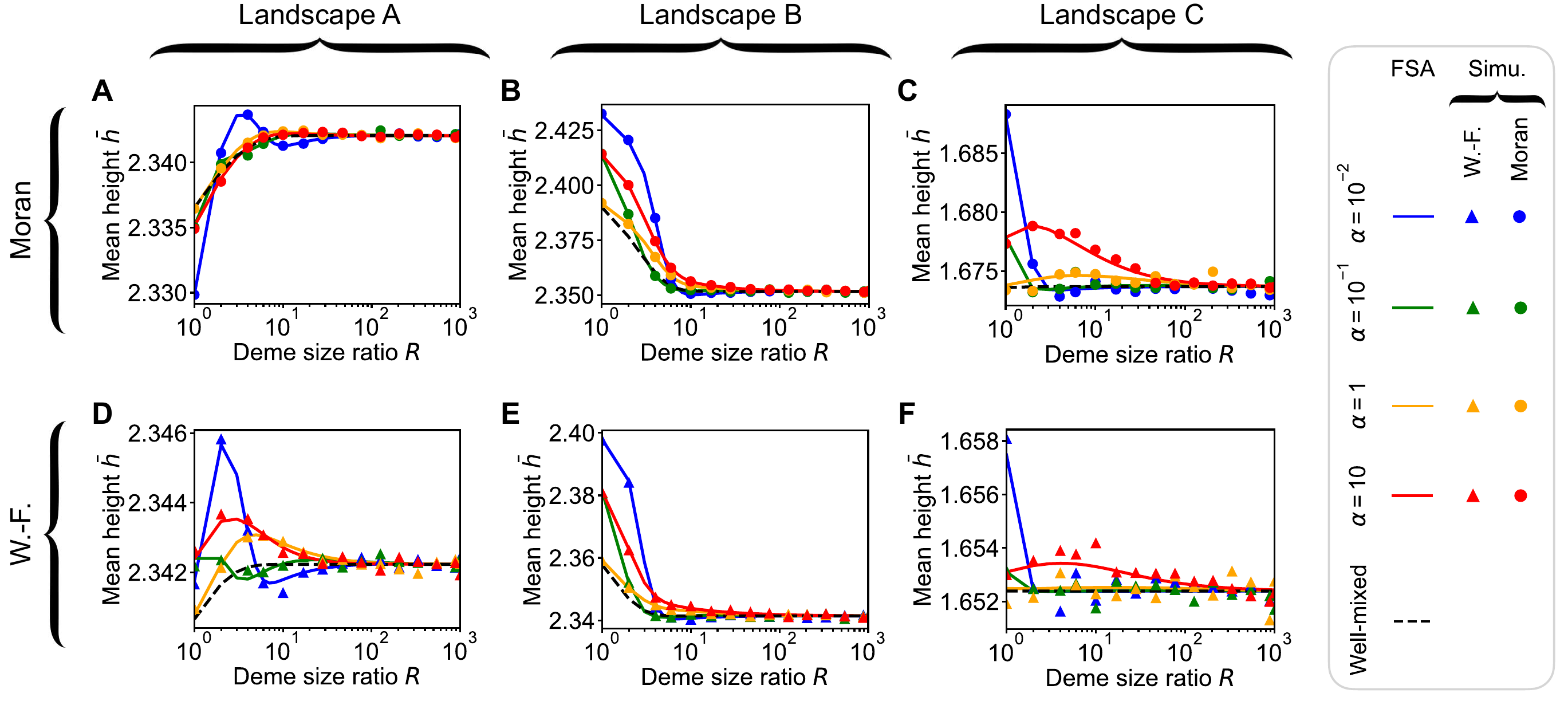}
 \caption{\textbf{Impact of the deme size ratio $R$ and migration asymmetry $\alpha$ of a doublet on early adaptation in  specific landscapes.} (A-C) Mean height $\bar{h}$ of the Moran doublet walk with rare migrations versus $R$ in landscapes A, B and C. (D-F) Same as (A-C) but for the Wright-Fisher doublet walk. Lines: numerical resolutions of the FSA~\cref{eq:systemfsa,eq:barh}; markers: simulation results averaged over $10^5$ walks per starting genotype. Dashed black line: well-mixed case with population size $DC(1-g/f_W)$ where $f_W$ is the fitness of the wild type, shown as reference. Parameter values (all panels): $C = 20$ and $ g = 0.01$.}
\label{fig:h_vs_D_doublet}
\end{figure}

\cref{fig:h_vs_D_doublet} shows the impact of the ratio $R$ of the sizes of the demes and of migration asymmetry $\alpha = m_S / m_L$ on the mean height $\bar{h}$ of the first fitness peak reached in landscapes A, B and C. Here, $R$ is varied by varying $D$ in the doublet with deme sizes $C$ and $C(D-1)$, i.e.\ by varying the size of the large deme while keeping the size of the small deme constant.

\newpage

\section{Beyond rare migrations}

\begin{figure}[h!]
 \centering
 \includegraphics[width=0.9\textwidth]{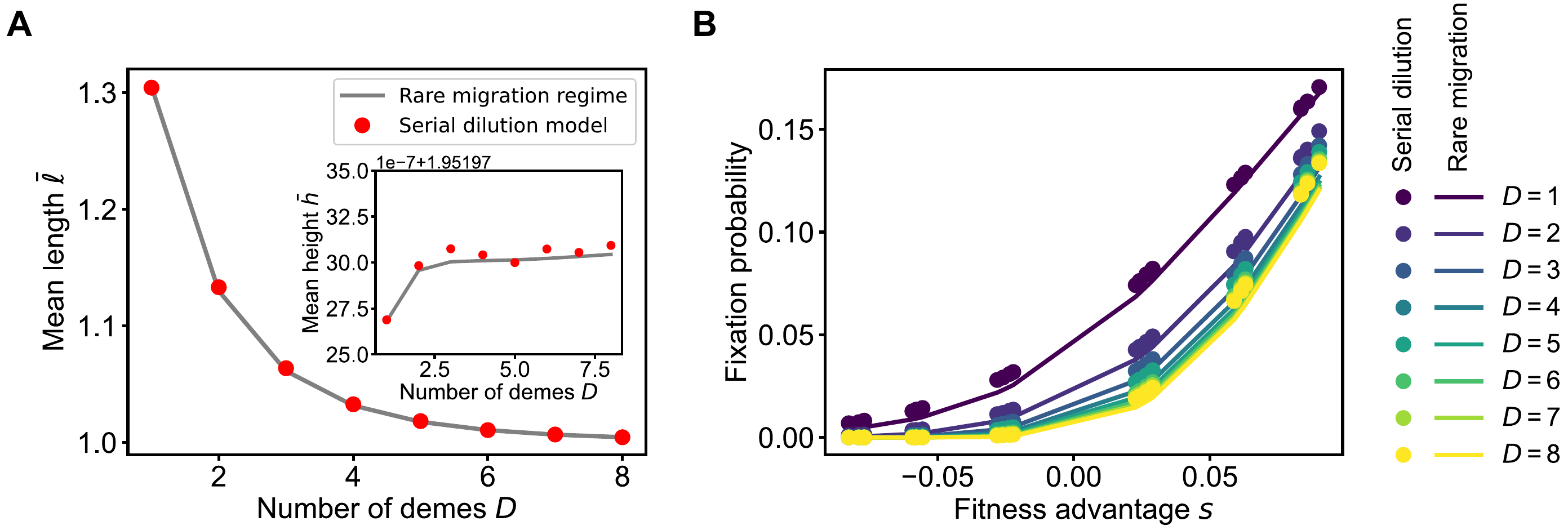}
 \caption{\textbf{Comparison between the Wright-Fisher star walk and the star walk in the serial dilution model, when migrations are rare.} (A) The mean length $\bar{\ell}$ (and mean height $\bar{h}$, see inset) of a walk before reaching a fitness peak when starting from a uniformly chosen initial genotype are shown versus $D$ in an $LK$ landscape with $L = 3$ and $K=1$. This landscape was chosen among the generated ones because all mutations have small fitness effects (absolute relative selective advantages $|s|<0.1$ for all possible mutations), so that the diffusion approximation is appropriate. Recall that this approximation is used in our treatment of the Wright-Fisher walk with rare migrations, see main text. Gray lines correspond to numerical resolutions of the FSA \cref{eq:systemfsa,eq:barh} for the Wright-Fisher star walk with rare migrations (the main approach employed in this work), while red markers correspond to the walk under our structured Wright-Fisher model, i.e.\ the serial dilution model of Ref.~\cite{abbara2023frequent} with growth time $t=1$. Specifically, the markers are obtained via numerical resolutions of the FSA~\cref{eq:systemfsa,eq:barh}, using direct simulations of the structured Wright-Fisher model to estimate fixation probabilities (at least $10^6$ replicates). Recall that in our usual approach, the analytical formula of the fixation probability in the structured population in the rare migration regime~\cite{Marrec21}, using the diffusion approximation to analytically express the fixation probability in a deme under the Wright-Fisher model. (B) Fixation probabilities versus relative mutant fitness advantage $s$, for different values of $D$. Lines and symbols are obtained as in (A). In addition, here, each marker corresponds to one value of $s$ in the landscape considered in panel A. In all panels, parameter values are $C = 20$, $\alpha = 0.1$ and $g = 0.01$. In the structured Wright-Fisher model, we chose migration rates $m_I=10^{-5}$ and $m_O=10^{-4}$.}
\label{fig:landscape_with_small_s}
\end{figure}

\begin{figure}[h!]
 \centering
 \includegraphics[width=0.9\textwidth]{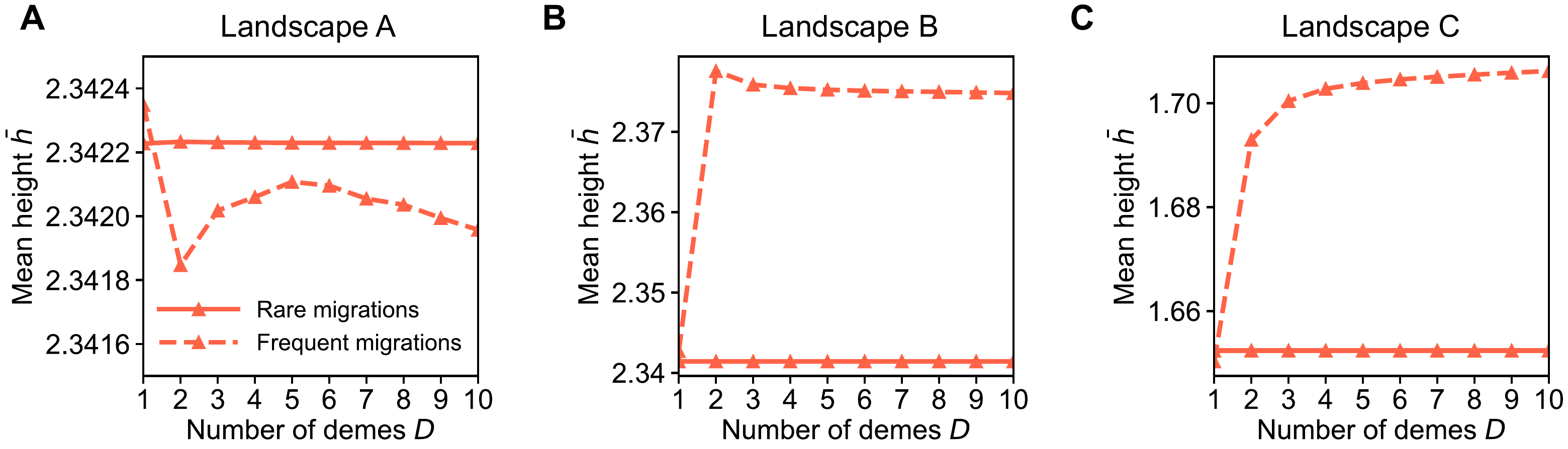}
 \caption{\textbf{Impact of the number of demes $D$ of a star on early adaptation in three $LK$ landscapes with $L=3$ and $K=1$ beyond the rare migration regime.} Same as \cref{fig:serial_dilution} in the main text, but with larger demes: $C = 200$ instead of $C = 20$, and larger asymmetry: $\alpha = 10$.}
\label{fig:serial_dilution_C=200}
\end{figure}

In \cref{fig:landscape_with_small_s}, we check that, for small migration rates, our results from the structured Wright-Fisher model from~\cite{abbara2023frequent} are consistent with those we obtained in the rare-migration regime, assuming the Wright-Fisher model in the diffusion approximation within demes. Next, \cref{fig:serial_dilution_C=200} shows the mean height $\bar{h}$ versus the number of demes $D$, as \cref{fig:serial_dilution} in the main text, but with larger demes and larger asymmetry.

\newpage

\section{Beyond small model landscapes}
\label{sec:beyondsmallLK}

\begin{figure}[h!]
 \centering
 \includegraphics[width=0.7\textwidth]{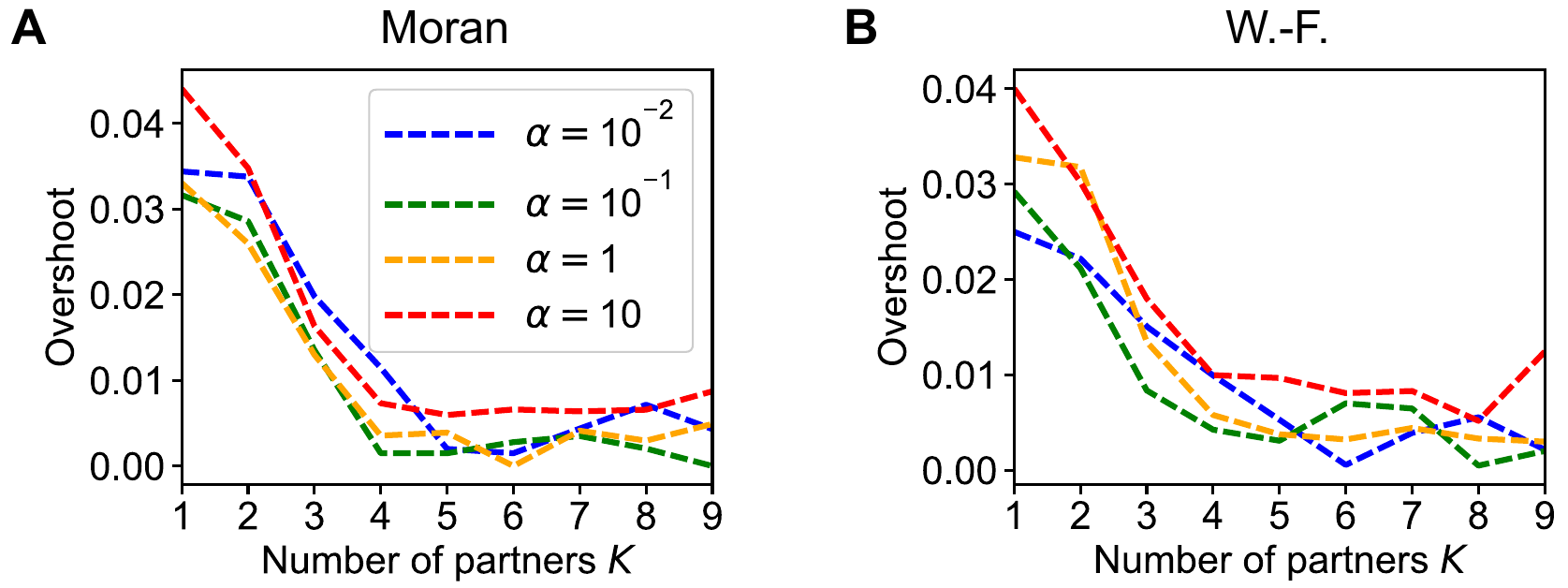}
 \caption{\textbf{Impact of the number of partners $K$ on early adaptation in $LK$ landscapes with $L = 10$.} (A) The overshoot of the ensemble mean height $\left<\bar{h}\right>$ of the first fitness peak reached when starting from a uniformly chosen initial genotype is shown versus $K$ for the Moran star walk with rare migrations, with various values of $\alpha$. The overshoot of $\left<\bar{h}\right>$ is defined as $\left<\bar{h}\right>_\textrm{max}-\left<\bar{h}\right>_{D\to\infty}$, where $\left<\bar{h}\right>_\textrm{max}$ is the maximum value of $\left<\bar{h}\right>$ versus $D$, while $\left<\bar{h}\right>_{D\to\infty}$ is the large-$D$ limit of $\left<\bar{h}\right>$ (evaluated at $D = 10^5$ in practice), see~\cite{servajean2023}. (B) Same as (A) but for the Wright-Fisher star walk. In both cases, $L = 10$, $C = 20$ and $g = 0.01$.}
\label{fig:impact_of_K}
\end{figure}

In \cref{fig:impact_of_K}, we consider fitness landscapes in the $LK$ model with $L=10$ and variable number of partners $K$.

\begin{figure}[h!]
 \centering
 \includegraphics[width=0.99\textwidth]{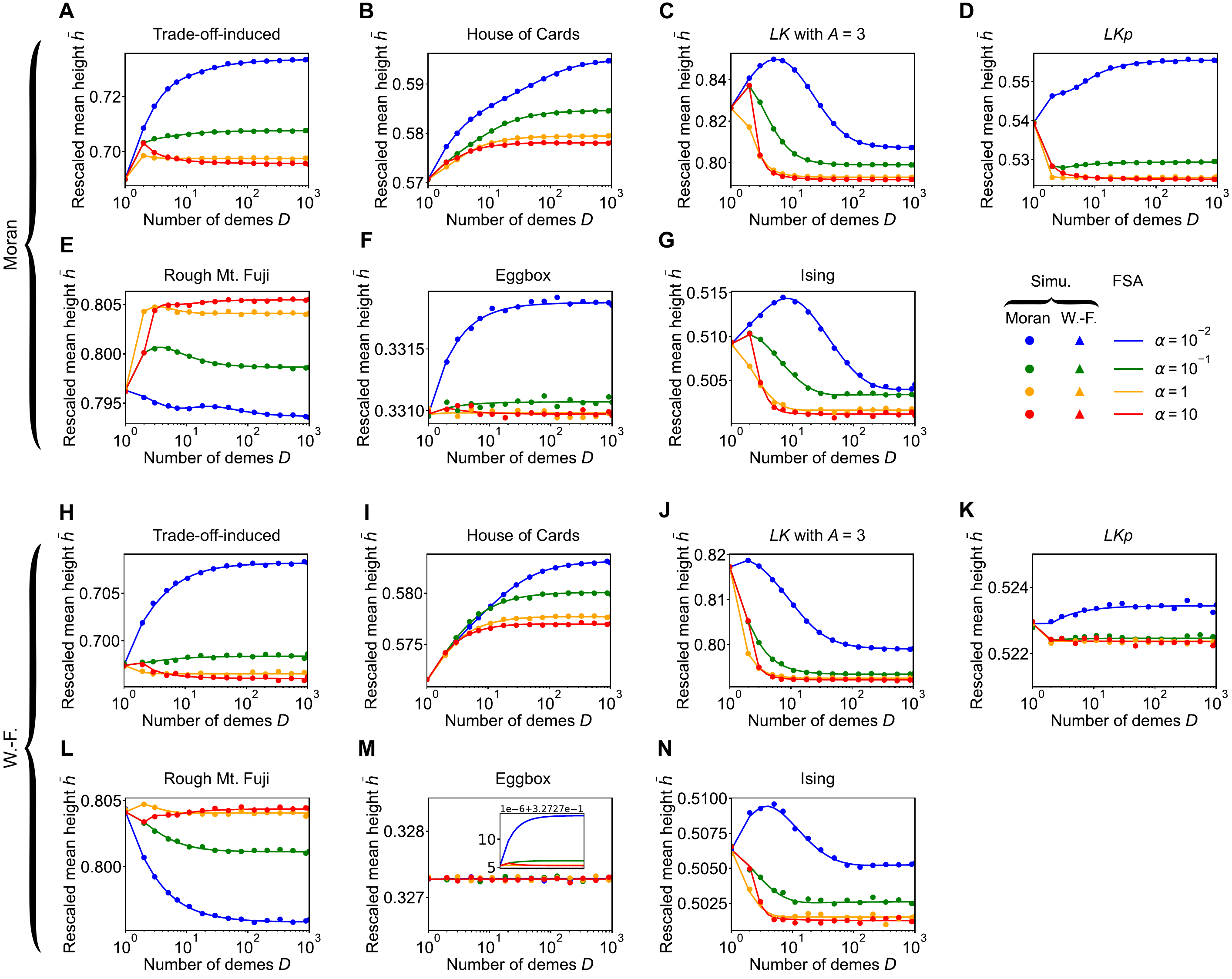}
 \caption{\textbf{Impact of the number of demes $D$ and migration asymmetry $\alpha$ of a star on early adaptation in various model landscapes.} (A) The rescaled mean height of the first fitness peak reached when starting from a uniformly chosen initial genotype is shown versus $D$ for the Moran star walk under rare migrations for various values of $\alpha$, in a tradeoff-induced landscape \cite{das2020, das2022driven}. The rescaled mean height is defined as $(\bar{h}-h_\textrm{min})/(h_\textrm{max}-h_\textrm{min})$, where $h_\textrm{min}$ (resp.\ $h_\textrm{max}$) denotes the fitness of the lowest peak (resp.\ highest peak). Recall that $h_\textrm{min}<\bar{h}<h_\textrm{max}$. As in \cref{fig:beyond_LK_L=3}, the tradeoff-induced landscape was generated with $c = 50$, $a = 1$, and $r_i$ and $m_i$ drawn from the same distribution as in \cite{servajean2023}. (B-G) Same as (A) but in a landscape generated by (B) the House of Cards model \cite{Kauffman1987, kingman1978}, (C) the $LK$ model with non-binary genotypes (with alphabet size $A = 3$) \cite{zagorski2016}, (D) the modified $LKp$ model with $p = 0.5$ and $q = 0.1$ \cite{barnett1998}, (E) the Rough Mount Fuji model \cite{aita2000, szendro2013b} with $f_0' = 6$, $C = 0.5$, and where the epistatic contributions to fitness are drawn from a standard normal distribution, (F) the Eggbox model \cite{ferretti2016} where fitness are drawn from a Gaussian with mean $ 6 \pm 2$ and a standard deviation of 1, and (G) the Ising model \cite{ferretti2016, diu1989} with $B = 6$ \cite{servajean2023} and where the couplings are drawn from a standard normal distribution. See section S4 in the supplementary material of Ref.~\cite{servajean2023} for more details on these models. (H-N) Same as (A-G) but for the Wright-Fisher star walk with rare migrations. In all panels, lines are numerical resolution of the FSA~\cref{eq:systemfsa,eq:barh}, while markers are simulation results averaged over at least $10^5$ walks per starting point. Parameter values in all panels: $L = 6$, $C = 20$ and $ g = 0.01$.  }
\label{fig:beyond_LK}
\end{figure}

In \cref{fig:beyond_LK}, we consider large fitness landscapes in other models than the $LK$ model.
We show the rescaled mean height $\bar{h}$ of the first fitness peak reached versus the number $D$ of demes in landscapes generated by various models for genomes of length $L = 6$. 
In most cases, smaller $\alpha$ yield larger values of $\bar{h}$ for large $D$. This is in line with our findings for small landscapes and for larger $LK$ landscapes (see main text), and confirms their broad validity. As in the case of small landscapes (see \cref{fig:beyond_LK_L=3}), there are exceptions, for example the Rough Mount Fuji landscape of \cref{fig:beyond_LK}(E,L). 
Besides, as for small landscapes and for larger $LK$ landscapes, we find that the value of $D$ from which $\bar{h}$ plateaus becomes larger as $\alpha$ decreases, i.e.\ suppression of selection enhances finite-size effects. When $\alpha = 10^{-2}$, in two landscapes, $\bar{h}$ versus $D$ displays a broad maximum (see \cref{fig:beyond_LK}(C,J) and (G,N), which respectively corresponds to a non-binary $LK$ landscape and an Ising landscape). Thus, in these landscapes, there is a finite value of $D$ that makes the search for high fitness peaks more efficient.

\clearpage
\newpage
\section{Steady-state distributions}\label{sec:effective_pop}

In this section, we focus on the steady state of the walks of structured populations on fitness landscapes. As mentioned in Section~``Model and methods'' of the main text, the Markov chains corresponding to our walks in fitness landscape possess a unique steady state towards which they converge from any starting genotype, since they are irreducible, aperiodic and positive recurrent \cite{aldous2002, norris}. 
By definition, the stationary probability $\pi_\theta$ that the population is in state $\theta$ satisfies
\begin{equation}  \label{eq:global-balance}
\sum_{\beta\in H} \pi_\theta P_{\theta\beta}=\sum_{\beta\in H}\pi_\beta P_{\beta\theta}\,,
\end{equation}

where $P_{\theta\beta}$, which denotes an element of the transition matrix $P$, corresponds to the transition probability from $\theta$ to $\beta$ upon one mutation event, and $H$ is the set of all genotypes (i.e.\ all vertices of the hypercube). 

We start by recalling the steady-state distribution for a well-mixed population in the Moran model and its derivation~\cite{sella2005}. Then, we extend this derivation to our spatially structured populations on graphs in the rare migration regime in some limits. We obtain explicit steady-state distributions in these limits. Comparing them to the well-mixed reference allows us to define effective population sizes associated to these steady-state distributions.

\subsection{Reference: well-mixed population} 

For a well-mixed population of size $N$ in the Moran model, Ref.~\cite{sella2005} showed that the stationary distribution can be expressed explicitly as: 
\begin{equation}
    \pi_\theta^{\textrm{Moran}} = \frac{f_\theta^{N-1}}{\sum_{\delta\in H} f_\delta^{N-1}}\,,
    \label{eq:stat_dist_wm}
\end{equation}
where $f_\theta$ denotes the fitness of genotype $\theta$. 

For completeness, let us recall the proof that $\pi_\theta^{\textrm{Moran}}$ in \cref{eq:stat_dist_wm} is the unique stationary distribution (see Ref.~\cite{sella2005}). Consider two different genotypes $\theta$ and $\beta$, and denote genome dimension by $L$. Upon a given mutation event, we have 
\begin{equation}
    P_{\theta\beta} = \frac{1}{L}\times \rho_{\theta\beta}\,,
    \label{eq:tran-wm}
\end{equation}
if $\theta$ and $\beta$ are neighboring genotypes, and $P_{\theta\beta} =0$ otherwise. 
The first term of this product is the probability that the mutation that occurs is toward genotype $\beta$. Indeed, there are $L$ possible mutations starting from a given genotype, and they are all equally likely. 
The second term, $\rho_{\theta\beta}$, is the probability that a mutant of type $\beta$ fixes in a well-mixed population where the wild type is $\theta$, see \cref{eq:fix_deme_moran} in the main text. Hence, for two neighboring genotypes $\theta$ and $\beta$, we can write: 
\begin{equation}
 \frac{P_{\theta\beta}}{P_{\beta\theta}}=\frac{\rho_{\theta\beta}}{\rho_{\beta\theta}}=\frac{f_\beta^{N-1}}{f_\theta^{N-1}}\,. 
 \label{eq:ratio-deme}
\end{equation}
Therefore, for any two neighboring genotypes $\theta$ and $\beta$, the probability distribution $\pi_\theta^{\textrm{Moran}}$ in \cref{eq:stat_dist_wm} satisfies 
\begin{equation}
\pi_\theta P_{\theta\beta}= \pi_\beta P_{\beta\theta} \,.
   \label{eq:balance}
\end{equation}
Furthermore, \cref{eq:balance} also holds for two non-neighboring genotypes, since both terms vanish, and for two identical genotypes, since both terms are identical. It thus holds for any pair of genotypes $(\theta,\beta)$. Summing \cref{eq:balance} over all genotypes $\beta$ shows that $\pi_\theta^{\textrm{Moran}}$ in \cref{eq:stat_dist_wm} satisfies \cref{eq:global-balance}, and is thus the unique stationary distribution, as announced. 

\cref{eq:balance} is known as the detailed balance condition, and is stronger than the global balance condition in \cref{eq:global-balance}. Since \cref{eq:balance} is satisfied by the stationary distribution, the Markov chain is reversible \cite{sella2005, mccandlish2018}. Note however that the Wright-Fisher origin-fixation dynamics beyond the diffusion approximation (which we do not consider here) is not reversible~\cite{manhart2012}.

\subsection{Structured populations with rare and strongly asymmetric migrations}

Let us consider our spatially structured populations on graphs in the rare migration regime. We will show that, in some limits, we can follow the line of reasoning presented in Ref.~\cite{sella2005} for the well-mixed population, and summarized in the previous section. This allows us to obtain explicit formulas for stationary distributions. Each spatial structure is associated with a specific matrix $P$. We focus on Moran walks in these structures (i.e., the fixation probability within a deme is the Moran fixation probability; see Section~``Model and methods'' in the main text).

\paragraph{Circulations.} Let us consider a population with current genotype $\theta$, structured as a circulation with $D$ demes that each have carrying capacity $C$. In the rare migration regime, the transition probability from genotype $\theta$ to a neighboring genotype $\beta$ upon one mutation event can be written as:
\begin{equation}
    P_{\theta\beta} = \frac{1}{L}\times \phi_{\theta\beta}\,.
\end{equation}
This equation is analogous to \cref{eq:tran-wm}, except that $\phi_{\theta\beta}$ is the probability that a mutant of type $\beta$ fixes in the circulation. As mentioned in Section~``Model and methods'' in the main text, for rare migrations, it can be expressed as~\cite{Marrec21}: 
\begin{equation}
    \phi_{\theta\beta}=\rho_{\theta\beta}\times \frac{1-\gamma}{1-\gamma^D}\,,\,\,\,\textrm{with}\,\,\,\gamma = \frac{\rho_{\beta\theta}}{\rho_{\theta\beta}}\,,
\end{equation}
where $\rho_{\theta\beta}$ denotes the probability that a mutant of type $\beta$ fixes in a deme where all other individuals are of type $\theta$. 
Assuming that $g\ll f_\alpha$ for any genotype $\alpha$, where $g$ is the death rate, we can approximate the stationary population sizes of demes by their carrying capacity $C$. We neglect the fluctuations around steady-state population sizes, which are small in the regime of interest where $C\gg 1$ and $g\ll f_\alpha$, and we use the Moran process to describe fixation in the demes (see Section~``Model and methods'' in the main text). This yields:
\begin{equation}
    \rho_{\theta\beta} = \frac{1-r}{1-r^C}\,,\,\,\,\textrm{with}\,\,\,r=\frac{f_\theta}{f_\beta}\,,
\end{equation}
and $\gamma = r^{C-1}$. 
Thus, we can write:
\begin{equation}
    \frac{P_{\theta\beta}}{P_{\beta\theta}} = \gamma^{-1} \frac{1-\gamma}{1-\gamma^{-1}} \frac{1-\gamma^{-D}}{1-\gamma^D} = \gamma^{-D}=r^{-(C-1)D}=\frac{f_\beta^{(C-1)D}}{f_\theta^{(C-1)D}}\,.
    \label{eq:ratio_clique}
\end{equation}
Therefore, the probability distribution given by
\begin{equation}
    \pi_\theta^\textrm{circ.} = \frac{f_\theta^{(C-1)D}}{\sum_{\delta\in H} f_\delta^{(C-1)D}}\,,
    \label{eq:stat_dist_clique}
\end{equation}
for each genotype $\theta$, satisfies the detailed balance condition \cref{eq:balance}, and is the stationary distribution.

\paragraph{Star.} Let us now consider the star in the rare migration regime. 

First, in the limit $\alpha \rightarrow 0$, where $\alpha$ denotes migration asymmetry (see Section~``Model and methods'' in the main text), the transition probability between two neighboring genotypes $\theta$ and $\beta$ simplifies to:
\begin{equation}
   P_{\theta\beta} = \frac{1}{L}\times \rho_{\theta\beta} \times \frac{1}{D}\,.
    \label{eq:star_alpha_small}
\end{equation}
Indeed, the fixation probability starting from one fully mutant deme then tends to $1/D$ \cite{Marrec21}. Thus, the ratios of transition probabilities are given by \cref{eq:ratio-deme} with $N=C$. Following the same line of reasoning as above leads to a stationary distribution
\begin{equation}
    \pi_\theta^{\textrm{star,} \,\, \alpha \rightarrow 0} = \frac{f_\theta^{C-1}}{\sum_{\delta\in H} f_\delta^{C-1}}\,.
    \label{eq:stat_dist_star_small_alpha}
\end{equation}
This coincides with the stationary distribution for a well-mixed population with size $C$ in the Moran model, see \cref{eq:stat_dist_wm}.

Second, if $\alpha \rightarrow \infty$, using the corresponding limit of the fixation probability starting from one fully mutant deme \cite{Marrec21} leads to the transition probability
\begin{equation}
   P_{\theta\beta} = \frac{1}{L} \times \rho_{\theta\beta} \times \frac{D-1}{D}\frac{1-\gamma^2}{1-\gamma^{2(D-1)}}\,,
    \label{eq:star_alpha_big}
\end{equation}
which leads to
\begin{equation}
   \frac{P_{\theta\beta}}{P_{\beta\theta}} = \frac{f_\beta^{C-1}}{f_\theta^{C-1}}\,\gamma^{4-2D}=\frac{f_\beta^{(C-1)(2D-3)}}{f_\theta^{(C-1)(2D-3)}}\,,
    \label{eq:ratio_starinf} 
\end{equation}
where we used $\gamma = r^{C-1}$. Therefore, the probability distribution given by

\begin{equation}
        \pi_\theta^{\textrm{star,} \,\, \alpha \rightarrow \infty} = \frac{f_\theta^{(C-1)(2D-3)}}{\sum_{\delta\in H} f_\delta^{(C-1)(2D-3)}}\,,
        \label{eq:stat_dist_star_big_alpha}
\end{equation}
for each genotype $\theta$, satisfies the detailed balance condition \cref{eq:balance}, and is the stationary distribution.

\paragraph{Line.} We showed above that, in the line with rare migrations in the limit of strong migration asymmetries, the fixation probability starting from one fully mutant deme tends to $1/D$, see \cref{eq:symmetry_line}. Thus, the stationary distribution is given by \cref{eq:stat_dist_star_small_alpha}, as for the star in the limit $\alpha \rightarrow 0$.

\paragraph{Doublet.}

Let us now consider the doublet structure (see \cref{sec:doublet}) in the limits of extreme migration asymmetries. 

When $\alpha = m_S / m_L \rightarrow 0$, Eq.~S92 in the supplementary material of Ref.~\cite{Marrec21} shows that the doublet has the same mutant fixation probability as the star in the limit $\alpha \rightarrow 0$. Therefore, the stationary distribution for the doublet in the limit $\alpha \rightarrow 0$ is given by \cref{eq:star_alpha_small}, and the effective size associated to it is that of the small deme, $C$.

When $\alpha \rightarrow \infty$, Eq.~S93 in the supplementary material of Ref.~\cite{Marrec21} shows that the fixation probability in the doublet, starting from one mutant, reduces to $(D-1) \rho_M^L/D$, where $\rho_M^L$ is the fixation probability within the large deme, starting from one mutant. As in \cref{sec:effective_pop}, we make the approximation that the size of the small deme is exactly $C$ and that the size of the large deme is exactly $C(D-1)$. This yields the following transition probability between two neighboring genotypes $\theta$ and $\beta$:
\begin{equation}
    P_{\theta \beta} = \frac{1}{L} \times \frac{D-1}{D} \frac{1-r}{1-r^{C(D-1)}}\,,
\end{equation}
which leads to
\begin{equation}
   \frac{P_{\theta\beta}}{P_{\beta\theta}} =\frac{f_\beta^{C(D-1)-1}}{f_\theta^{C(D-1)-1}}\,.
\end{equation}
Therefore, the probability distribution given by
\begin{equation}
        \pi_\theta^{\textrm{doublet,} \,\, \alpha \rightarrow \infty} = \frac{f_\theta^{C(D-1)-1}}{\sum_{\delta\in H} f_\delta^{C(D-1)-1}}\,,
        \label{eq:stat_dist_doublet_large_alpha}
\end{equation}
for each genotype $\theta$, satisfies the detailed balance condition \cref{eq:balance}, and is the stationary distribution.

\subsection{Steady-state effective population sizes for rare and strongly asymmetric migrations} 
\label{subsec:effsize}

\paragraph{Calculation of steady-state effective population sizes.} Comparing \cref{eq:stat_dist_clique,eq:stat_dist_star_small_alpha,eq:stat_dist_star_big_alpha,eq:stat_dist_doublet_large_alpha} to the well-mixed case in \cref{eq:stat_dist_wm} reveals a strong formal similarity. Specifically, only the exponent of the fitness differs. This leads us to define an effective population size for the steady state, as the size of a well-mixed population that would display the same stationary distribution. Table~1 in the main text lists the effective population sizes corresponding to the main spatial structures we studied here. In addition, for the doublet, the effective population size is $C$ (resp.\ $C(D-1)$) in the limit $\alpha \rightarrow 0$ (resp.\ $\alpha \rightarrow \infty$), which corresponds to the size of the small (resp.\ large) deme.

While we focused on Moran walks here, similar derivations could be performed for Wright-Fisher walks by taking the approximation from Ref.~\cite{sella2005} for the fixation probability within a deme, instead of the classical expression used in the main text (see Ref.~\cite{sella2005}).

\paragraph{Consequence for the mean steady-state fitness.} Working with the steady-state distribution for a well-mixed population of fixed size $N$, given by \cref{eq:stat_dist_wm}, Ref.~\cite{servajean2023} showed that the mean steady-state fitness $F$ in a given fitness landscape, given by
\begin{equation}
 F =\sum_{\delta \in H} f_\delta \pi_\delta\,,
\label{eq:stationary_mean_fit}
\end{equation}
monotonically increases with population size $N$ (see Supplementary Material section S1.2 of~\cite{servajean2023}). Therefore, the mean steady-state fitness is larger for those of our spatial structures that have large effective population sizes than for those that have smaller ones. 

\subsection{Impact of migration asymmetry on steady-state effective sizes}

\begin{figure}[h!]
 \centering
 \includegraphics[width=0.95\textwidth]{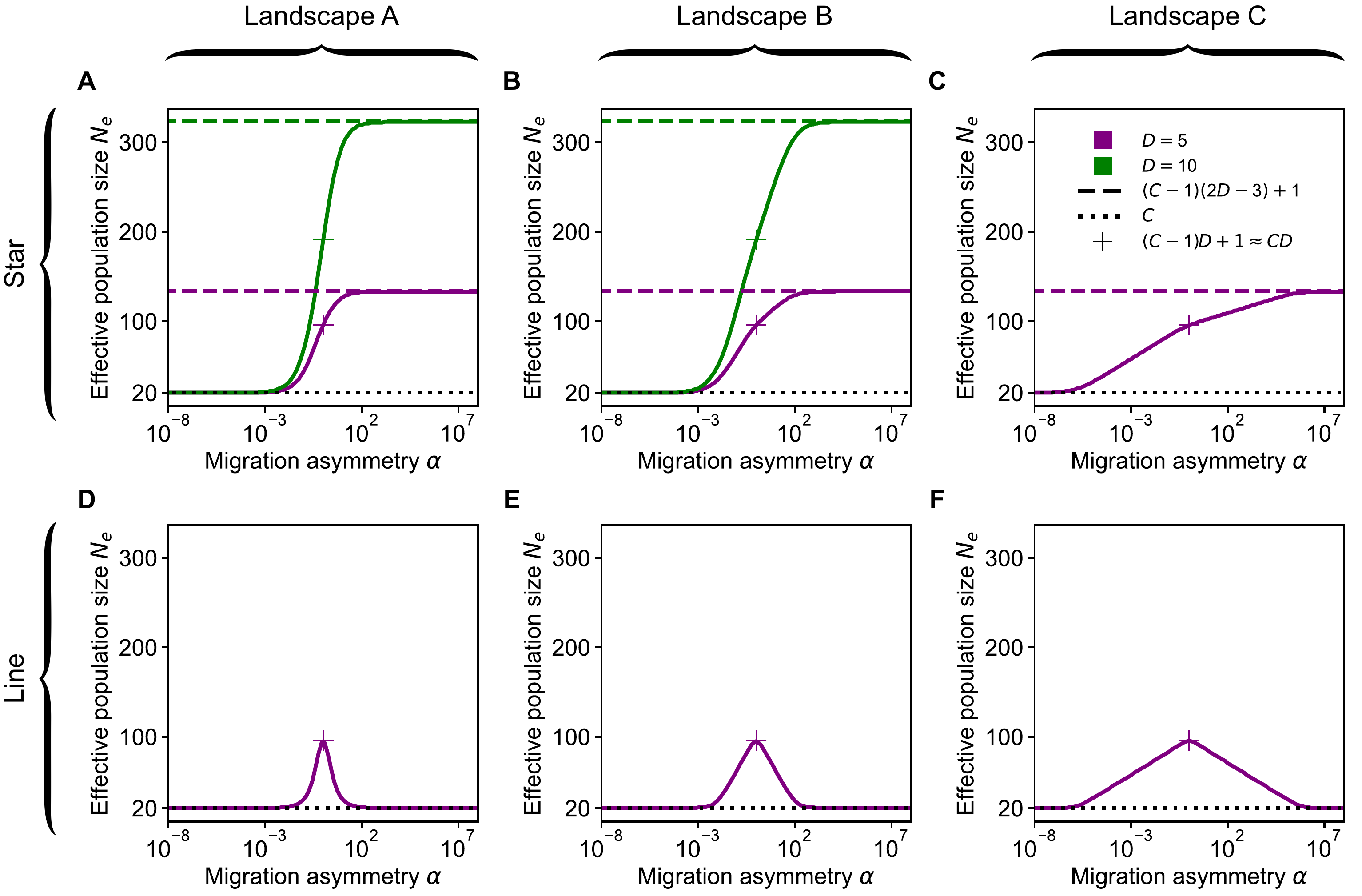}
 \caption{\textbf{Impact of migration asymmetry on steady-state effective population size in three $LK$ landscapes with $L=3$ and $K=1$.} (A-C) Steady-state effective population size $N_e$ of the star with rare migrations versus migration asymmetry $\alpha$ in landscape A (A), B (B) and C (C) (see \cref{fig:h_vs_D}(A-C)) with $D = 5$ (green) and $D = 10$ (purple). (D-F) Same as (A-C) but for the line with $D = 5$.  Horizontal dashed and dotted lines: analytical asymptotic values; plus markers: analytical prediction for circulations (see Table~1 in the main text). Solid curves: numerical results. To obtain them, we developed the following method. First, in the fitness landscape of interest, we compute the steady-state distribution, using the eigenvector associated with eigenvalue 1 of the transition matrix of the Markov process describing the population's evolution \cite{o1993}. We do this (i) for the structured population we consider, and (ii) for well-mixed populations with various population sizes $N$. Second, we compare the distributions obtained for (i) and (ii), using the Jensen-Shannon divergence. Finally, we estimate the steady-state effective population size $N_e$ as the value of $N$ that minimizes the Jensen-Shannon divergence. 
 Note that in landscape C, for the star with $D=10$ and large $\alpha$, the population is almost always on the large peak, which makes it numerically challenging to estimate $N_e$. This is why this case is not shown in panel (C).  In addition, in panels (D-F), because computing the fixation probability in the line is computationally challenging when $D$ is large (see \cref{eq:p_fixation_line_expanded}), we only show the results for $D = 5$.  Parameter values (in all panels): $C = 20$ and $ g = 0.01$.
  }
\label{fig:Ne_vs_alpha}
\end{figure}

While our analytical approach only provides results for extreme values of $\alpha$ in the star and in the line, and there is no guarantee that a steady-state effective population size rigorously exists beyond these cases, we numerically estimate it for various values of $\alpha$.  
\cref{fig:Ne_vs_alpha} shows the numerically estimated steady-state effective population size $N_e$ of the star and the line versus $\alpha$. The asymptotic values are in very good agreement with our analytical predictions for the star and for the line, and the result for $\alpha = 1$ is in good agreement with our analytical prediction for circulations (see \cref{table:effective_pop}). For the star, we observe that $\alpha>1$ and amplification of selection are associated to steady-state effective population sizes larger than population size. The opposite holds when $\alpha<1$ both in the star and in the line, which is associated to suppression of selection.
We further observe in \cref{fig:Ne_vs_alpha} that the range of values of $\alpha$ leading to intermediate effective population sizes depends on the landscape. In particular, both in the star and in the line, the transition between the small-$\alpha$ and the large-$\alpha$ regime is broader in landscape C than in landscapes A and B.

\clearpage
\newpage

\bibliographystyle{unsrt} %NATBIB
\bibliography{bib}

\end{document}